\documentclass[fleqn,usenatbib]{mnras}

\usepackage{newtxtext,newtxmath}

\usepackage[T1]{fontenc}

\DeclareRobustCommand{\VAN}[3]{#2}
\let\VANthebibliography\thebibliography
\def\thebibliography{\DeclareRobustCommand{\VAN}[3]{##3}\VANthebibliography}

\usepackage{graphicx,xspace}	
\usepackage{amsmath}	
\usepackage{verbatim}
\usepackage{hyperref}
\usepackage{pifont}
\usepackage{array}
\usepackage{multirow}
\usepackage[utf8]{inputenc}

\newcommand{\ic}{IceCube\xspace}
\newcommand{\icG}{IceCube-Gen2\xspace}
\newcommand{\ant}{ANTARES\xspace}
\newcommand{\arca}{KM3NeT/ARCA\xspace}
\newcommand{\gx}{GX~339--4\xspace}
\newcommand{\cyg}{Cyg~X--1\xspace}

\newcommand{\gr}{$\gamma$-ray\xspace}
\newcommand{\grs}{$\gamma$-rays\xspace}
\newcommand{\g}{$\gamma$\xspace}
\newcommand{\pg}{p$\gamma$\xspace}
\newcommand{\integral}{\textit{INTEGRAL}\xspace}
\newcommand{\hess}{H.E.S.S.\xspace}

\newcommand{\bh}{BHXB\xspace}
\newcommand{\bhs}{BHXBs\xspace}
\newcommand{\hm}{HMXB\xspace}
\newcommand{\lm}{LMXB\xspace}
\newcommand{\hms}{HMXBs\xspace}
\newcommand{\lms}{LMXBs\xspace}

\defcitealias{kantzas2020cyg}{K21}
\newcommand{\kcyg}{{\citetalias{kantzas2020cyg}}\xspace}
\defcitealias{kantzas2022gx}{K22}
\newcommand{\kgx}{{\citetalias{kantzas2022gx}}\xspace}
\defcitealias{cooper2020xrbcrs}{C20}
\newcommand{\coop}{{\citetalias{cooper2020xrbcrs}}\xspace}

\title[CRs and neutrinos from \bh jets]{Possible contribution of X-ray binary jets to the Galactic  cosmic ray and neutrino flux
}

\author[D. Kantzas et al.]{
D. Kantzas,$^{1,2,3}$\thanks{E-mail: dimitrios.kantzas@lapth.cnrs.fr}
S. Markoff,$^{2,3}$
A. J. Cooper,$^{4,2}$
D. Gaggero,$^{5}$
M. Petropoulou$^{6,7}$ \&
P. De La Torre Luque$^{8}$
% ...
\\
$^{1}$Laboratoire d'Annecy-le-Vieux de Physique Th\'{e}orique (LAPTh), CNRS, USMB, F-74940 Annecy, France\\
$^{2}$Anton Pannekoek Institute for Astronomy (API), University of Amsterdam, Science Park 904, 1098 XH Amsterdam, the Netherlands\\
$^{3}$GRavitational AstroParticle Physics Amsterdam (GRAPPA), University of Amsterdam, Science Park 904, 1098 XH Amsterdam, the Netherlands\\
$^4$Astrophysics, The University of Oxford, Keble Road, Oxford OX1 3RH, UK\\
$^5$INFN Sezione di Pisa, Polo Fibonacci, Largo B. Pontecorvo 3, 56127 Pisa, Italy\\ 
$^6$Department of Physics, National and Kapodistrian University of Athens, University Campus Zografos, GR 15783, Athens, Greece\\ 
$^7$Institute of Accelerating Systems \& Applications, University Campus Zografos, Athens, Greece\\
$^{8}$Stockholm University and The Oskar Klein Centre for Cosmoparticle Physics, Alba Nova, 10691 Stockholm, Sweden\\
}

\date{Accepted XXX. Received YYY; in original form ZZZ}

\pubyear{2022}

\begin{document}
\label{firstpage}
\pagerange{\pageref{firstpage}--\pageref{lastpage}}
\maketitle

\begin{abstract}

For over a century, the identification of high-energy cosmic ray (CR) sources remains an open question. For Galactic CRs with energy up to $10^{15}$\,eV, supernova remnants (SNRs) have traditionally been thought the main candidate source. However, recent TeV \gr observations have questioned the SNR paradigm. Propagating CRs are deflected by the Galactic magnetic field, hence, \grs and neutrinos produced via inelastic hadronic interactions are the only means for unveiling the CR sources. In this work, we study the \gr and neutrino emission produced by CRs accelerated inside Galactic jets of stellar-mass black holes in X-ray binaries (\bhs). We calculate the intrinsic neutrino emission of two prototypical \bhs, Cygnus~X-–1 and GX~339-–4, for which we have high-quality, quasi-simultaneous multiwavelength spectra. Based on these prototypical sources, we discuss the likelihood of the 35 known Galactic \bhs to be efficient CR accelerators. Moreover, we estimate the potential contribution to the CR spectrum of a viable population of \bhs that reside in the Galactic plane. When these \bhs go into outburst, they may accelerate particles up to 100s of TeV that contribute to the diffuse \gr and neutrino spectra while propagating in the Galactic medium. Using HERMES, an open-source code that calculates the hadronic processes along the line of sight, we discuss the contribution of BHXBs to the diffuse \gr and neutrino fluxes, and compare these to their intrinsic \gr and neutrino emissions. Finally, we discuss the contribution of \bhs to the observed spectrum of Galactic CRs.

\end{abstract}

\begin{keywords}
cosmic rays -- neutrinos -- acceleration of particles 
\end{keywords}



\section{Introduction}
For more than a hundred years, the origin of cosmic rays (CRs) has remained an unsolved mystery. CRs are accelerated atomic nuclei of extraterrestrial origin that populate a power law that covers more than ten orders of magnitude in energy. Energetic protons of Galactic origin dominate the low-energy regime of the CR spectrum that shows a break at around 1\,PeV ($10^{15}$\,eV), known as the ``knee''. The CR composition beyond the ``knee'' is not certain, but it is likely that the CR spectrum becomes dominated by heavier elements at the highest energy regime up to the ``ankle'' feature at around 1\,EeV ($10^{18}$\,eV; \citealt[][]{buitink2016large,Auger2021spectrum}). Beyond the ``ankle'', different experiments favour different CR composition, and it is not clear yet whether CRs are proton or iron dominated \citep[see e.g.][]{Deligny_2023Auger,ta2022Highlights}.

The astrophysical sources capable of accelerating CRs up to (and beyond) 1\,PeV are the so-called PeVatrons. Based on energetic arguments, supernovae (SNe) and young supernova remnants (SNRs) in particular, have so-far been considered the dominant candidate sources of Galactic CRs \citep[][]{baade1934cosmic,Ginzburg1964,hillas1984origin,Blasi2013origin}. When CRs are accelerated in the SN/SNR shock front, they interact with the dense gas clouds to initiate inelastic collisions that will lead to the formation of secondary particles, such as \grs and neutrinos \citep[e.g.][]{stephens1981production,Dermer1986secondary,BEREZINSKY1991375,mannheim1993proton,mannheim1994interactions,rachen1998photohadronic,Aharonian2000broadband,Blattnig2000Parametrizations,MUCKE2003protonBLLac}. These secondary \grs carry energy that is approximately one tenth of the energy of the primary protons. If CRs were accelerated up to PeV energies, we would hence expect to observe SNe and SNRs up to ultra-high energies above $100\,$TeV (if the sources were optically thin to photon-photon pair production at these energies). However, current TeV observations show that SNRs exhibit a softening or even a cutoff in the \gr spectrum before the $\sim$ hundreds of TeV regime \citep[e.g.][]{Aharonian2006detailed,Aharonian2007primary,Aharonian_2007_observations,Aharonian_2009_discovery,Abramowski2011new,Ackermann2013detection,Abramowski2014HESSJ1640_465,Archambault_2017observations}. Such a spectral feature questions the SN paradigm, and hence theoretical and observational investigation into new candidate sources is ongoing. 

Recent TeV \gr observations suggest that various other Galactic sources can accelerate particles up to TeV and PeV energies. Clusters of young massive stars \citep[][]{Ackermann2011coccon,Abramowski2012discovery,aharonian2018massive},
pulsars \citep{HESSCollaboration2018Vela,MAGICCollaboration2020Geminga}, pulsar wind nebulae \citep[][]{Amenomori2019first,lopez2022gamma}, \gr binaries \citep{Aharonian2005discoveryXRB,Hinton_2008} are among the prime examples of capable sources. The latest observations of the Galactic plane from HAWC \citep[][]{Albert_2020HAWC}, LHAASO \citep[][]{Cao2021PeV} and Tibet-AS$_{\gamma}$ \citep[][]{Amenomori2021first} found no association with SNRs and instead suggest the existence of numerous sources capable of producing \gr emission beyond 1\,PeV with the aforementioned sources being in the list, as well as many of yet unknown nature.

The relativistic jets launched by stellar-mass black holes in X-ray binaries (\bhs) also shine up to \grs. Cygnus~X--3 and Cygnus~X--1 (\cyg) are known to emit GeV radiation \citep[][]{Tavani_2009CygnusX3, Sabatini_2013,malyshev2013high,zanin2016detection}, but it remains unclear whether the origin of this emission is due to the acceleration of electrons \citep[][]{zdziarski2014jet} or protons \citep[][]{pepe2015lepto}. 
There is no persistent TeV emission detected yet from these two sources \citep[][]{Albert_2020HAWC,Albert_2021_HMmicroquasars}, making it impossible to argue against or in favour of any proton acceleration in their jets. Any TeV detection however would be hard to be explained by a leptonic scenario due to Klein-Nishina suppression. 
The well-known case of SS433 was moreover the first BH-candidate found to emit in the TeV regime with the acceleration sites associated with the interaction of its jets with the surrounding medium \citep[][]{abeysekara2018very,Safi-Harb_2022}. More \bhs are now detected by \hess with the TeV emission to be along the jets (Olivera-Nieto~et~al.~in~prep). 

As suggested for the case  of large-scale jets of Active Galactic Nuclei (AGN), when protons accelerate to high energies above a few PeV, they also produce astrophysical neutrinos and antineutrinos via inelastic collisions that carry energy of the order of TeV-PeV \citep[e.g.][]{Murase2014diffuse,Petropoulou2015photohadronic,Aartsen2018Multimessenger,Keivani_2018}. Astrophysical neutrinos are hence the smoking gun of CR acceleration. The current state-of-the-art neutrino detectors \ic and \ant, in the South Pole and the Mediterranean Sea respectively, have not yet confirmed the Galactic origin of any neutrino to favour any particular source as a potential PeV accelerator (also known as PeVatron; \citealt[][]{aartsen201910years}). The diffuse neutrino observations seem to be isotropic, implying that extragalactic sources dominate the detected spectrum at these energies (however see \citealt{Kovalev_2022} and \citealt{Luque2022prospects} for a diffuse neutrino emission from the Galactic plane). The accumulated data after ten years of operation of the \ic experiment, as well as the expected data from the next-generation facilities, such as \arca \citep[][]{aiello2019arca} and \icG \citep[][]{IceCube-Gen2,Aartsen_2021gen2}, offer not only a unique opportunity to set new constraints on individual sources but might also start revealing Galactic and dimmer neutrino point sources.

\bhs have been suggested as neutrino sources in the past \citep[e.g.][]{levinson2001probing,romero2003hadronic,romero2005misaligned,Torres_2005,BEDNAREK2005galactic,Romero2008proton,reynoso2008ss433,Zhang2010neutrinoLMXRBs,carulli2021neutrino,cooper2020xrbcrs} but were neglected for numerous years, mainly due to the lack of \gr and neutrino constraints. 
The increasing number of newly discovered \bhs, which may not be bright enough to be detected in the \gr regime, makes the \bh population a viable candidate to contribute to the Galactic neutrino spectrum, and hence the CR spectrum. In particular, recent X-ray observations of the Galactic centre suggest the existence of hundreds-to-thousands of \bhs in the pc-scale of the Galactic centre \citep{Hailey2018cusp,Mori_2021}. Based on these observations, \citet[][hereafter \coop]{cooper2020xrbcrs} suggest that of the order of a few thousands of \bhs suffice to contribute significantly to the CR spectrum, and possibly dominate the spectrum slightly above the ``knee'', before the transition to the extragalactic origin.

In this work, we revisit the idea of \bhs being Galactic CR sources and hence potential neutrino sources. We tie the multiwavelength emission of \bh jets to the dynamical properties of the acceleration site, and hence we can derive more self-consistently predictions of the neutrino counterparts. We use, in particular, the multizone jet model of \citet[][hereafter \kcyg]{kantzas2020cyg} that accounts for acceleration of both leptons and hadrons. The leptons accelerate to non-thermal energies to emit synchrotron radiation along the jets, being able to reproduce the flat-to-inverted radio spectra detected by multiple \bhs \citep[][]{blandford1979relativistic,hjellming1988radio,falcke1995jet,markoff2001jet,Markoff2005,lucchini2022BHJet}. The accelerated protons interact inelastically with protons in the bulk flow and the jet radiation to produce secondary particles via inelastic proton-proton (pp) and proton-photon (\pg) collisions that contribute to the high-energy regime of the spectrum. We further develop this particular jet model to account for the intrinsic neutrino counterpart due to the above processes.
We further investigate the contribution of \bhs to the CR spectrum, utilising a physical jet model whose parameters we can now better constrain via the multiwavelength observations of ``canonical'' \bhs, such as \cyg (\kcyg) and \gx \citep[][hereafter \kgx]{kantzas2022gx}. In particular, we use the up-to-date \bh WATCHDOG catalogue of \cite{Tetarenko2016} to study the possibility of the intrinsic neutrino emission of these sources to contribute to the observed spectrum, and discuss the likelihood of future facilities, such as \arca, to be able to detect any Galactic neutrinos from \bhs. We further study the contribution of these sources to the CR spectrum when the accelerated CRs escape the acceleration sites and propagate through the interstellar medium towards Earth. We finally discuss the interaction of the propagated CRs with the intergalactic medium to produce diffuse \grs and neutrinos.

In Section~\ref{sec: model}, we discuss the jet model we utilise to produce the intrinsic neutrino fluxes, that we present in Section~\ref{sec: intrinsic neutrino emission} for the known \bhs. In Section~\ref{sec: diffuse emission}, we present our new analysis for a population of 1000 \bhs, in Section~\ref{sec: discussion} we discuss the contribution of \bhs to the CR spectrum, and in Section~\ref{sec: summary and conclusions} we summarise our results and conclude.

\section{Neutrinos from \bh jets}\label{sec: model}

Our physical jet model is based originally on work presented in \cite{Markoff2005} that explains the flat radio spectrum detected in \bhs \citep[see, e.g.,][]{Corbel2002NIR,Fender2000flat,Fender2009jets}. In a multizone stratified jet, particles accelerate to non-thermal energies to produce the multiwavelength spectrum from radio to X-rays \citep[][]{blandford1979relativistic, falcke1995jet,markoff2001jet,markoff2003exploring,Markoff2005,corbel2013formation,russell2019maxij1535571,tetarenko2021measuring,cao2021evidence}. The population of primary leptons accelerates continuously along the jets beyond some specific distance (see below). The accelerated particles form a power-law distribution (in energy) and emit non-thermal synchrotron radiation that is further upscattered to higher energies via inverse Compton scattering \citep[][]{Markoff2005,lucchini2019breaking,lucchini2022BHJet}. \kcyg included for the first time hadronic acceleration along the jets to account for the inelastic pp and \pg interactions that lead to the production of secondary electrons/positrons and \grs. In this work, we further self-consistently calculate the neutrino distributions that form due to the above interactions, as we describe below.

\subsection{Jet dynamics and particle acceleration}\label{sec: jet quantities}
Details of our multizone jet model can be found in, e.g., \citet[][]{Markoff2005,crumley2017symbiosis,lucchini2022BHJet} and \kcyg, but we briefly cover the basic assumptions here for consistency. In particular, during the so-called hard X-ray spectral state, \bhs launch bipolar, mildly relativistic and collimated jets. We assume the jets are launched at some height from the black holes, assuming a relativistically thermal lepton and cold proton composition. We do not account for any particular jet launching mechanism, as both Blandford-Znajek \citep[][]{blandford1977extraction} and Blandford-Payne \citep[][]{Blandford1982hydromagnetic} are viable. We assume that the leptonic population of the jet base follows a Maxwell-J\"{u}ttner distribution that is characterized by its peak temperature. The two particle distributions share the same number density, which we calculate assuming particle-number conservation along the jets. Our multi-zone jet model can account for different jet compositions including leptonic pairs \citep{Kantzas2023MassLoading}, however in this work for simplicity we assume a charge neutral electron/proton jet composition throughout. Finally, we calculate the bulk velocity profile of the jets, as well as the magnetic energy density, by solving the 1D relativistic Bernoulli equation along the jet axis \citep{Koenigl1980,crumley2017symbiosis,lucchini2022BHJet}.  

Whereas the particles are thermal at the jet base and remain thermal for some distance from the launching point, we assume that a fraction of them accelerates into a power law at some specific distance $z_{\rm diss}$ from the black hole and beyond. We assume that particles are accelerated up to some maximum energy that we calculate self-consistently by equating the acceleration timescale to the radiation loss timescales. The power-law index of the non-thermal particle distributions has been the subject of much research for decades, and it strongly depends on the nature of the acceleration site and the acceleration mechanism. Using a simple jet model we cannot capture the exact acceleration mechanism. We therefore take the power-law index $p$ to be a free parameter. We examine in particular two different cases, first assuming a soft power-law index of $p=2.2$, and compare it to a harder one with $p=1.7$. We choose these two more extreme values as they could represent different acceleration mechanisms, where the soft case may favour diffusive shock acceleration \citep[e.g.][]{drury1983introduction,malkov2001nonlinear,sironi2009particle,Caprioli2012nldsa,Caprioli2014i,park2015simultaneous} and the hard case magnetic reconnection in strongly magnetized plasmas (\citealt[][]{Sironi_2014,Guo2014relativisticreconnection,Guo2016}; although also see \citealt[][]{petropoulou2018steady} where they find evidence for a softening in the power law slope at a magnetisation $\sim 50$). Finally, we assume for simplicity that both electrons and protons accelerate into distributions with the same power-law index.

\subsection{Lepto-Hadronic Processes}\label{hadronic processes}
We refer the interested reader to \cite{lucchini2022BHJet} for a detailed description of how we calculate the leptonic radiative processes; i.e., the cyclo-synchrotron emission of either thermal or non-thermal leptons, as well as inverse Compton scattering (ICS). In this work, we describe only the hadronic processes which are responsible for the neutrino counterpart and CR acceleration.

When protons accelerate to high energy, they can interact with particles in the bulk flow and the associated radiation to initiate particle cascades that lead to the formation of secondary electrons/positrons, \grs and neutrinos \citep[][]{rachen1993extragalactic,mannheim1993proton,mannheim1994interactions,rachen1998photohadronic,MUCKE2003protonBLLac}. An alternative hadronic contribution to the electromagnetic spectrum is when accelerated protons emit synchrotron radiation \citep[][]{aharonian2002proton,MUCKE2001121}.

In this work, we self-consistently calculate the neutrino distributions as they emerge from the inelastic pp collisions following the semi-analytical formalism of \cite{kelner2006energy}. We account for both the bulk thermal protons of the flow and the cold protons of the stellar wind of the companion (\kcyg, see Appendix~\ref{app: neutrino flux along the jet}). We use the cross-section of the pp interactions based on \cite{kafexhiu2014parametrization}:
\begin{equation}\label{cross section pp}
\begin{split}
    \sigma_{\rm{pp}}\left(T_{\rm p}\right) = \left[ 30.7 - 0.96\log \left( \frac{T_{\rm p}}{T_{\rm{thr}}} \right) + 0.18\log ^2 \left( \frac{T_{\rm p}}{T_{\rm{thr}}} \right) \right]\\ \times \left[ 1 - \left( \frac{T_{\rm{thr}}}{T_{\rm p}} \right)^{1.9} \right]^3 \,\rm{mb},
\end{split}
\end{equation}
where $T_{\rm p}$ is the proton kinetic energy in the laboratory frame and $T_{\rm{thr}} =2{\rm m_{\pi}+m_{\pi}^2/2m_p} \simeq 0.2797\,$GeV is the threshold for this interaction to take place. The formalism of \cite{kelner2006energy} leads directly to steady state distributions of secondary electrons/positrons, \grs and neutrinos. Therefore, the underlying assumption is that intermediate unstable charged muons and pions immediately decay without radiating. This is an accurate approximation as long as the maximum energy (or Lorentz factor $\gamma_p$ in fact) of the initial accelerated protons fulfils the criterion \citep[][]{boettcher2013leptohadronic}: 
\begin{equation}
    \gamma_p\ll 
    \begin{cases}
    7.8\times 10^{11}/B \rm{~for~pions} \\
    5.6\times 10^{10}/B \rm{~for~muons,} 
    \end{cases}
\end{equation}
where $B$ is the strength of the magnetic field in the jet rest frame in units of Gauss. In the case of \bhs where the magnetic field does not exceed the $10^5\,\rm G$ at the particle acceleration region, and the maximum proton energy does not go far beyond $\sim$PeV ($\gamma_p \simeq 10^6$), we find that these two conditions are always satisfied. 

The accelerated protons in the jet can also interact with the intrinsic radiation in the jet flow, and in particular with the X-ray photons emitted as synchrotron radiation of the primary electrons. We account for this inelastic \pg interactions that lead to the formation of secondary electrons/positrons, \grs and neutrinos. We follow the semi-analytical formalism of \cite{kelner2008energy}. 

In Section~\ref{fig: intrinsic neutrino emission and contribution to total spectrum} we present the distributions of muon and electron neutrinos and anti-neutrinos from both pp and \pg interactions, and discuss the contribution of each distribution.

\begin{figure*}
\begin{minipage}{2\columnwidth}
    \includegraphics[width=\columnwidth]{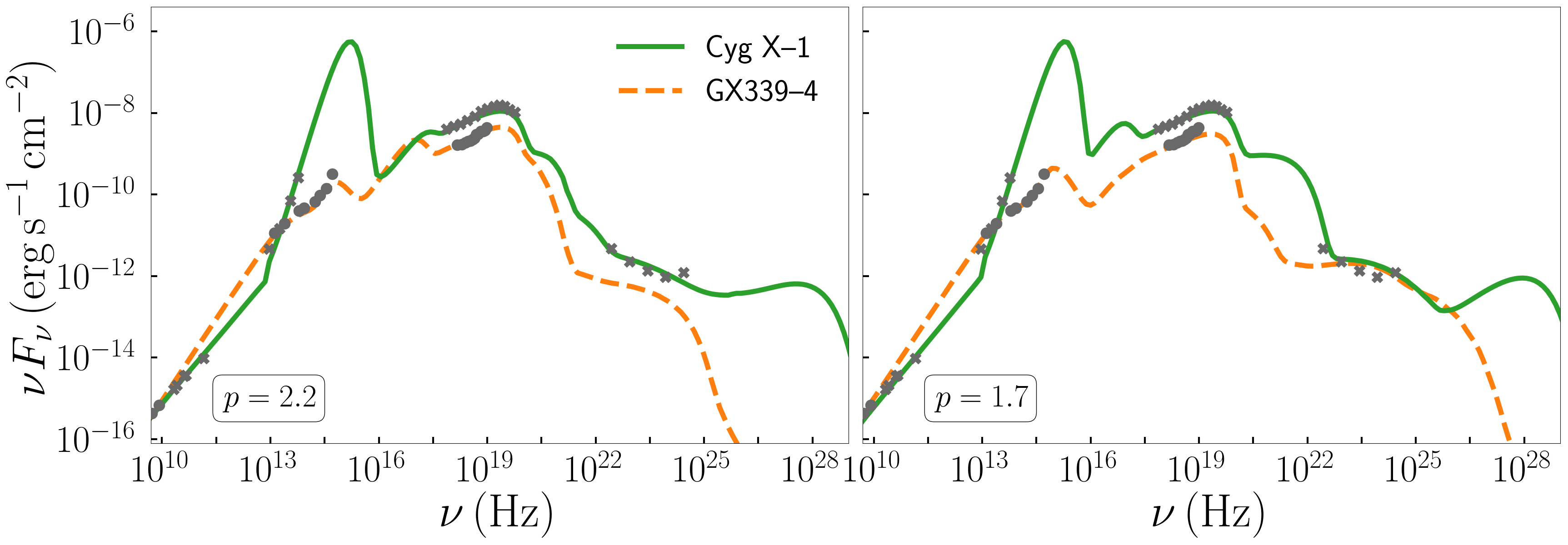}
\end{minipage}
    \caption{The predicted multiwavelength energy spectra of \cyg and \gx, and the data-points we fit for (see \kcyg and \kgx, respectively, for figures with the full datasets and fits). In the \textit{left} panel we show the soft case where the non-thermal particles accelerate in a power-law with index $p=2.2$, and in the \textit{right} panel we show the hard case where $p=1.7$.
    }\label{fig: SEDs from cyg and gx339-4}
\end{figure*}

\subsection{Electromagnetic spectrum}
We use the broadband photon spectra computed for 
the canonical high-mass \bh (HMXB) \cyg (\kcyg), and the canonical low-mass \bh (LMXB) \gx \citep[][hereafter \kgx]{kantzas2022gx} -- see Fig.~\ref{fig: SEDs from cyg and gx339-4}. Based on the jet parameters used for these spectra, we can compute
self-consistently 
the neutrino distributions. As we discussed in the previous section, the different acceleration mechanisms may lead to different power-law indexes, so we choose to examine two different cases similar to \kcyg and \kgx. In particular, we study the case where the accelerated proton population is a soft power law with $p=2.2$ (left panel of Fig.~\ref{fig: SEDs from cyg and gx339-4}), and compare it to a harder power law with $p=1.7$ (right panel of Fig.~\ref{fig: SEDs from cyg and gx339-4}). The former case of $p=2.2$ explains well the flat-to-inverted radio spectrum, but recent results on modelling the spectral energy distribution of \cyg (see, e.g., \citealt[]{zdziarski2017high}; \kcyg) suggest that a hard power law of accelerated electrons is necessary to explain the polarized emission in the MeV band as detected by \integral \citep[][]{2011Sci...332..438L,jourdain2012separation,0004-637X-807-1-17,Cangemi2020longterm}. Due to lack of similar observations on other sources, however, we examine both cases similar to \kcyg and \kgx.

\section{Intrinsic neutrino emission}\label{sec: intrinsic neutrino emission}

We calculate the neutrino flux for each jet segment beyond $z_{\rm diss}$ for both pp and p\g interactions. We use in particular the formalism of \cite{kelner2006energy} following equations 71 and 78 but for neutrinos instead. 
For the case of p\g interactions, we use the formalism of \cite{kelner2008energy} following equation 30 and the tabulated parameters of tables II and III updated by \cite{kelner2008erratum}.

\subsection{Neutrino emission from \cyg and \gx}
\begin{figure*}
\begin{minipage}{\columnwidth}
    \includegraphics[width=1.0\columnwidth]{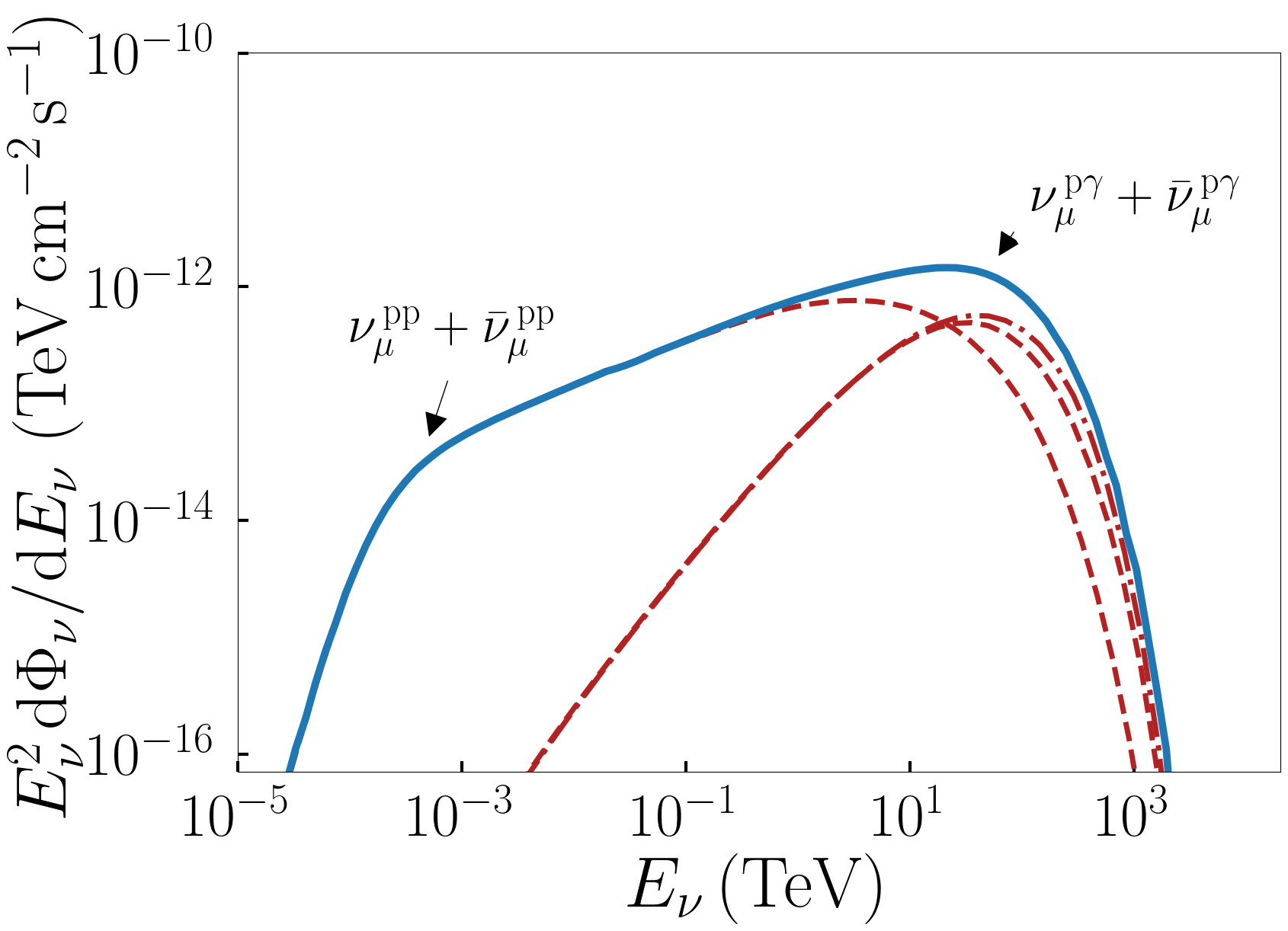}
\end{minipage}\begin{minipage}{\columnwidth} 
    \includegraphics[width=1.0\columnwidth]{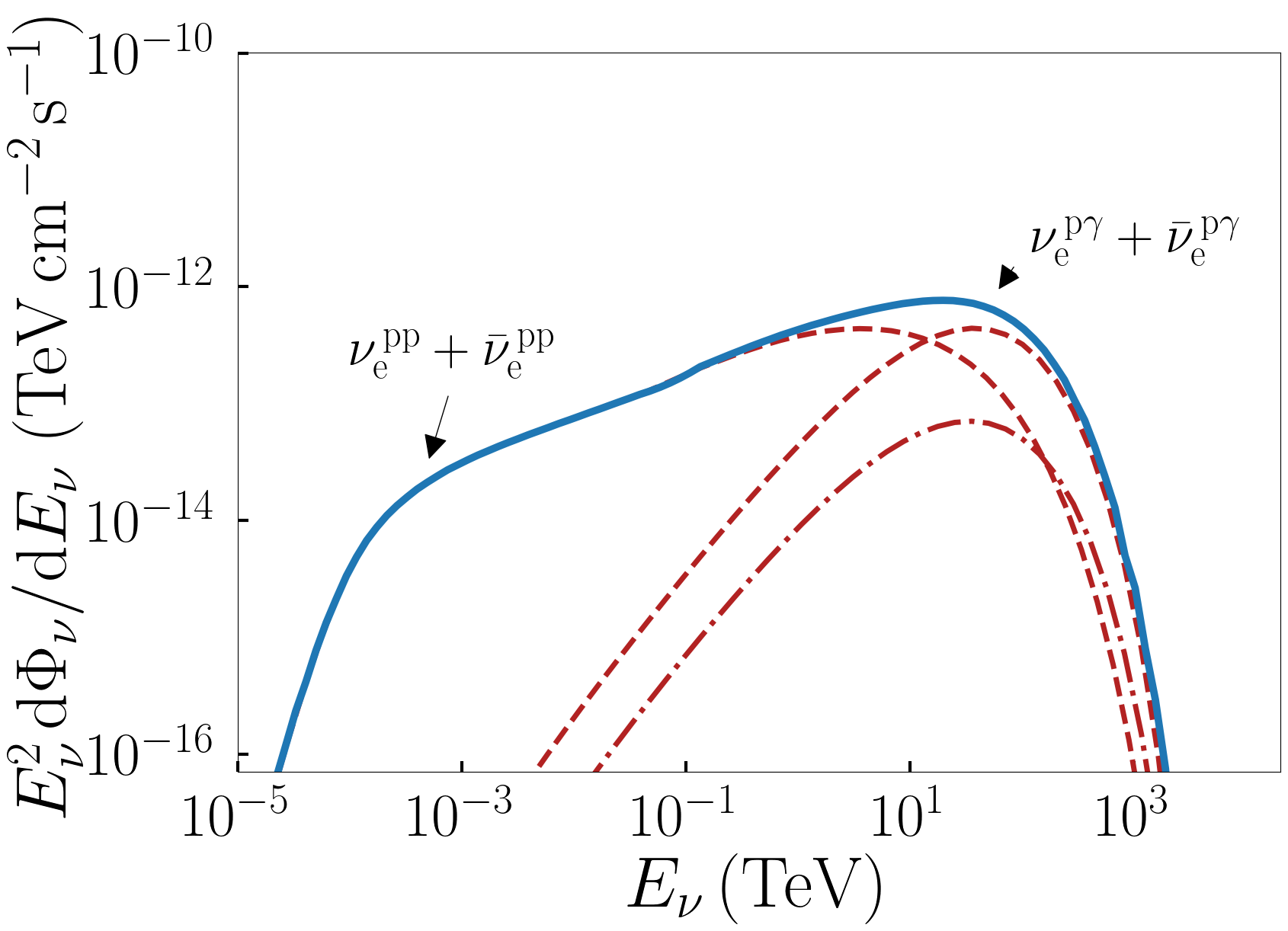}
\end{minipage}\\ \vspace{-0.10cm}
\begin{minipage}{\columnwidth}
    \includegraphics[width=1.0\columnwidth]{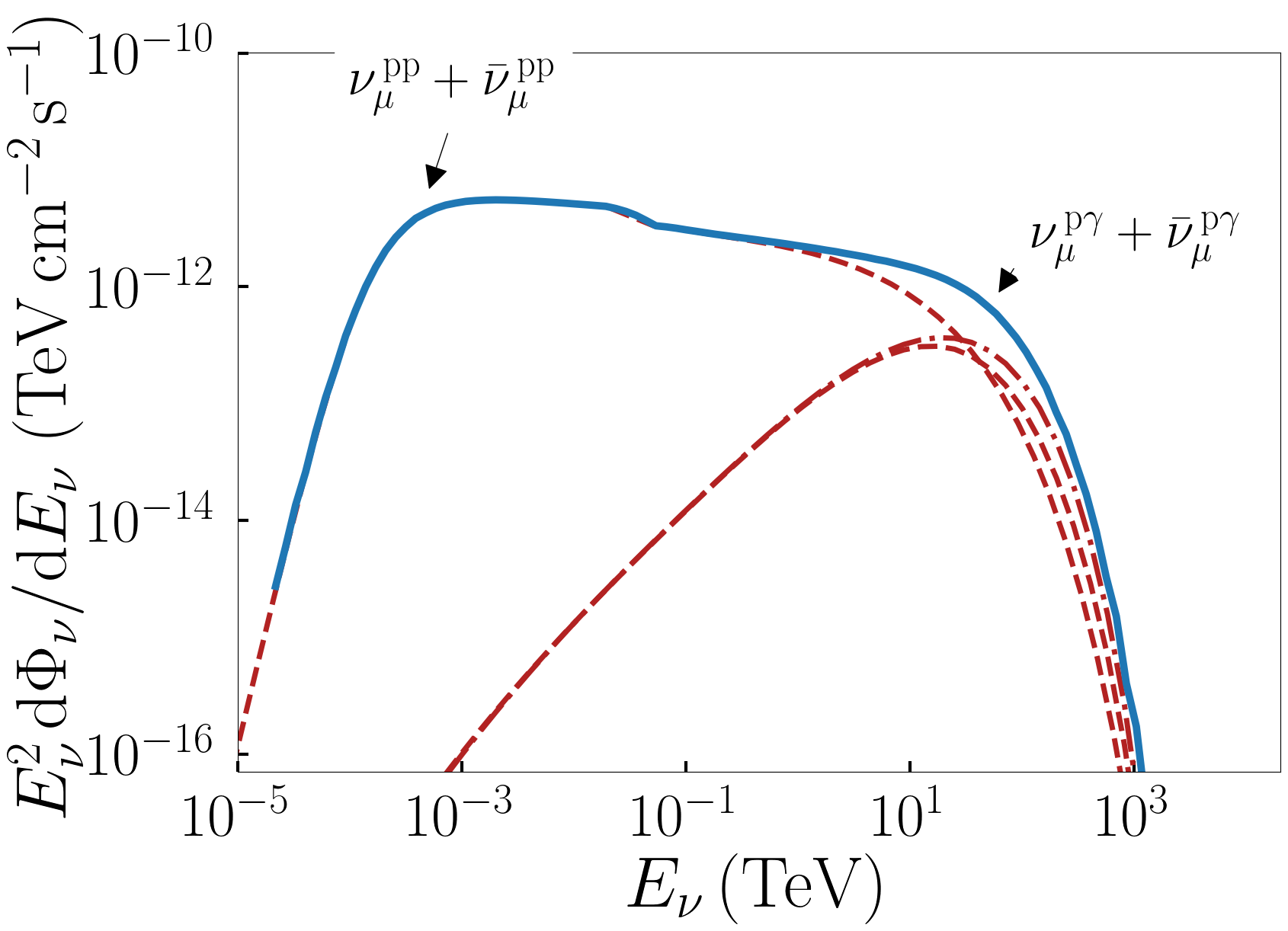}
\end{minipage}\begin{minipage}{\columnwidth}
    \includegraphics[width=1.0\columnwidth]{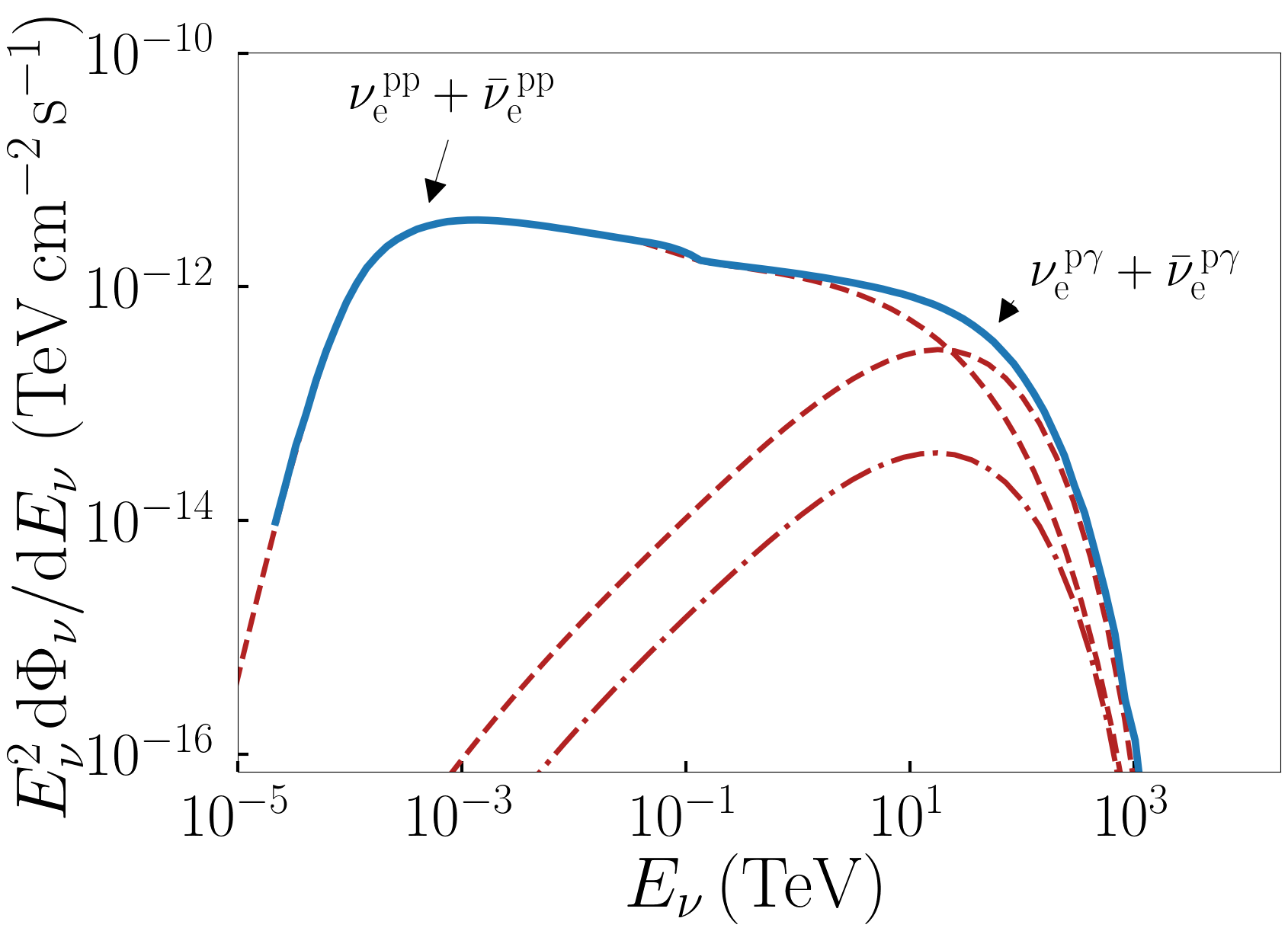}
\end{minipage}

    \caption{The intrinsic neutrino and antineutrino energy flux of the jest of \cyg. In the \textit{top} panels, the neutrinos originate in a hard power law of non-thermal protons with an index $p=1.7$, and in the \textit{bottom} panels, we assume $p=2.2$. In the \textit{left} panels, we show the distributions of muon neutrinos, and in the \textit{right} panels, we show the electron neutrinos. The neutrinos from pp interactions always dominate the low-energy regime of the spectrum, while the neutrinos from \pg interactions dominate in the maximum energy of the order of 100\,TeV. For the secondaries of \pg, we plot the muon/electron neutrinos with dashed lines, and the muon/electron antineutrino with dash-dotted lines. The solid line is for the total neutrino/antineutrino spectrum. 
    }\label{fig: neutrinos Cyg X-1 with components}
\end{figure*}

In Fig.~\ref{fig: neutrinos Cyg X-1 with components} we plot the integrated neutrino energy flux along the jets of \cyg, for both muon neutrinos/antineutrinos (left), and electron neutrinos/antineutrinos (right). In all panels, we highlight the contribution of pp 
and \pg interactions 
to disentangle the contribution of each process in the GeV-PeV regime. In Appendix~\ref{app: neutrino flux along the jet} we discuss the neutrino flux due to pp interactions for the case of \cyg.

In all four panels, we see that the pp-initiated neutrinos form a power law that follows the parent power law of protons \citep{kelner2006energy}, and dominates the total neutrino spectrum until tens of TeV. In particular, they share the same slope that extends to some maximum energy that is approximately one tenth of the maximum proton energy. The \pg neutrinos, on the other hand, form a distribution that peaks close to the maximum energy, which once again is approximately one tenth of the maximum proton energy. For the case of a hard power law of protons, the \pg neutrinos can dominate the high-energy regime of the spectrum, whereas for the case of a soft power law, the \pg neutrinos dominate in the cutoff of the spectrum.

\begin{figure*}
\begin{minipage}{\columnwidth}
    \includegraphics[width=1.\columnwidth]{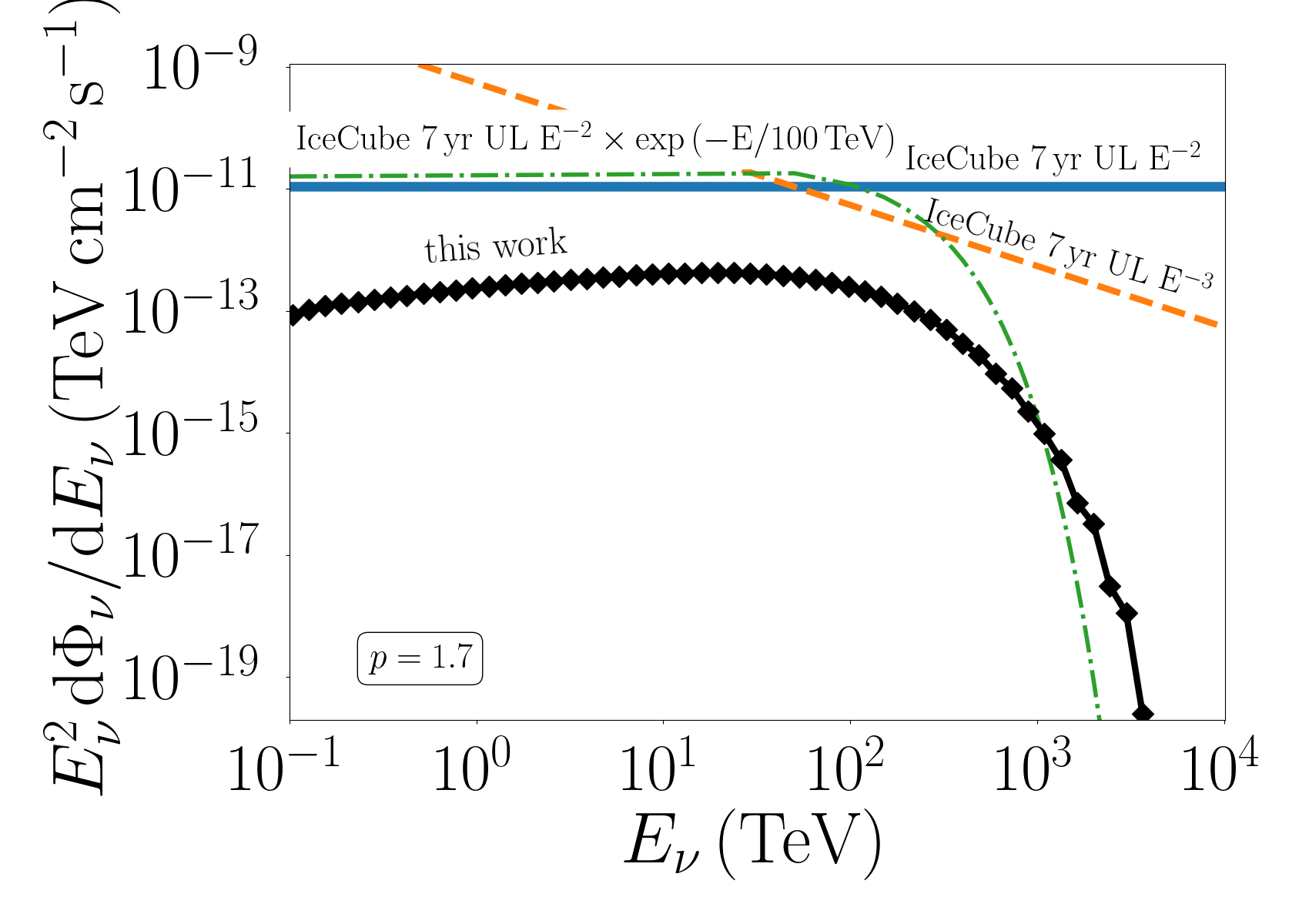}
\end{minipage}\begin{minipage}{\columnwidth}
    \includegraphics[width=1.\columnwidth]{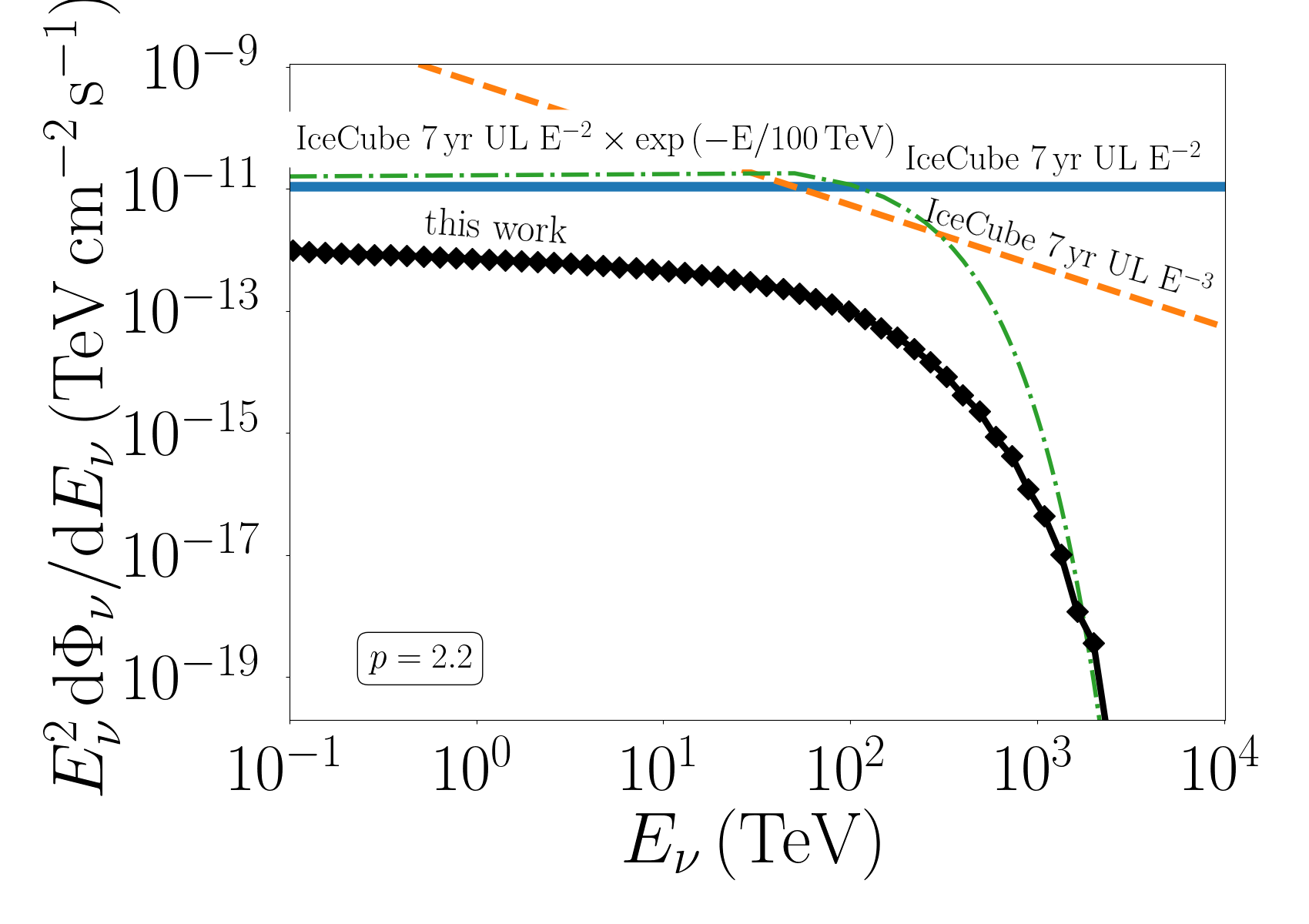}
\end{minipage}\vspace{-0.2cm}
\begin{minipage}{\columnwidth}
    \includegraphics[width=1.\columnwidth]{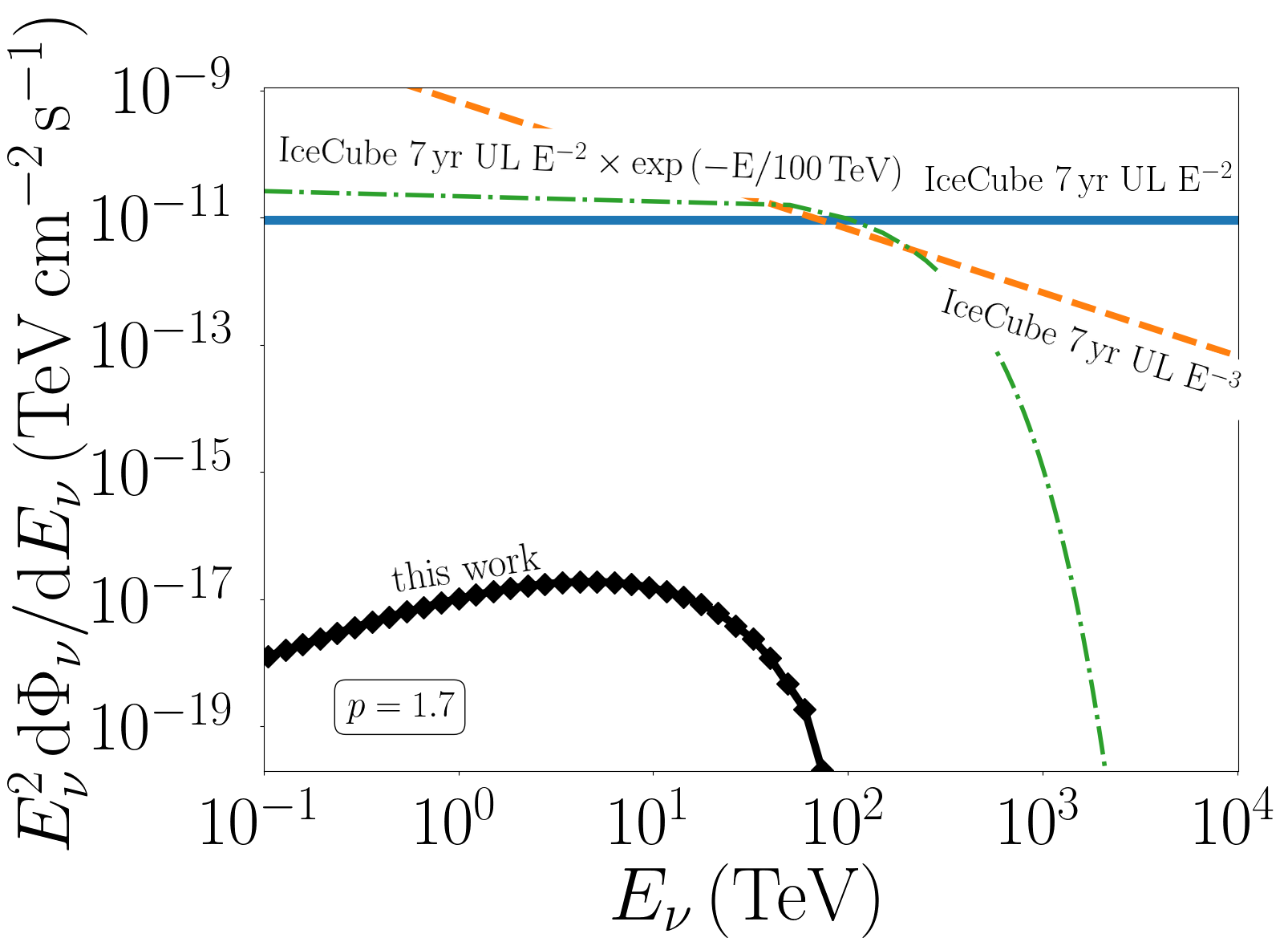}
\end{minipage}\begin{minipage}{\columnwidth}
    \includegraphics[width=1.\columnwidth]{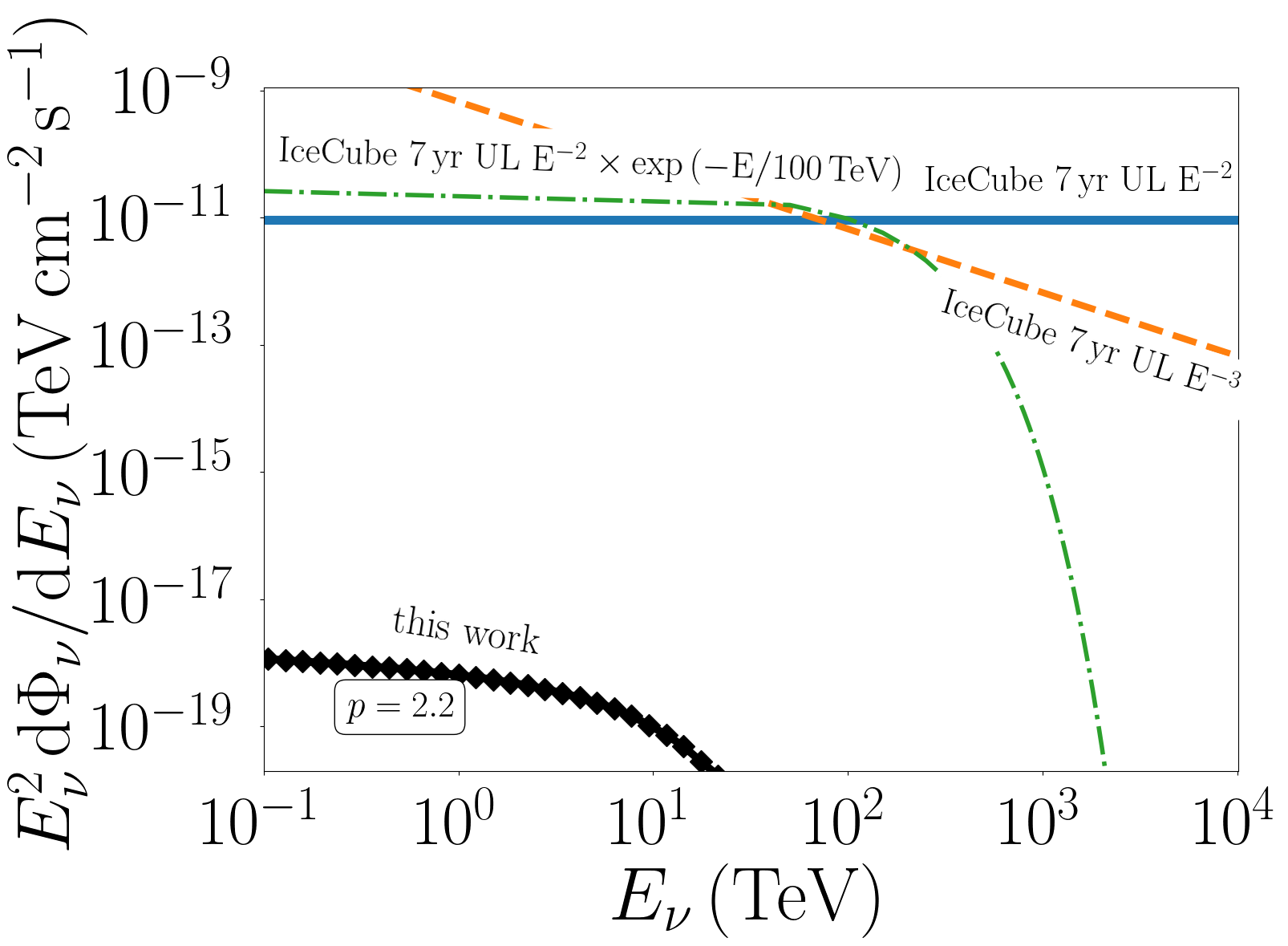}
\end{minipage}
    \caption{The total predicted $\nu_e+\bar{\nu}_e$ flux of the jets of \cyg (upper panels) and \gx (lower panels). We assume a hard power-law of accelerated protons in the \textit{left} ($p=1.7$) and for a soft power-law in the \textit{right} ($p=2.2$). We compare the predicted neutrino flux to the \ic upper limits of cascade-like events after 7\,yr of operation \citep[][]{Aartsen_2019} for the three different assumed spectra, as indicated in the plots.
    }\label{fig: electron neutrinos Cyg X-1}
\end{figure*}

In the upper panels of Fig.~\ref{fig: electron neutrinos Cyg X-1}, we plot the electron neutrino emission of \cyg for the two power-law slopes, $p=1.7$ (left) and $p=2.2$ (right), and compare it to the upper limits of the cascade-like events of \ic after 7\,yr of operation \citep[][]{Aartsen_2019}. The three different lines in both plots, correspond to the three different energy spectra as assumed by \ic. We see that the predicted emission is below the sensitivity of \ic by an order of magnitude, and hence the detection of any electron neutrino from \cyg is unlikely with the current generation of neutrino observatories, at least based on the typical hard state behaviour.

In the lower panels of Fig.~\ref{fig: electron neutrinos Cyg X-1}, we show the total predicted electron neutrino and antineutrino emission of the jets of \gx during an outburst similar to the bright outburst in 2010. The produced neutrinos share the same properties as the parent particles.
The neutrino flux of \gx is too low to be detected, which is expected based on the estimated power of the jets (\kgx). We also compare the predicted neutrino flux to the upper limits as set by \ic \citep[][]{Aartsen_2019}. We see that \gx is also incapable of producing enough electron neutrinos to be detected by \ic. In both panels, we account for neutrino oscillation such that we convert the produced rate of $\nu_{\mu}:\nu_e:\nu_{\tau}=1:2:0$ to a detected rate at Earth $\nu_{\mu}:\nu_e:\nu_{\tau}=1:1:1$.

\subsection{Neutrino rates}

The current neutrino detectors are not sensitive enough to detect a neutrino spectrum from any particular Galactic source, we can thus detect only single events. We convert the neutrino spectra to rate of events 
\begin{equation}\label{eq: neutrino rate}
    R =  \int 4\pi \dfrac{{\rm d}\Phi_{\nu}}{{\rm d}E_{\nu}}\,A_{\rm{eff}}(E_{\nu})\,{\rm d}E_{\nu},
\end{equation}
where ${\rm{d}}\Phi_{\nu}/{\rm{d}}E_{\nu}$ is the neutrino differential flux, $A_{\rm{eff}}$ is the effective area of the detector, and we integrate between $0.1\,$TeV and $10\,$PeV. We use the effective areas of \ic \citep{aartsen2013evidence,Aartsen_2019}, \ant \citep{albert2017ANTARES} and the simulated effective area of \arca \citep{AdrianMartinez2016KM3NeT2} (see Appendix~\ref{app: neutrino effective areas}). 

In Table~\ref{table:neutrino rates}, we show the predicted event rates (per year) for the three different detectors accounting for neutrino oscillation. The number of track-like events ($\nu_{\mu} $ and/or $\bar{\nu}_{\mu}$) for both \ic and \arca is similar for the case of \cyg and around 1--1.3 neutrinos per year when the source launches jets for both $p=2.2$ and $p=1.7$ power-law slopes. The number of shower-like events ($\nu_e$ and/or $\bar{\nu}_e$) for \ic is significantly reduced due to the effective area of the detector and the veto forced by the \ic collaboration, namely the subtraction of the outer volume of the detector \citep{aartsen2013evidence}. 

In agreement with what we expect from the neutrino energy spectrum, we see that \gx\ cannot produce a detectable rate of events, neither for \ic nor \arca. 
\begin{table*}
\renewcommand{\arraystretch}{1.4} 
\begin{tabular}{cl|rrr|rrr}
                                     &                             & \multicolumn{3}{c}{soft proton power law ($p=2.2$)}                                                                         & \multicolumn{3}{c}{hard proton power law ($p=1.7$)}                                                                         \\
                                     &                             & \multicolumn{1}{c}{\ic} & \multicolumn{1}{c}{\ant} & \multicolumn{1}{c}{\arca} & \multicolumn{1}{c}{\ic} & \multicolumn{1}{c}{\ant} & \multicolumn{1}{c}{\arca} \\
\hline
\multirow{2}{*}{\cyg} & $\nu_{\mu}+\bar{\nu}_{\mu} \ (\rm yr^{-1})$ & 1.1                               & 0.006                                & 1.3                                 & 1.0                               & 0.007                                & 1.3                                 \\
                                     & $\nu_{e}+\bar{\nu}_{e} \ (\rm yr^{-1})$     & 0.002                               & 0.001                                & 0.6                                 & 0.005                               & 0.001                                & 0.4                                 \\
\\
\multirow{2}{*}{\gx}  & $\nu_{\mu}+\bar{\nu}_{\mu} \ (\rm yr^{-1})$ & $7\times 10^{-10}$                           & $3\times 10^{-7}$                            & $3\times 10^{-5}$                                 & $6\times 10^{-10}$                           & $3\times 10^{-7}$                            & $3\times 10^{-5}$                                 \\

                                     & $\nu_{e}+\bar{\nu}_{e} \ (\rm yr^{-1})$     & $3\times 10^{-9}$                           & $3\times 10^{-7}$                            & $1\times 10^{-5}$                                 & $3\times 10^{-9}$                           & $3\times 10^{-7}$                            & $1\times 10^{-5}$                                

\end{tabular}
\caption{The predicted neutrino rate (per year) for \cyg and \gx for the various detectors, and for two different proton power-law slopes. It is worth mentioning that \ic and \arca could detect of the order of one muon neutrino from \cyg if it launched relativistic jets for a year. } 
\label{table:neutrino rates}
\end{table*}

\subsection{Contribution from all known \bhs}

Although we have now established that \cyg and \gx on their own cannot be detected as neutrino sources for the current generation of facilities, it is interesting to examine whether the entire population of BHXBs together would be a significant contributor to the diffuse neutrino background.
We utilise the most recent XRB catalogue of \cite{Tetarenko2016} that includes all the \bhs and \bh-candidates detected until 2016. Out of these, we select those sources that are known \bhs or \bh-candidates, and known to emit in radio bands described as jet emission. This is a very conservative estimate because XRBs with neutron stars have been observed to launch jets that sometimes may be as powerful as \bhs (see, e.g., \citealt{Eijnden2018evolving,Coriat2019twisted} but also see \citealt{Chatzis2022radio} for an alternative scenario). 

From the 70~sources in the catalogue of \citealt{Tetarenko2016}, we consider 31 for this analysis. From these sources, 4 are \hms, and 27 are \lms. We use the neutrino spectra computed for \cyg and \gx, as representative for the high-mass and \lms, respectively, due to the lack of better quality simultaneous observations of \bhs. However we rescale the neutrino spectra for each source according to their distance, the mass of the black hole and the inclination. For many sources, the exact distance, the inclination and/or the mass of the black hole are not well constrained. In these cases, \cite{Tetarenko2016} assumed fiducial values: a distance of 8\,kpc, the mass of the black hole equal to 8\, M$_{\odot}$, and an inclination angle of 60$^{\circ}$, which we also adopt.
We finally include four more \lms/\bh-candidates discovered after 2016 based on the Faulkes Telescopes project \citep[][]{lewis2008continued}. We show the exact values of these sources in Table~\ref{table: 35 known sources} of Appendix~\ref{app: known sources}.

In Fig.~\ref{fig: sensitivity}, we plot the differential $\nu_{\mu}+\bar{\nu}_{\mu}$ energy flux at 1\,TeV for the 35~known \bhs as a function of the declination angle $\delta$. We compare our predicted values to the 90\% median sensitivity of \ic \citep{aartsen201910years} and the predicted sensitivity of \arca \citep{aiello2019arca} assuming that the flux scales as $E_{\nu}^{-2}$ (solid lines) or $E_{\nu}^{-3}$ (dashed line). For simplicity, we use the publicly available sensitivity curves of \ic with blue lines, and with a thin solid gray line the interpolated sensitivity for $p=2.2$. Comparing to this sensitivity curve, we see that \cyg is slightly below the threshold for detection. 
We indicate the other \hms V4641~Sgr, MWC~656 and Cyg~X--3 that contribute significantly to the overall predicted spectrum, as we discuss in the Appendix~\ref{app: contribution from individual sources}. The other four \lms we highlight are the sources \cite{abbasi2022search} discuss in their analysis.
Moreover, the \lm with the strongest neutrino flux is MAXI~J1836-194, which launches a jet with an inclination angle between 5 and 15$^{\circ}$ \citep[][]{Russell2014J1836} making it a good candidate source for further spectral analysis, in particular in the case of lepto-hadronic jets \citep[see][for a purely leptonic scenario]{Lucchini2021correlation}. Finally, in Fig.~\ref{fig: sensitivity}, we indicate the upper limits of those microquasars as derived by the \ic collaboration \citep[][]{Aartsen_2019,abbasi2022search} for track-like events that are in the list of sources we examine in this work. The above works present multiple upper limits based on different assumptions on the energy dependence of the neutrino flux. We show only the upper limits derived under the assumption of a neutrino flux that scales as $E^{-2}$, because in our two models we assume a proton power-law index close to 2, and hence the produced secondaries follow a similar distribution (because the pp contribution dominates the majority of the neutrino spectrum).

\begin{figure}
	\includegraphics[width=1.05\columnwidth]{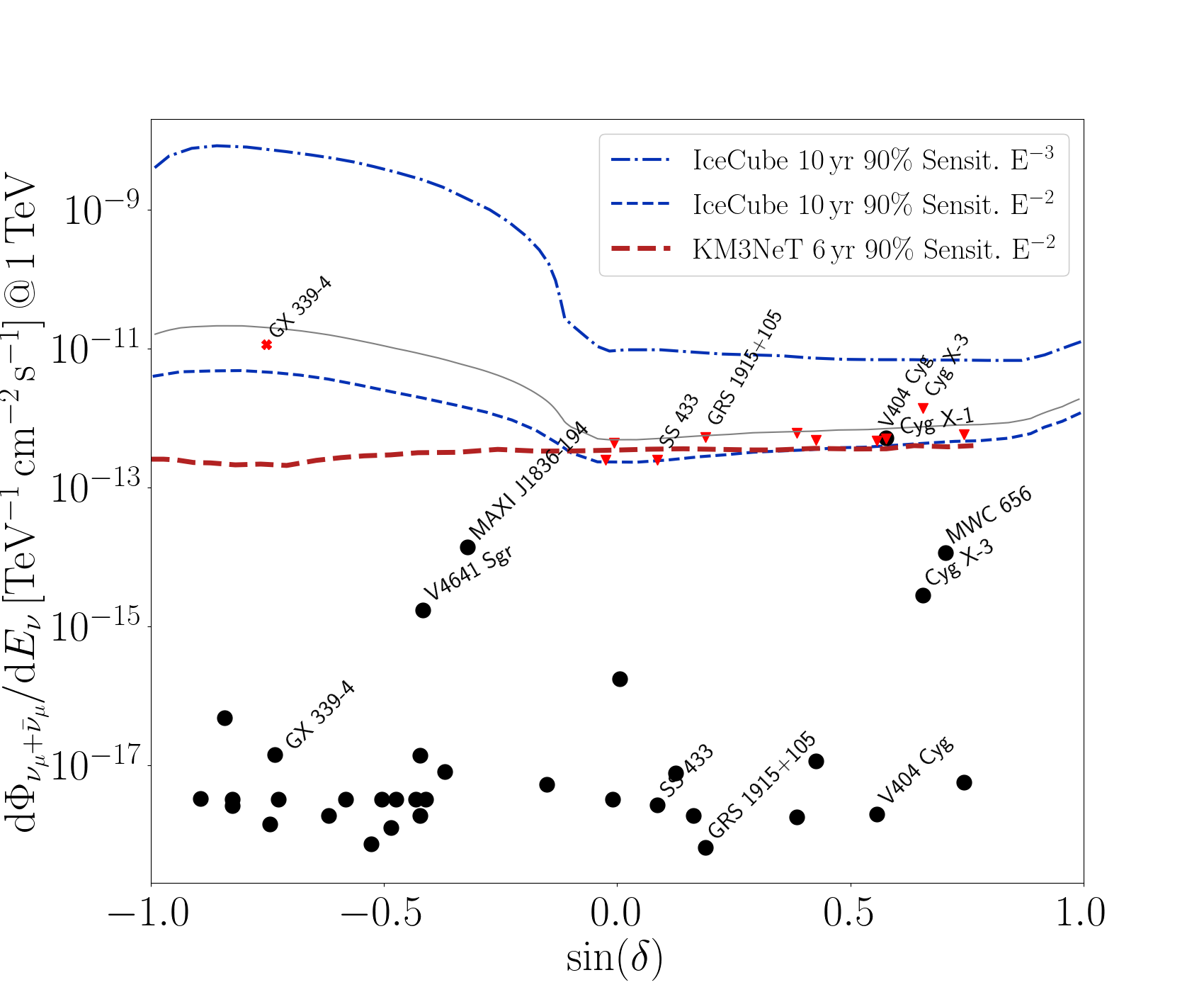}
    \caption{The predicted $\nu_{\mu}+\bar{\nu}_{\mu}$ differential flux at 1\,TeV for the 35 known \bhs we study here versus the $\sin$ of their declination ($\delta$). We use the results for the case of $p=2.2$. We also show the 90 per cent median sensitivity of \ic (dot-dashed blue line for $E_{\nu}^{-3}$ and dashed blue line for $E_{\nu}^{-2}$; \citealt{aartsen201910years}) and \arca (thick, dashed red line; \citealt{aiello2019arca}) for reference for $E^{-2}$ neutrino spectrum. We note that we cannot directly compare to these sensitivity curves because in our work we assume an $E_{\nu}^{-2.2}$ dependence. We hence plot the interpolated flux for $E_{\nu}^{-2.2}$ with a thin solid gray line. The red cross (\citealt{Aartsen_2019}) and triangles (\citealt{abbasi2022search}) correspond to the 90 per cent upper limits for $E_{\nu}^{-2}$.}
    \label{fig: sensitivity}
\end{figure}

\begin{figure}
	\includegraphics[width=1.1\columnwidth]{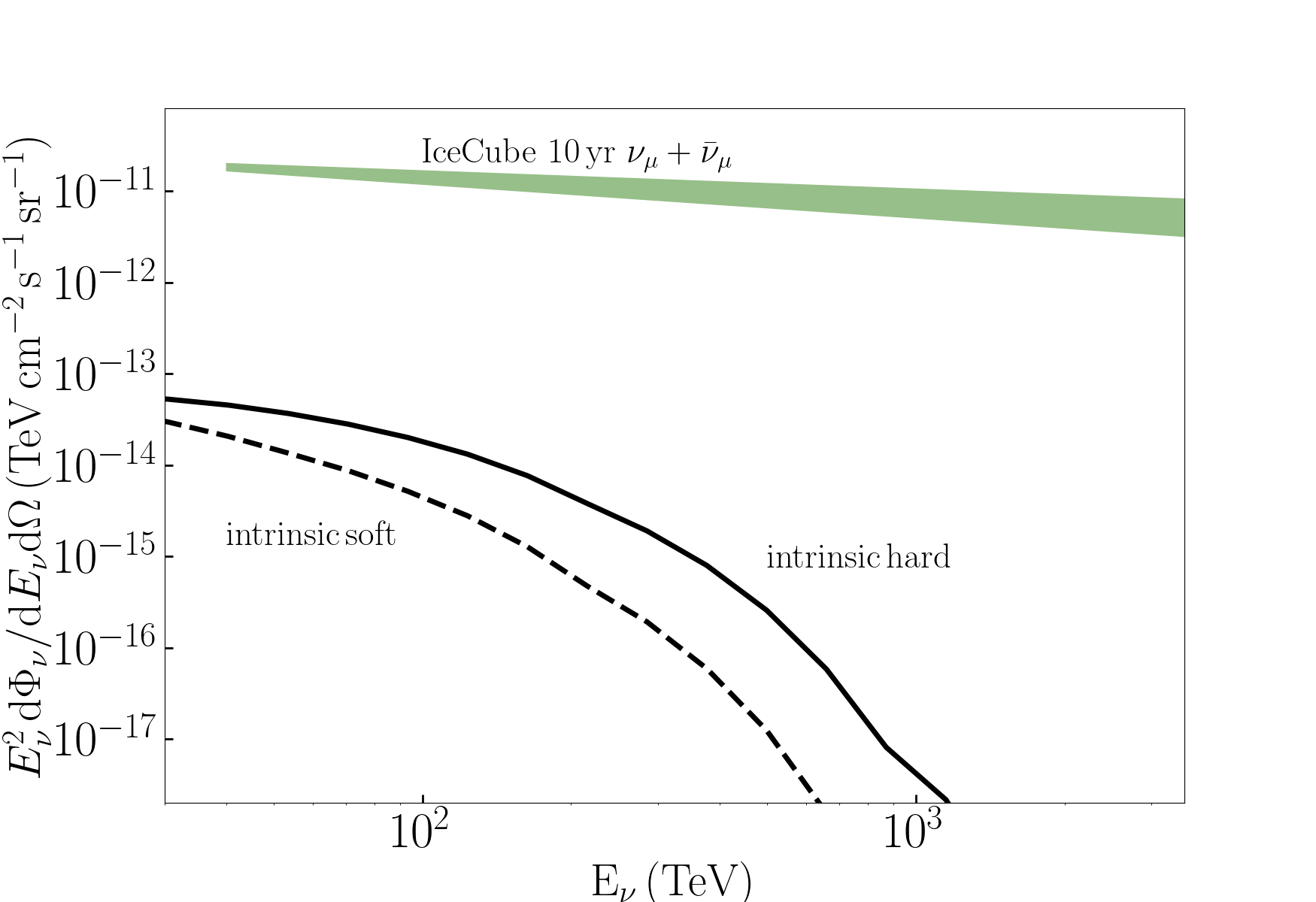}
    \caption{The predicted intrinsic $\nu_{\mu}+\bar{\nu}_{\mu}$ energy flux for the 35 known \bhs assuming a soft power law of protons ($p=2.2$; dashed line) or a hard power law ($p=1.7$; solid line), assuming that the sources have a one-percent duty cycle. For comparison purposes, we also plot the astrophysical neutrino flux as observed by \ic in 10\,yr \citep[][]{stettner2019measurement}.}
    \label{fig: intrinsic neutrino emission and contribution to total spectrum}
\end{figure}

In Fig.~\ref{fig: intrinsic neutrino emission and contribution to total spectrum}, we sum the intrinsic $\nu_{\mu}+\bar{\nu}_{\mu}$ emission from all 35 known \bhs assuming either a soft power law of accelerated protons with $p=2.2$ for every source, or a hard power law with $p=1.7$. We further assume an average duty cycle of 1 per cent (\citealt[][]{Tetarenko2016}; \coop; but also see \citealt[][]{Deegan2009duty} who suggest 0.1-0.5 per cent for the particular case of GRS~1915+105) to account for the fact that \bhs do not launch persistent jets, but they transit between different spectral states \citep[][]{Belloni2010}. For the unique case of \cyg in particular, we assume a duty cycle of 25~per cent because in the past ten years, it has spent 1/4th of its time in the jet-launching state \citep[][]{Cangemi2020longterm}. During the so-called soft state however, the jets of \hms do not entirely quench but rather become less powerful, hence they may still produce neutrinos during the rest 75 per cent of their time, but at a lower rate. Comparing our results of Fig.~\ref{fig: intrinsic neutrino emission and contribution to total spectrum} to the detected astrophysical neutrino signal of \ic \citep[][]{stettner2019measurement}, we see that the known population of \bhs cannot contribute more than $\sim0.2$ per cent in the energy band around 40\,TeV, assuming a hard proton power law. The contribution in the case of a soft power law of protons is even less than 0.1 per cent.

\section{Diffuse secondary emission}\label{sec: diffuse emission}
\subsection{\bhs as CR sources}

In the above, we calculated the intrinsic neutrino emission of 35 known \bh jets. In this section, we use \texttt{DRAGON2} \citep[][]{Evoli_2017,Evoli_2018} to examine the contribution of a broader population of \bhs to the CR spectrum. In particular, we conservatively assume the existence of 1000 Galactic sources that follow a spatial distribution similar to the observed distribution of pulsars \citep[][]{Lorimer2006}, but we also account for a pc-scale spike in the number of sources close to the Galactic centre, as indicated by \cite{Mori_2021}\footnote{The spike of \bhs around the Galactic centre peaks at 1\,pc with a full width half maximum of 10\,pc}. Based on the detected \bhs so far, we assume that 90 per cent of them are low-mass and 10~per cent are \hms \citep[][]{Tetarenko2016}. Similar to our previous analysis, the \bhs follow the behaviour of the prototypical source, therefore the jets of low \bhs accelerate protons up to 100\,TeV, and the high-mass sources accelerate protons up to 1\,PeV. 

Using \texttt{DRAGON2}, we inject CR protons which propagate in the Galactic plane in a 2D grid of radius 12\,kpc and height 4\,kpc. We assume in particular that the 
protons that escape the jets to propagate in the interstellar medium carry 10 per cent of the proton power \citep{Blasi2013} and neglect any neutrons that could convert to protons \citep{Atoyan_2003}.
The results we present here scale linearly with the power carried away by the escaping particles. 

In Fig.~\ref{fig: timescales for Cyg} we show the inverse timescales of the various cooling processes comparing them to escape. The escape rate is at least three orders of magnitudes greater than any cooling process, which means that the protons manage to escape freely, carrying the majority of their energy. The cooling processes we account are: synchrotron, pp, p\g, and Bethe-Heitler processes, but none of them can compare to escape, at least in the energy range we are interested in. 
We only propagate the population of protons as we have not included heavier elements in our previous studies to be able to further constrain their spectral properties. 
We adopt a spatially constant diffusion coefficient that depends on  the momentum of the particle as $D = D_0 (p/p_0)^\delta$, with $D_0=2.7\times 10^{28}\,\rm cm^2\,s^{-1}$ and $\delta = 0.45$~\citep[][]{Fornieri2020,Luque_2021Markov}.

\begin{figure}
\includegraphics[width=1.1\columnwidth]{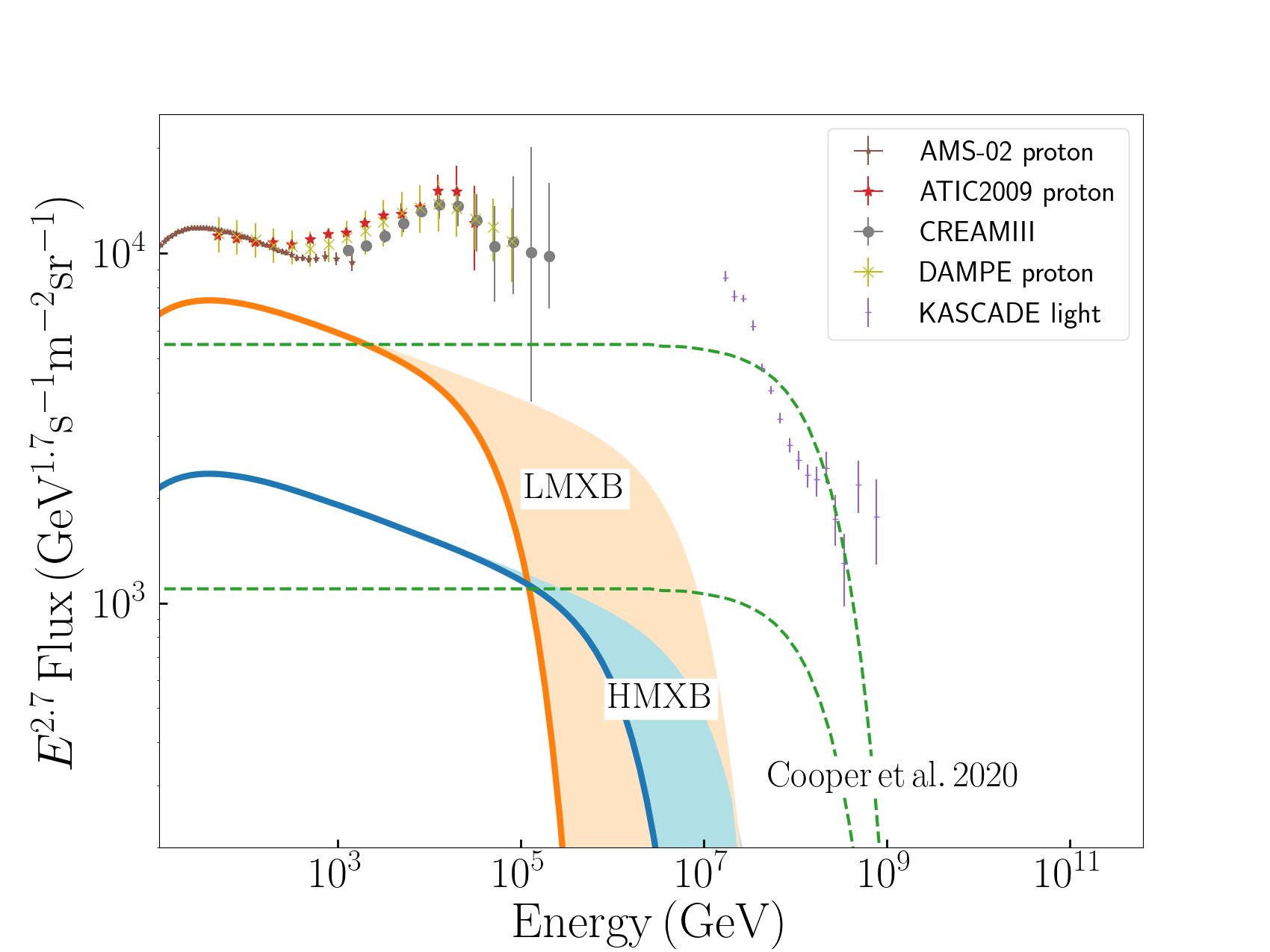}
    \caption{The contribution of a 1000 \bhs to the CR spectrum calculated with \texttt{DRAGON2}. The population consists of 90 per cent \lms and 10 per cent of \hms.  We assume that all the sources accelerate protons to a soft power law with an index $p=2.2$ up to some maximum energy that depends on whether it is a low-mass or a \hms (solid lines). We compare our results to the observational constraints as shown in the legend, and we over-plot the ``optimistic'' (uppermost) and ``pessimistic'' (lowermost) scenarios of \citealt[][]{cooper2020xrbcrs} for comparison. We show for comparison an extreme scenario (dashed lines) where the particles follow similar power-laws in energy but with larger maximum energy of the order of $10^7$\,GeV. The observational data are as labelled from ATIC2009 \citep[][]{chang2008excess}, CREAMIII \citep[][]{Yoon_2017}, DAMPE \citep[][]{dampe2019measurement} and KASCADE \citep[][]{Apel2011kascade}.
    }
    \label{fig: CR spectrum for 1000 sources}
\end{figure}

In Fig.~\ref{fig: CR spectrum for 1000 sources} we show the contribution of both \lms and \hms to the overall CR spectrum. We see that 900 \lms can contribute up to 50~per cent of the spectrum at the TeV energy range. The contribution falls significantly though and at $\sim 100\,$TeV the spectrum drops exponentially (solid lines). The 100 \hms cannot significantly contribute to the CR spectrum even though they allow for proton energy that is 10 times greater than the one of \lms due to our assumption that \hms are merely 10 per cent of the entire population. We moreover include the ``optimistic'' and the ``pessimistic'' scenarios of \coop for comparison, and we see that despite the similar contribution in the low-energy regime at around TeV, the approach we follow here leads to different maximum CR energy (see the discussion in Section~\ref{sec: bhs and the cr spectrum}). Finally, we overplot the shaded regions up to the extreme scenario where the maximum energy of the CRs reaches values of the order of $10^7$\,GeV as particle acceleration theory allows for (see Section~\ref{sec: bhs and the cr spectrum}). Such a scenario is derived from the extreme values allowed from the best fits of \kcyg and \kgx.  

\subsection{Diffuse gamma-ray and neutrino emission from \bhs}

\begin{figure}
	\includegraphics[width=1.00\columnwidth]{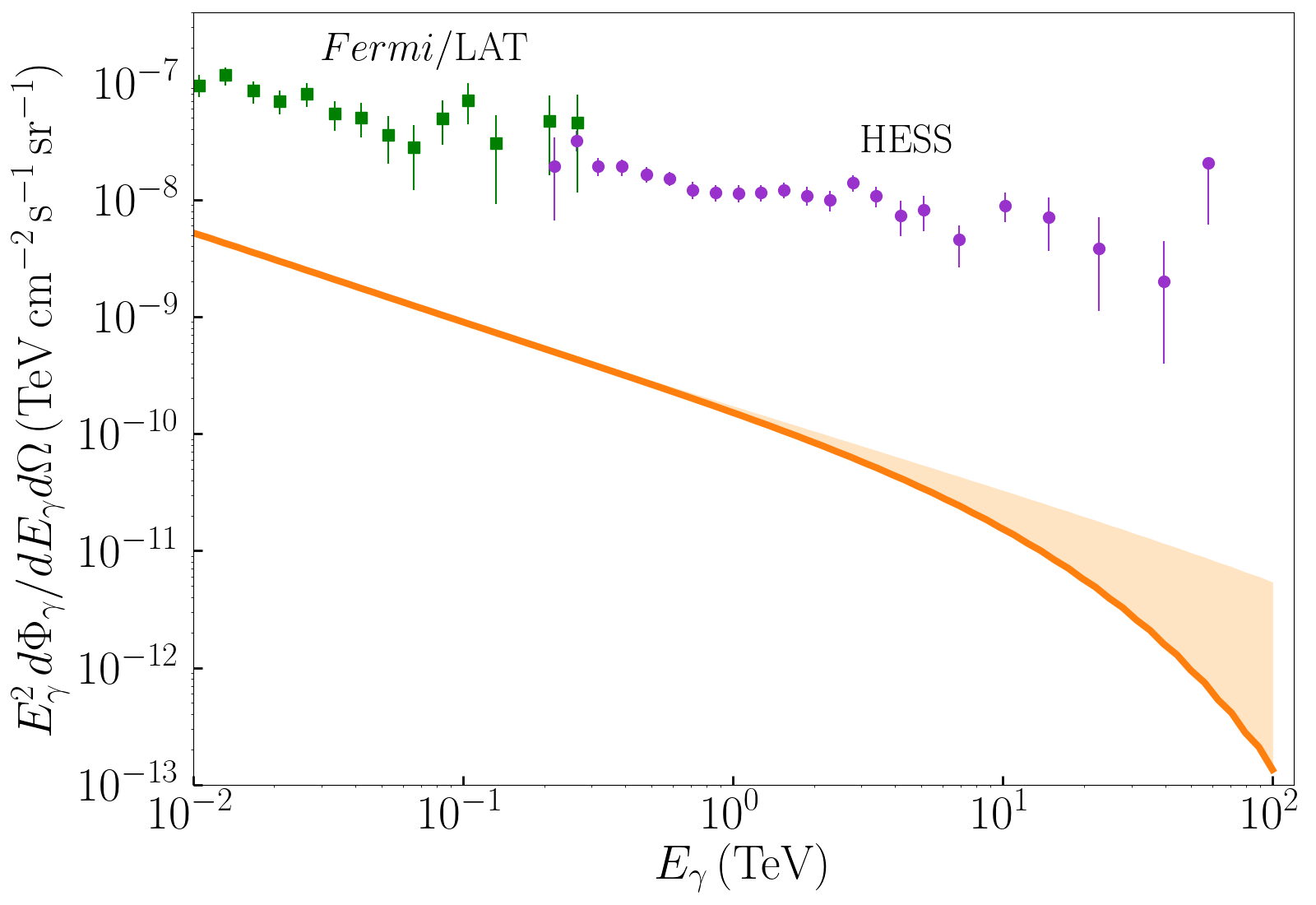}
    \caption{The diffuse \gr emission of CRs accelerated in \bh jets as calculated with \texttt{HERMES} for Galactic latitudes within 5$^\circ$ from the Galactic centre. We plot the diffuse emission in the shaded region for the extreme acceleration scenario (see Fig.~\ref{fig: CR spectrum for 1000 sources}). We compare the results of \textit{Fermi}/LAT (squares; \citealt[][]{Abdollahi_2020}) and the \hess collaboration (circles; \citealt[][]{HESS2018GalacticCenter}). 
    }
    \label{fig: diffuse gamma-ray emission of a 1000 sources}
\end{figure}

\begin{figure}
	\includegraphics[width=1.0\columnwidth]{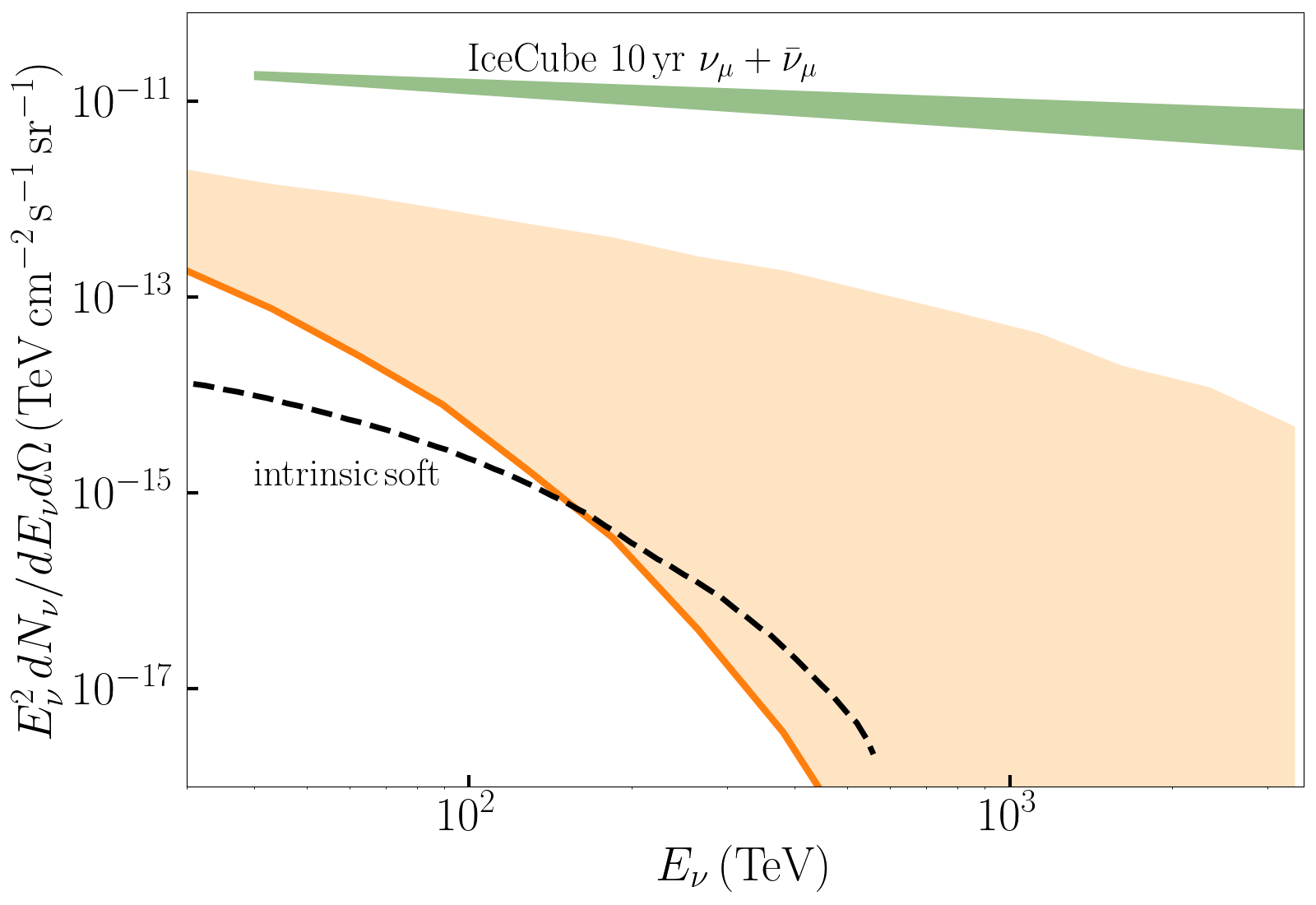}
    \caption{The diffuse neutrino emission of CRs accelerated in 1000 \bh jets while propagating in the Galactic plane (solid line). We only account for  $|b|\leq 5^{\circ}$ and $|l|\leq 180^{\circ}$, and plot the diffuse neutrino emission of \ic for comparison \citep[][]{stettner2019measurement}. The shaded region corresponds to the diffuse neutrino emission from an extreme scenario that allows for higher proton energies (see Fig.~\ref{fig: CR spectrum for 1000 sources}), and the dashed line shows for comparison the intrinsic emission for the 35 known sources.
    }
    \label{fig: diffuse neutrino emission of a 1000 sources}
\end{figure}

After CR protons escape the acceleration site, they interact with the interstellar medium while propagating through the Galactic plane. To account for the inelastic collisions between the propagating protons and the interstellar medium, we utilise \texttt{HERMES}. \texttt{HERMES} is 
a publicly available code designed to compute the emission originated from a variety of non-thermal processes including synchrotron and free-free radio emission, gamma-ray emission from bremsstrahlung and inverse-Compton scattering, \gr and neutrino emission from pion decay \citep[][]{Dundovic2021hermes}.
In Fig.~\ref{fig: diffuse gamma-ray emission of a 1000 sources}, we plot the diffuse hadronic \grs originating from the interaction of CR protons with atomic hydrogen, the density of which is estimated from the 21\,cm line emission and the molecular hydrogen indirectly traced by the CO molecular gas \citep[][]{luque2022galactic}. We compare our results to the diffuse \gr spectrum of \textit{Fermi}/LAT \citep[][]{Abdollahi_2020} and the \hess collaboration \citep[][]{HESS2018GalacticCenter}. In the aforementioned figure, we only plot the diffuse emission and do not account for the intrinsic emission of the 1000 sources because we will examine that in a forthcoming work. In Fig.~\ref{fig: diffuse neutrino emission of a 1000 sources}, we plot the diffuse neutrinos of the same processes and compare them to the astrophysical background as observed by \ic \citep[]{stettner2019measurement}.

\section{Discussion}\label{sec: discussion}
\subsection{\cyg and \gx as Galactic neutrino sources} \label{sec: cyg and gx as neutrino sources}

Recently, the \ic collaboration released the updated upper limits for the Galactic \bhs using 7.5\,yr of data, where they set the 90 per cent upper limit of \cyg at $0.5\times 10^{-12}\,\rm TeV^{-1}\,cm^{-2}\, s^{-1}$, a value very close to our predicted flux \citep[][]{abbasi2022search}. We convert the \ic flux to fluence by multiplying with 7.5\,yr to obtain $1.2\times 10^{-4}\,\rm TeV^{-1}\,cm^{-2}$. The neutrino fluence of \cyg according to our prediction is $2\times 10^{-6}\,\rm TeV^{-1}\,cm^{-2}$ assuming the neutrino production lasts for an orbital period of 6.5\,days. This assumption allows us to derive the lower limit of the fluence. Despite accounting for the persistent jets \cyg launches \citep[][]{rushton2012weak}, it is difficult to identify any astrophysical neutrinos. This is in part because the TeV energy range in which we expect emission is dominated by atmospheric neutrinos \citep[][]{Aartsen_2017}. It is also worth mentioning that \cite{abbasi2022search} assume an energy dependence of the neutrino flux with $E_{\nu}^{-1.25}$, which is significantly harder even compared to our hard power-law index of $p=1.7$, inherited by the non-thermal protons. This difference in the energy dependence may relax the fact that our prediction is so close to the \ic upper limits. In other words, if the \ic collaboration uses a softer index, then the upper limits will increase.

In Table~\ref{table:neutrino rates}, we show the neutrino rate per year from the region of \cyg. The muon neutrino rate after accounting for the effective area of \ic and \arca are of the order of one neutrino per year, close to previous studies, such as \citealt[][]{Anchordoqui2014estimating} where the authors used LS~5039 as a prototype. If we account for the duty cycle of \cyg, the neutrino rate may decrease fourfold based on the duty cycle of \cyg the past 20\,yr \citep[][]{Cangemi2020longterm}. Furthermore, we base our neutrino estimates on only the steady-state emission of the jets of \cyg when the system was in the inferior conjunction, such that the companion star was behind the black-hole-jet system on the line of sight. A more detailed analysis, where the orbital attenuation of the non-thermal emission is properly accounted for, is outside the scope of this work \citep[see, e.g.,][for a neutrino prediction during a flaring state]{vieyro2012magnetized}. Finally, when deriving the neutrino rate from equation~\ref{eq: neutrino rate}, we integrate in the energy range between 0.1\,TeV and 1\,PeV, thus we do not know the exact expected energy of the incoming neutrino(s). As aforementioned, the low-energy regime of the detected neutrino spectrum is dominated by the atmospheric background, therefore detectable astrophysical neutrinos from \cyg will occur in the energy range between $\sim 10\,$TeV and $\sim 100\,$TeV.
 
Due to the high number of outbursts and the plethora of observations, \gx has attracted the attention of the community and several previous works have predicted the neutrino emission based on different models and/or physical processes \citep[see, e.g.,][]{Distefano_2002,Zhang2010neutrinoLMXRBs}. The muon and electron neutrino spectra we present here, and consequently the neutrino rates, are intrinsically connected to the most up-to-date multiwavelength constraints of \gx (\kgx). 
We find that \gx likely fails to be a potential source of Galactic neutrinos. 
The neutrinos are mainly produced in the particle acceleration region and then the flux drops along the jet. The particle acceleration region 
is set by the break frequency in the observed spectrum, which is variable for the case of XRBs \citep[see e.g.][]{Russell2014J1836,russell2019maxij1535571}. 
For the case of \gx there is evidence that the synchrotron break is at $\sim 2600\ R_g$, corresponding to 40000\,km for a 10 solar mass black hole \citep{Gandhi2011synchbreak}. This offset from the jet base is also confirmed by the 100\,ms lag between X-ray and optical bands \citep[see e.g.,][]{Kalamkar2016detection}.
In the case of \cyg, on the other hand, there is no such evidence for the synchrotron break because \cyg is persistent and the cooling break is in the IR band that is dominated by the emission of the massive companion star. We hence used $z_{\rm diss}$ as a free parameter while fitting the multi-wavelength spectrum. The best-fit value is at $\sim 100\ R_g \simeq 3000\,$km for a $21.4\, \rm M_{\odot}$ black hole. Moreover, the jet base radius of \gx is much larger than that of \cyg ($\sim 100$ and 5$R_g$, respectively), hence the density of the target particles is significantly decreased for the case of \gx assuming a similar injected power (of the order of $5\times 10^{37}\, \rm erg\, s^{-1}$).

\subsection{\bhs as Galactic neutrino sources}

Comparing the total neutrino spectrum of the 35~known \bhs to the astrophysical background of \ic \citep{stettner2019measurement}, we find that \bhs cannot contribute more than 0.1~percent at 40\,TeV even if all the sources accelerate protons into a hard non-thermal power law with index $p=1.7$ (see Fig.~\ref{fig: intrinsic neutrino emission and contribution to total spectrum}). The contribution of \bhs significantly drops at larger energies and eventually cuts off at approximately 1\,PeV. 
On the other hand, we used some very conservative assumptions and totally neglected neutron star X-ray binaries, as well as we only account for the intrinsic emission and neglect the impact the jets may have on the interstellar medium \citep[see e.g. the case of SS~433][]{abeysekara2018very}.

\subsection{Gamma-ray emission counterpart}
\begin{figure}
    	\includegraphics[width=1.05\columnwidth]{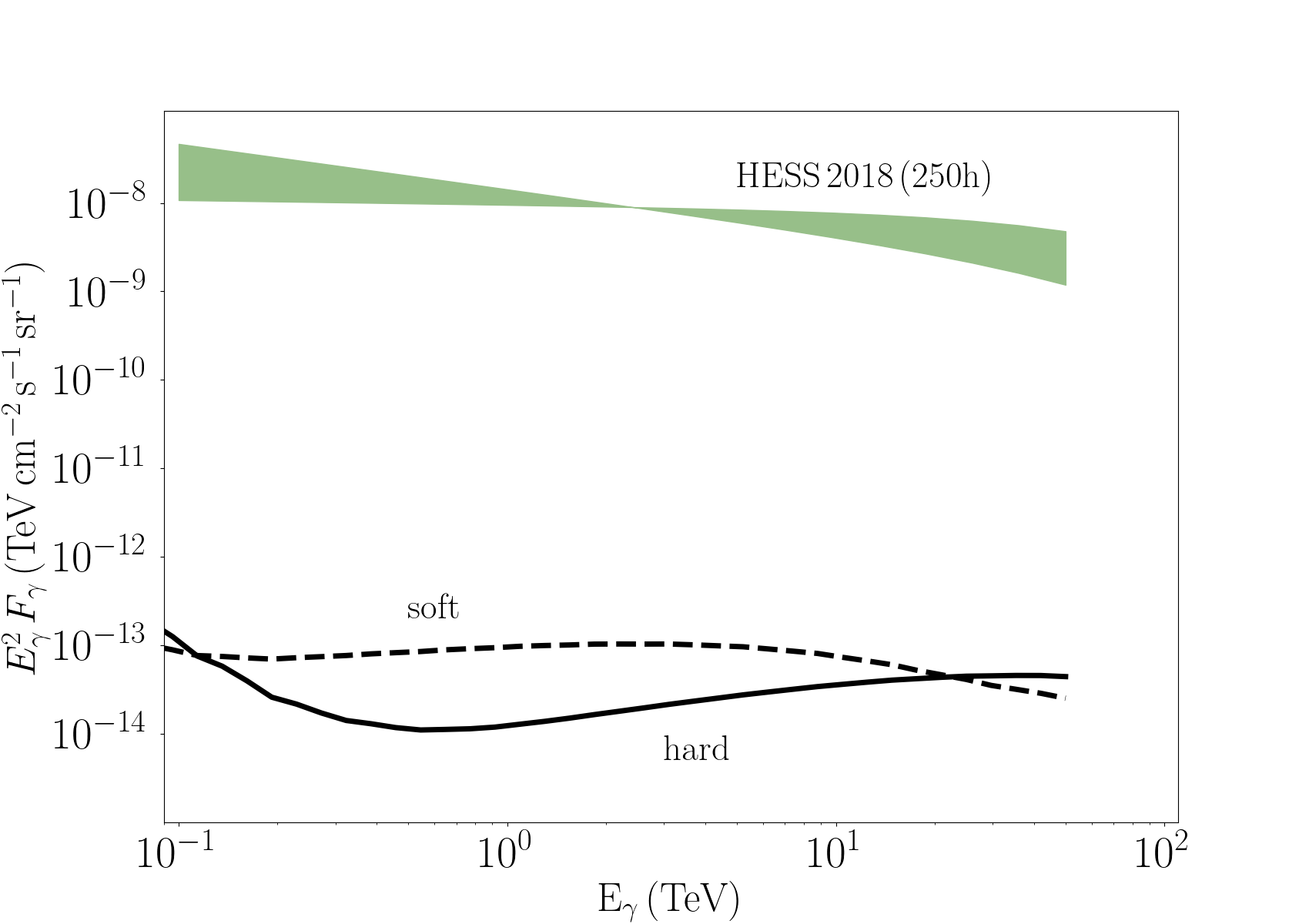}
    \caption{The overall intrinsic \gr emission of the 35 \bhs of this work. We assume that the accelerated protons follow either a soft power-law of $p=2.2$ (dashed line) or a hard power-law of $p=1.7$ (solid line). We also plot the diffuse \gr emission detected after 250 hours by \hess \citep[][]{HESS2018GalacticCenter} as a reference to our predicted \gr emission. 
    }
    \label{fig: gamma rays total}
\end{figure}

We compare the overall GeV-to-TeV emission of the 35 \bh sources for the two different power-law indices to the observed flux (see Fig.~\ref{fig: gamma rays total}). The \hess collaboration has detected the Galactic plane in the same energy range, we hence use these observations for the comparison. We use the result of the best fit of the observational data that is a power-law with an exponential cutoff at 100\,TeV, a photon index of $2.28\pm 0.02_{\rm stat}\pm 0.2_{\rm syst}$, and a normalisation at 1\,TeV of $1.2\pm 0.04_{\rm stat}\pm 0.2_{\rm syst}\times 10^{-8}\,\rm TeV\,cm^{-2}\,s^{-1}\,sr^{-1}$ \citep[][]{HESS2018GalacticCenter}. This diffuse \gr emission is from the inner 200\,pc of the Galactic Center, but not all the sources we study here are located in this region. These observations, however, originate in a \gr bright region of the sky and hence we use them as an upper limit for the overall contribution of the so-far known \bhs. We see that the predicted intrinsic high-energy emission from the \bhs is of the order of five orders of magnitude below the \hess observations and thus do not violate the \gr constraints. We stress once more though that we neglect any jet impacting on the surrounding medium that could lead to further \gr and neutrino emission. 

\begin{figure}
	\includegraphics[width=1.05\columnwidth]{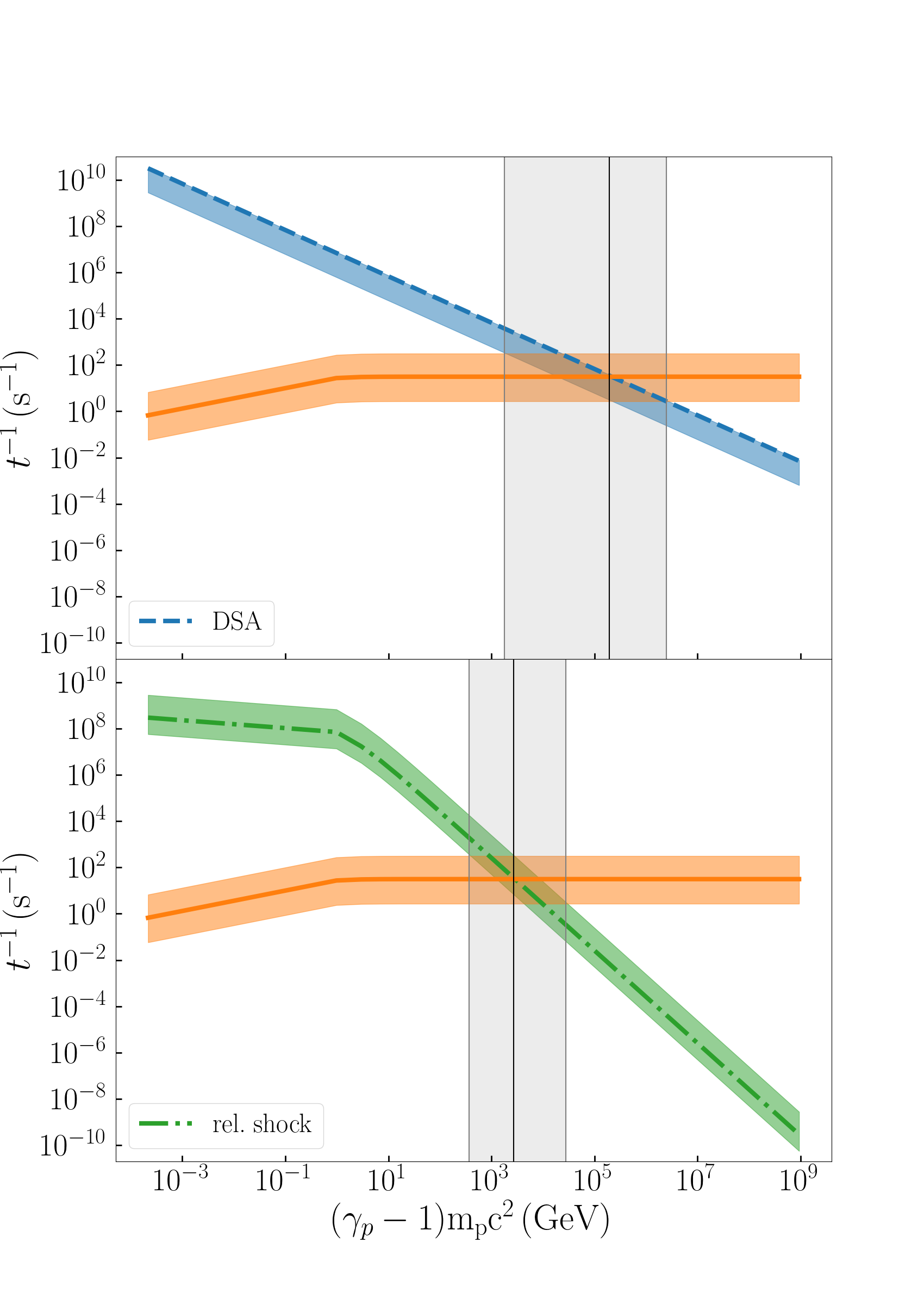}
    \caption{The acceleration and escape rates of CRs in \lms. When the two timescales are equal, the accelerated protons reach their maximum energy, as we show with the vertical black line. We show in both plots with a solid orange line the confinement timescale. In the \textit{top} panel, we assume a Bohm-like, DSA-type acceleration such that $\gamma_p \propto t$ (dashed line), and in the \textit{bottom} we assume a relativistic shock acceleration where $\gamma_p \propto t^{1/2}$ (dot-dashed line). The shaded regions show the range of the maximum energy within 1\,$\sigma$ uncertainty.
    }
    \label{fig: proton acceleration}
\end{figure}

\subsection{\bhs and the CR spectrum}\label{sec: bhs and the cr spectrum}

Recent population synthesis results \citep[see, e.g.,][]{Olejak2019synthesis}, and in agreement with new X-ray observations of the Galactic centre \citep{Hailey2018cusp,Mori_2021} suggest it is likely that of the order of 1000 accreting \bhs exist in the Galaxy capable of launching jets. Such systems may have faint radio jets currently in quiescence, or their outbursts may otherwise be obscured. 
In this work, we use \texttt{DRAGON2} to find their likely contribution to the CR spectrum while properly accounting for propagation in the Galactic disc (see Fig.~\ref{fig: CR spectrum for 1000 sources}). We find that \bhs can contribute up to maximum 50~per cent in the low-energy regime of the spectrum close to TeV energies. This percentage can be decreased by a factor of 3 if the power carried away by the escaping protons is 3 per cent instead of 10 per cent of the total proton power as we assume above \citep{Bell2013escape}, leading to a contribution of up to $\sim$ 17 per cent. This contribution is in broad agreement with the findings presented in \coop below the knee. 

There are two notable differences between the CR contribution calculated in this work and the one in \coop: the larger CR flux in the $\sim$TeV regime of the spectrum compared to the pessimistic model of \coop, and the lack of contribution at higher energies. The first can be explained by the fact that a softer power law is assumed, resulting in the steeper slope observed in Fig.~\ref{fig: CR spectrum for 1000 sources}. Secondly,
\coop did not perform multiwavelength fitting to obtain the model parameters, but used instead some simple scaling arguments. In \kcyg and \kgx we demonstrated that by fitting the entire spectrum we get additional constraints on the jet physical parameters that rather drastically reduce the maximum energy. Within the scope of our assumptions, we find that 
\cyg \& \gx require a high-energy hadronic cut-off for accelerated protons at a few hundreds of TeV. Given this, we find that, at least for these prototype sources, \bhs cannot reproduce this prominent excess in the 100\,PeV energy band that would distinguish them from other Galactic or extragalactic sources, as suggested by \coop. Further multi-wavelength studies of \bhs may aid in estimating the contribution of \bhs in the CR spectrum.

The maximum energy of accelerated CRs is an important parameter for the CR contribution, and also the neutrino spectrum. In this work, we assume that all the \lms follow the behaviour of the prototypical source \gx for which, based on the results of \kgx, protons get accelerated only up to $\sim 100$\,TeV. In \kcyg and \kgx, we see that the maximum energy is driven by the escape and not the losses, and hence we calculate this maximum energy by equating the acceleration timescale to the confinement timescale \citep[][]{hillas1984origin}. For a Bohm-like acceleration where particles diffusively cross a shock front (diffusive shock acceleration; DSA, also known as first-order Fermi acceleration), $t_{\rm acc} \propto E_p/f_{sc}$, where $E_p$ is the proton energy and $f_{sc}$ is the ratio of the scattering mean free path to the gyroradius and is usually considered the acceleration efficiency \citep[]{jokipii1987rate,aharonian2002proton}. The confinement timescale $t_{\rm esc} \propto R_{\rm acc}$, where $R_{\rm acc}$ is the radius of the acceleration site. 
$f_{sc}$ is the most impactful free parameter, whose exact value depends on the acceleration mechanism and is not well constrained yet. \kgx set $f_{sc}=0.01$, a value that allows for the primary electrons to reach high enough energy to radiate in the GeV regime through ICS without overshooting the \textit{Fermi}/LAT upper limits (\coop assume $f_{sc}=0.1$). Similar to \kgx, we assume that the acceleration efficiency is the same for both primary electrons and protons because we do not have a more robust constraint for the $f_{sc}$ of protons. Recent numerical simulations of particle acceleration though suggest that the efficiency of electrons and protons may differ, and in fact the proton acceleration may be more efficient depending on the kinematic conditions of the acceleration site \citep[][]{Caprioli2014i,crumley2019kinetic}. 
A more efficient proton acceleration would allow for protons to accelerate to energies beyond 100\,TeV, and hence allowing for the \bh population to contribute more significantly in the CR spectrum, as in \coop.

The efficiency of the particle acceleration also depends on the magnetisation of the shock and its medium, as well as the obliquity of the system, which is defined as the angle between the upstream magnetic field vector and the shock normal. In relativistic, magnetised media as we expect in the jets we examine here, when the magnetic field is almost aligned with the shock front, DSA is thought to be inefficient
\citep[][]{Sironi2013}. In that case in particular, the acceleration timescale based on particle-in-cell numerical simulations (PIC) is 
$t_{\rm acc,pic}= 4\gamma_p^2/\Gamma^2 \omega_{pi}$ \citep[][]{Sironi2013}. Here, $\omega_{pi} = \sqrt{\rm {m_e/m_p}}\,\omega_{pe}$ and $\omega_{pe} = (4\pi e^2n_e/\Gamma \rm m_e)^{1/2}$ are the plasma frequencies for the protons and the electrons, respectively. In the plasma frequency, $e$ is the elementary charge, $n_e$ is the number density of the particles at the acceleration region, and $\Gamma$ is its bulk Lorentz factor.

In Fig.~\ref{fig: proton acceleration}, we plot the acceleration and escape rates as a function of the proton energy $(\gamma_p-1)\rm m_pc^2$. The intersection of the two lines, indicates the maximum energy that protons can attain. In the top panel, we use the Bohm acceleration timescale, and in the bottom, we use the relativistic shock acceleration according to \citet[][]{Sironi2013} with $n_e = 10^{11}\, \rm cm^{-3}$ and $\Gamma = 2$. 
In both panels of Fig.~\ref{fig: proton acceleration}, we show the exact timescales (solid line for escape, dashed line for Bohm acceleration, dot-dashed line for relativistic shock) and indicate the maximum energy with a black vertical line. We show the possible values of the maximum energy within the shaded regions that are within the 1\,$\sigma$ uncertainty of the free parameters (see Table 3 of \kgx) and for $0.001\le f_{sc}\le 0.1$, where the former boundary indicates a relatively inefficient acceleration and the latter indicates an efficient particle acceleration. The exact dynamical values we use here correspond to the first particle acceleration region of the jet, but beyond that region, the maximum energy remains almost constant (see Fig.~4 of \coop). Overall, we see that protons in BHXBs can in principle accelerate to a maximum energy that ranges between 400\,GeV and 2.5\,PeV. 

An alternative plausible acceleration mechanism which could be more efficient, is magnetic reconnection \citep[][]{Drenkhahn2002,Uzdensky2011magnetic,Guo2014relativisticreconnection,Sironi_2014,sironi2015relativistic,kagan2015relativistic}. The acceleration timescale of magnetic reconnection scales as $t_{\rm rec}\propto E_p$ and based on PIC simulations, $t_{\rm rec} = E_p/(0.1ecB)$, where the factor 0.1 comes for the reconnection rate of the field lines $r_{\rm rec} = v_{in}/v_A$ \citep[][]{Sironi_2014} In the case of a strongly magnetised medium, the Alfv\'en velocity $v_A = c \sqrt{\sigma/(1+\sigma)}\simeq c$, and hence the inflow velocity $v_{in}\simeq 0.1c$. The inflow velocity defines the electric field responsible for the acceleration $E_{\rm rec} \simeq 0.1 B$ \citep[][]{Uzdensky2011magnetic,Sironi_2014}. Comparing the reconnection timescale to the Bohm acceleration timescale, we see that for $f_{sc}\simeq 0.1$, the two timescales are equal if Bohm acceleration is very efficient. 
In strong magnetised regions, however, one has to consider the effect of the synchrotron radiation on the accelerated particles, and hence a more self-consistent treatment is necessary \citep[][]{Hakobyan_2019, Ripperda_2022}. 
Consequently, in the efficient acceleration regime of DSA, or in the case of magnetic reconnection where particles do not radiate away their energy immediately, it is likely for \lm jets to accelerate particles to slightly larger energies. If such a scenario is feasible for the majority of the Galactic \lms, then that would increase their contribution to the CR spectrum (Fig.~\ref{fig: CR spectrum for 1000 sources}). In Fig.~\ref{app:fig: proton acceleration for high-mass bh} we show the maximum CR energy for the case of \hms based on \kcyg.

\subsection{Heavier element contribution}
The non-thermal acceleration of heavier nuclei in shocks is more efficient because the acceleration timescale scales inversely with the charge of the accelerated element \citep[see e.g.][for the case of SNRs]{CAPRIOLI2011nonlinear}. However, there is no evidence for non-thermal acceleration of heavier elements in `normal' \bhs. There is only the case of SS433 that shows iron line emission \citep{Migliari2002iron} but this is a unique system embedded in a nebula and seemingly quite slow mass loaded. To properly account for heavier elements in our approach, we require the exact abundances that based on the Galactic CR proton-to-helium ratio is less than 10 per cent for energies above $\sim$10GeV \citep{Adriani2011pamela}, and the proton-to-iron ratio drops way below 10 per cent for CR energies below 1TeV \citep{Thoudam2016spectrum}. Moreover, to calculate the products of inelastic interactions of heavier elements, we need the cross-sections that are less constrained than proton cross-sections. Hence, the assumption of only protons is fair within the astrophysical uncertainties.

\section{Summary and Conclusions}\label{sec: summary and conclusions}
Despite detailed studies over many decades, we still do not fully understand the origin of CRs, even those of Galactic nature. CRs are deflected after leaving their acceleration sites due to the Galactic magnetic field, making it impossible to use their location on the sky to pinpoint their sources. We can however use indirect, multi-messenger means to examine the CR acceleration in astrophysical sources by studying the intrinsic \grs and neutrinos produced, as well as the diffuse emission when accelerated CRs interact with the surrounding medium and/or radiation.

In this work,
we have presented a multizone, lepto-hadronic model for \bh jets (in the hard state), which accounts for neutrino distributions due to the inelastic pp and \pg collisions (\citealt[][]{kelner2006energy} and \citealt[][]{kelner2008energy}, respectively). We first produce the intrinsic neutrino distributions for the two canonical \bhs, namely the high-mass \cyg, and the low-mass \gx. We constrain the neutrino emission based on the work of \kcyg for \cyg and \kgx for \gx, who constrain the proton acceleration in the \bh jets by reproducing the multiwavelength radio-to-\gr spectrum. We expand our predictions on the intrinsic neutrino 
emission for 35~known \bhs discovered so far. 
We find that \cyg is a promising candidate neutrino source ($\sim$1 muon neutrino per year while in hard state for \ic), and in fact, our predicted muon neutrino flux is close to the most up-to-date upper-limits set by \ic. 
From the other 34~sources, only one more source (MAXI~J1836-194) might warrant further, more detailed investigation.

Future imaging Cherenkov telescopes, such as the Cherenkov Telescope Array \citep[][]{ACHARYA20133} will be up to five orders of magnitude more sensitive to transient BHXB jet outburst than current facilities. 
Cubic kilometer neutrino detectors, such as \arca \citep[][]{AdrianMartinez2016KM3NeT2} in the Mediterranean Sea, Baikal Deep Underwater Neutrino Telescope in Baikal \citep{WISCHNEWSKI2005Baikal} and \icG in the South Pole, which will be 5 times more sensitive than the current facilities \citep[][]{IceCube-Gen2,Aartsen_2021gen2}, will be able to further constrain the diffuse emission, and hence lead to a better understanding of CR acceleration.

Recent X-ray observations of the Galactic centre support the existence of hundreds to thousands \bhs in quiescence that remain undetected due to their duty cycle. In this work, we assume the existence of 1000 such sources distributed in the Galaxy capable of accelerating CRs following the behaviour of the two prototypical sources in the hard state. When CRs escape the acceleration sites, they interact with the Galactic molecular gas while they propagate in the Galactic plane, to produce further \grs and neutrinos that contribute to the diffuse spectra we detect on Earth. To properly study the CR propagation we use \texttt{DRAGON2} \citep[][]{Evoli_2017,Evoli_2018}, and to self-consistently calculate the \gr and neutrino counterparts, we use the publicly available code \texttt{HERMES} \citep[][]{Dundovic2021hermes}. These 1000 \bhs are able to contribute up to 50 per cent in the TeV regime of the CR spectrum, but whether they can contribute in the PeV regime strongly depends on the particle acceleration properties that are still poorly constrained. This contribution of 50 per cent in the TeV regime would decrease if the power of protons that escape the jets is less than 10 per cent of the total proton power as we assumed in this work. Finally, the exact number of the Galactic \bhs is not well constrained but according to the early data release of the third \textit{Gaia} catalogue \citep[][]{GaiaDR3201early,Gomel2022dr3}, which includes over 30 million stars, \textit{Gaia} will be able to classify non-single stars and provide a more accurate number of \bhs in quiescence, which are all candidate CR sources when they go into outbursts. If these sources can also contribute to the non-thermal \gr and neutrino emission, remains an open question that can we can tackle with the development of better theoretical infrastructure similar to this work.

The field of high-energy astrophysics with Galactic transient sources, especially those that launch relativistic jets, is in rapid development. Now that we know that the jets of several \bhs interact with the surrounding medium leading to hundreds of TeV \grs (\citealt{abeysekara2018very}, Olivera-Nieto~et~al.~in~prep) we may need to add another location for particle acceleration in the interstellar medium that would have different conditions but also implications on our understanding of CR sources.

\section*{Acknowledgements}
We would like to thank the anonymous reviewer for their fruitful comments to improve the quality of the manuscript. DK acknowledges funding from the French Programme d’investissements d’avenir through the Enigmass Labex. SM and DK are grateful for support from the Dutch Research Council (NWO) VICI grant (no. 639.043.513). 
PTL is supported by the Swedish Research Council under contract 2019-05135. DG acknowledges support from the project “Theoretical Astroparticle Physics (TAsP)” funded by the INFN. 

\section*{Data availability}
No new data were generated or analysed in support of this research.




\bibliographystyle{mnras}
\bibliography{neutrinos} 

\begin{thebibliography}{}
\makeatletter
\relax
\def\mn@urlcharsother{\let\do\@makeother \do\$\do\&\do\#\do\^\do\_\do\%\do\~}
\def\mn@doi{\begingroup\mn@urlcharsother \@ifnextchar [ {\mn@doi@}
  {\mn@doi@[]}}
\def\mn@doi@[#1]#2{\def\@tempa{#1}\ifx\@tempa\@empty \href
  {http://dx.doi.org/#2} {doi:#2}\else \href {http://dx.doi.org/#2} {#1}\fi
  \endgroup}
\def\mn@eprint#1#2{\mn@eprint@#1:#2::\@nil}
\def\mn@eprint@arXiv#1{\href {http://arxiv.org/abs/#1} {{\tt arXiv:#1}}}
\def\mn@eprint@dblp#1{\href {http://dblp.uni-trier.de/rec/bibtex/#1.xml}
  {dblp:#1}}
\def\mn@eprint@#1:#2:#3:#4\@nil{\def\@tempa {#1}\def\@tempb {#2}\def\@tempc
  {#3}\ifx \@tempc \@empty \let \@tempc \@tempb \let \@tempb \@tempa \fi \ifx
  \@tempb \@empty \def\@tempb {arXiv}\fi \@ifundefined
  {mn@eprint@\@tempb}{\@tempb:\@tempc}{\expandafter \expandafter \csname
  mn@eprint@\@tempb\endcsname \expandafter{\@tempc}}}

\bibitem[\protect\citeauthoryear{Aartsen et~al.,}{Aartsen
  et~al.}{2013}]{aartsen2013evidence}
Aartsen M.~G.,  et~al., 2013, \mn@doi [Science] {10.1126/science.1242856},
  \href {https://science.sciencemag.org/content/342/6161/1242856} {342,
  1242856}

\bibitem[\protect\citeauthoryear{Aartsen et~al.,}{Aartsen
  et~al.}{2015}]{aartsen2015combined}
Aartsen M.~G.,  et~al., 2015, \mn@doi [ApJ] {10.1088/0004-637x/809/1/98}, \href
  {https://doi.org/10.1088%2F0004-637x%2F809%2F1%2F98} {809, 98}

\bibitem[\protect\citeauthoryear{Aartsen et~al.,}{Aartsen
  et~al.}{2016}]{aartsen2016observation}
Aartsen M.~G.,  et~al., 2016, \mn@doi [ApJ] {10.3847/0004-637x/833/1/3}, \href
  {https://doi.org/10.3847%2F0004-637x%2F833%2F1%2F3} {833, 3}

\bibitem[\protect\citeauthoryear{Aartsen et~al.,}{Aartsen
  et~al.}{2017a}]{Aartsen_2017}
Aartsen M.~G.,  et~al., 2017a, \mn@doi [ApJ] {10.3847/1538-4357/835/2/151},
  \href {https://doi.org/10.3847/1538-4357/835/2/151} {835, 151}

\bibitem[\protect\citeauthoryear{Aartsen et~al.,}{Aartsen
  et~al.}{2017b}]{aartsen2017constraints}
Aartsen M.~G.,  et~al., 2017b, \mn@doi [ApJ] {10.3847/1538-4357/aa8dfb}, \href
  {https://doi.org/10.3847%2F1538-4357%2Faa8dfb} {849, 67}

\bibitem[\protect\citeauthoryear{Aartsen et~al.,}{Aartsen
  et~al.}{2018}]{Aartsen2018Multimessenger}
Aartsen M.,  et~al., 2018, \mn@doi [Science] {10.1126/science.aat1378}, \href
  {https://www.science.org/doi/abs/10.1126/science.aat1378} {361, 1378}

\bibitem[\protect\citeauthoryear{Aartsen et~al.,}{Aartsen
  et~al.}{2019}]{Aartsen_2019}
Aartsen M.~G.,  et~al., 2019, \mn@doi [ApJ] {10.3847/1538-4357/ab4ae2}, \href
  {https://doi.org/10.3847%2F1538-4357%2Fab4ae2} {886, 12}

\bibitem[\protect\citeauthoryear{Aartsen et~al.,}{Aartsen
  et~al.}{2020}]{aartsen201910years}
Aartsen M.~G.,  et~al., 2020, \mn@doi [Phys. Rev. Lett.]
  {10.1103/PhysRevLett.124.051103}, \href
  {https://link.aps.org/doi/10.1103/PhysRevLett.124.051103} {124, 051103}

\bibitem[\protect\citeauthoryear{Aartsen et~al.,}{Aartsen
  et~al.}{2021}]{Aartsen_2021gen2}
Aartsen M.~G.,  et~al., 2021, \mn@doi [J. Phys. G: Nucl. Part. Phys.]
  {10.1088/1361-6471/abbd48}, \href {https://doi.org/10.1088/1361-6471/abbd48}
  {48, 060501}

\bibitem[\protect\citeauthoryear{Abbasi et~al.,}{Abbasi
  et~al.}{2022}]{abbasi2022search}
Abbasi R.,  et~al., 2022, \mn@doi [ApJ Letters] {10.3847/2041-8213/ac67d8},
  \href {https://arxiv.org/abs/2202.11722} {930, L24}

\bibitem[\protect\citeauthoryear{Abdo et~al.,}{Abdo
  et~al.}{2008}]{abdo2008measurement}
Abdo A.,  et~al., 2008, \mn@doi [ApJ] {10.1086/592213}, 688

\bibitem[\protect\citeauthoryear{Abdollahi et~al.,}{Abdollahi
  et~al.}{2020}]{Abdollahi_2020}
Abdollahi S.,  et~al., 2020, \mn@doi [ApJ Supplement Series]
  {10.3847/1538-4365/ab6bcb}, \href {https://doi.org/10.3847/1538-4365/ab6bcb}
  {247, 33}

\bibitem[\protect\citeauthoryear{Abeysekara et~al.,}{Abeysekara
  et~al.}{2018}]{abeysekara2018very}
Abeysekara A.,  et~al., 2018, \mn@doi [Nature] {10.1038/s41586-018-0565-5},
  \href {https://doi.org/10.1038/s41586-018-0565-5} {562, 82}

\bibitem[\protect\citeauthoryear{Abramowski et~al.,}{Abramowski
  et~al.}{2011}]{Abramowski2011new}
Abramowski A.,  et~al., 2011, \mn@doi [A\&A] {10.1051/0004-6361/201016425},
  \href {https://doi.org/10.1051/0004-6361/201016425} {531, A81}

\bibitem[\protect\citeauthoryear{Abramowski et~al.,}{Abramowski
  et~al.}{2012}]{Abramowski2012discovery}
Abramowski A.,  et~al., 2012, \mn@doi [A\&A] {10.1051/0004-6361/201117928},
  \href {https://doi.org/10.1051/0004-6361/201117928} {537, A114}

\bibitem[\protect\citeauthoryear{Abramowski et~al.,}{Abramowski
  et~al.}{2014}]{Abramowski2014HESSJ1640_465}
Abramowski A.,  et~al., 2014, \mn@doi [MNRAS] {10.1093/mnras/stu139}, \href
  {https://doi.org/10.1093/mnras/stu139} {439, 2828}

\bibitem[\protect\citeauthoryear{Abramowski et~al.,}{Abramowski
  et~al.}{2016}]{abramowski2016acceleration}
Abramowski A.,  et~al., 2016, \mn@doi [{Nature}] {10.1038/nature17147}, \href
  {http://hal.in2p3.fr/in2p3-01303680} {531, 476}

\bibitem[\protect\citeauthoryear{Abreu et~al.,}{Abreu
  et~al.}{2021}]{Auger2021spectrum}
Abreu P.,  et~al., 2021, \mn@doi [The European Physical Journal C]
  {10.1140/epjc/s10052-021-09700-w}, \href
  {https://doi.org/10.1140%2Fepjc%2Fs10052-021-09700-w} {81}

\bibitem[\protect\citeauthoryear{Acharya et~al.,}{Acharya
  et~al.}{2013}]{ACHARYA20133}
Acharya B.,  et~al., 2013, \mn@doi [Astroparticle Physics]
  {10.1016/j.astropartphys.2013.01.007}, \href
  {https://www.sciencedirect.com/science/article/pii/S0927650513000169} {43, 3}

\bibitem[\protect\citeauthoryear{Ackermann et~al.,}{Ackermann
  et~al.}{2011}]{Ackermann2011coccon}
Ackermann M.,  et~al., 2011, \mn@doi [Science] {10.1126/science.1210311}, \href
  {https://www.science.org/doi/abs/10.1126/science.1210311} {334, 1103}

\bibitem[\protect\citeauthoryear{{Ackermann} et~al.,}{{Ackermann}
  et~al.}{2012}]{fermilat2012galacticcentre}
{Ackermann} M.,  et~al., 2012, \mn@doi [\apj] {10.1088/0004-637X/750/1/3},
  \href {https://ui.adsabs.harvard.edu/abs/2012ApJ...750....3A} {750, 3}

\bibitem[\protect\citeauthoryear{Ackermann et~al.,}{Ackermann
  et~al.}{2013}]{Ackermann2013detection}
Ackermann M.,  et~al., 2013, \mn@doi [Science] {10.1126/science.1231160}, \href
  {https://www.science.org/doi/abs/10.1126/science.1231160} {339, 807}

\bibitem[\protect\citeauthoryear{Adri{\'{a}}n-Mart{\'{\i}}nez
  et~al.,}{Adri{\'{a}}n-Mart{\'{\i}}nez
  et~al.}{2016}]{AdrianMartinez2016KM3NeT2}
Adri{\'{a}}n-Mart{\'{\i}}nez S.,  et~al., 2016, \mn@doi [J. Phys. G: Nucl.
  Part. Phys.] {10.1088/0954-3899/43/8/084001}, \href
  {https://doi.org/10.1088%2F0954-3899%2F43%2F8%2F084001} {43, 084001}

\bibitem[\protect\citeauthoryear{Adriani et~al.,}{Adriani
  et~al.}{2011}]{Adriani2011pamela}
Adriani O.,  et~al., 2011, \mn@doi [Science] {10.1126/science.1199172}, \href
  {https://www.science.org/doi/abs/10.1126/science.1199172} {332, 69}

\bibitem[\protect\citeauthoryear{Adrián-Martínez et~al.,}{Adrián-Martínez
  et~al.}{2016}]{ADRIANMARTINEZ2016143}
Adrián-Martínez S.,  et~al., 2016, \mn@doi [Physics Letters B]
  {https://doi.org/10.1016/j.physletb.2016.06.051}, \href
  {http://www.sciencedirect.com/science/article/pii/S0370269316303112} {760,
  143 }

\bibitem[\protect\citeauthoryear{Aharonian}{Aharonian}{2002}]{aharonian2002proton}
Aharonian F.,  2002, \mn@doi [MNRAS] {10.1046/j.1365-8711.2002.05292.x}, \href
  {https://doi.org/10.1046/j.1365-8711.2002.05292.x} {332, 215}

\bibitem[\protect\citeauthoryear{Aharonian \& Atoyan}{Aharonian \&
  Atoyan}{2000}]{Aharonian2000broadband}
Aharonian F.~A.,  Atoyan A.~M.,  2000, \mn@doi [arXiv:0009009]
  {10.48550/ARXIV.ASTRO-PH/0009009}

\bibitem[\protect\citeauthoryear{Aharonian et~al.,}{Aharonian
  et~al.}{2005}]{Aharonian2005discoveryXRB}
Aharonian F.,  et~al., 2005, \mn@doi [Science] {10.1126/science.1113764}, \href
  {https://science.sciencemag.org/content/309/5735/746} {309, 746}

\bibitem[\protect\citeauthoryear{Aharonian et~al.,}{Aharonian
  et~al.}{2006}]{Aharonian2006detailed}
Aharonian F.,  et~al., 2006, \mn@doi [A\&A] {10.1051/0004-6361:20054279}, \href
  {https://doi.org/10.1051/0004-6361:20054279} {449, 223}

\bibitem[\protect\citeauthoryear{Aharonian et~al.,}{Aharonian
  et~al.}{2007a}]{Aharonian2007primary}
Aharonian F.,  et~al., 2007a, \mn@doi [A\&A] {10.1051/0004-6361:20066381},
  \href {https://doi.org/10.1051/0004-6361:20066381} {464, 235}

\bibitem[\protect\citeauthoryear{Aharonian et~al.,}{Aharonian
  et~al.}{2007b}]{Aharonian_2007_observations}
Aharonian F.,  et~al., 2007b, \mn@doi [ApJ] {10.1086/512603}, \href
  {https://doi.org/10.1086/512603} {661, 236}

\bibitem[\protect\citeauthoryear{Aharonian et~al.,}{Aharonian
  et~al.}{2009}]{Aharonian_2009_discovery}
Aharonian F.,  et~al., 2009, \mn@doi [ApJ] {10.1088/0004-637x/692/2/1500},
  \href {https://doi.org/10.1088/0004-637x/692/2/1500} {692, 1500}

\bibitem[\protect\citeauthoryear{Aharonian, Yang  \& de
  O{\~n}a~Wilhelmi}{Aharonian et~al.}{2019}]{aharonian2018massive}
Aharonian F.,  Yang R.,   de O{\~n}a~Wilhelmi E.,  2019, \mn@doi [Nature
  Astronomy] {10.1038/s41550-019-0724-0}, \href
  {https://doi.org/10.1038/s41550-019-0724-0} {3, 561}

\bibitem[\protect\citeauthoryear{Aiello et~al.,}{Aiello
  et~al.}{2019}]{aiello2019arca}
Aiello S.,  et~al., 2019, \mn@doi [Astroparticle Physics]
  {https://doi.org/10.1016/j.astropartphys.2019.04.002}, \href
  {http://www.sciencedirect.com/science/article/pii/S0927650518302809} {111,
  100 }

\bibitem[\protect\citeauthoryear{Albert et~al.,}{Albert
  et~al.}{2017}]{albert2017ANTARES}
Albert A.,  et~al., 2017, \mn@doi [Phys. Rev. D] {10.1103/PhysRevD.96.082001},
  \href {https://link.aps.org/doi/10.1103/PhysRevD.96.082001} {96, 082001}

\bibitem[\protect\citeauthoryear{Albert et~al.,}{Albert
  et~al.}{2018}]{Albert_2018}
Albert A.,  et~al., 2018, \mn@doi [ApJ] {10.3847/2041-8213/aaeecf}, \href
  {https://doi.org/10.3847%2F2041-8213%2Faaeecf} {868, L20}

\bibitem[\protect\citeauthoryear{Albert et~al.,}{Albert
  et~al.}{2020}]{Albert_2020HAWC}
Albert A.,  et~al., 2020, \mn@doi [ApJ] {10.3847/1538-4357/abc2d8}, \href
  {https://doi.org/10.3847/1538-4357/abc2d8} {905, 76}

\bibitem[\protect\citeauthoryear{Albert et~al.,}{Albert
  et~al.}{2021}]{Albert_2021_HMmicroquasars}
Albert A.,  et~al., 2021, \mn@doi [ApJ Letters] {10.3847/2041-8213/abf35a},
  \href {https://doi.org/10.3847/2041-8213/abf35a} {912, L4}

\bibitem[\protect\citeauthoryear{Aleksi\'{c} et~al.,}{Aleksi\'{c}
  et~al.}{2015}]{Aleksic2015MagicObservations}
Aleksi\'{c} J.,  et~al., 2015, \mn@doi [A\&A] {10.1051/0004-6361/201424879},
  \href {https://doi.org/10.1051/0004-6361/201424879} {576, A36}

\bibitem[\protect\citeauthoryear{Amenomori et~al.,}{Amenomori
  et~al.}{2019}]{Amenomori2019first}
Amenomori M.,  et~al., 2019, \mn@doi [Phys. Rev. Lett.]
  {10.1103/PhysRevLett.123.051101}, \href
  {https://link.aps.org/doi/10.1103/PhysRevLett.123.051101} {123, 051101}

\bibitem[\protect\citeauthoryear{Amenomori et~al.,}{Amenomori
  et~al.}{2021}]{Amenomori2021first}
Amenomori M.,  et~al., 2021, \mn@doi [Phys. Rev. Lett.]
  {10.1103/PhysRevLett.126.141101}, \href
  {https://link.aps.org/doi/10.1103/PhysRevLett.126.141101} {126, 141101}

\bibitem[\protect\citeauthoryear{Anchordoqui, Goldberg, Paul, da Silva  \&
  Vlcek}{Anchordoqui et~al.}{2014}]{Anchordoqui2014estimating}
Anchordoqui L.~A.,  Goldberg H.,  Paul T.~C.,  da Silva L. H.~M.,   Vlcek
  B.~J.,  2014, \mn@doi [Phys. Rev. D] {10.1103/PhysRevD.90.123010}, \href
  {https://link.aps.org/doi/10.1103/PhysRevD.90.123010} {90, 123010}

\bibitem[\protect\citeauthoryear{Apel et~al.,}{Apel
  et~al.}{2011}]{Apel2011kascade}
Apel W.~D.,  et~al., 2011, \mn@doi [Phys. Rev. Lett.]
  {10.1103/PhysRevLett.107.171104}, \href
  {https://link.aps.org/doi/10.1103/PhysRevLett.107.171104} {107, 171104}

\bibitem[\protect\citeauthoryear{Archambault et~al.,}{Archambault
  et~al.}{2017}]{Archambault_2017observations}
Archambault S.,  et~al., 2017, \mn@doi [ApJ] {10.3847/1538-4357/836/1/23},
  \href {https://doi.org/10.3847/1538-4357/836/1/23} {836, 23}

\bibitem[\protect\citeauthoryear{Atoyan \& Dermer}{Atoyan \&
  Dermer}{2003}]{Atoyan_2003}
Atoyan A.~M.,  Dermer C.~D.,  2003, \mn@doi [ApJ] {10.1086/346261}, \href
  {https://doi.org/10.1086/346261} {586, 79}

\bibitem[\protect\citeauthoryear{Atri et~al.,}{Atri
  et~al.}{2020}]{Atri2020parallax}
Atri P.,  et~al., 2020, \mn@doi [MNRAS: Letters] {10.1093/mnrasl/slaa010},
  \href {https://doi.org/10.1093/mnrasl/slaa010} {493, L81}

\bibitem[\protect\citeauthoryear{Baade \& Zwicky}{Baade \&
  Zwicky}{1934}]{baade1934cosmic}
Baade W.,  Zwicky F.,  1934, \mn@doi [Proceedings of the National Academy of
  Sciences] {10.1073/pnas.20.5.259}, \href
  {https://doi.org/10.1073/pnas.20.5.259} {20, 259}

\bibitem[\protect\citeauthoryear{Bednarek, Burgio  \& Montaruli}{Bednarek
  et~al.}{2005}]{BEDNAREK2005galactic}
Bednarek W.,  Burgio G.,   Montaruli T.,  2005, \mn@doi [New Astronomy Reviews]
  {https://doi.org/10.1016/j.newar.2004.11.001}, \href
  {http://www.sciencedirect.com/science/article/pii/S1387647304001538} {49, 1 }

\bibitem[\protect\citeauthoryear{Bell, Schure, Reville  \& Giacinti}{Bell
  et~al.}{2013}]{Bell2013escape}
Bell A.~R.,  Schure K.~M.,  Reville B.,   Giacinti G.,  2013, \mn@doi [Monthly
  Notices of the Royal Astronomical Society] {10.1093/mnras/stt179}, \href
  {https://doi.org/10.1093/mnras/stt179} {431, 415}

\bibitem[\protect\citeauthoryear{{Belloni}}{{Belloni}}{2010}]{Belloni2010}
{Belloni} T.,  ed. 2010, {The Jet Paradigm}  Lecture Notes in Physics, Berlin
  Springer Verlag Vol. 794, \mn@doi{10.1007/978-3-540-76937-8.
}

\bibitem[\protect\citeauthoryear{Belloni, Zhang, Kylafis, Reig  \&
  Altamirano}{Belloni et~al.}{2020}]{Belloni2020J1348}
Belloni T.~M.,  Zhang L.,  Kylafis N.~D.,  Reig P.,   Altamirano D.,  2020,
  \mn@doi [MNRAS] {10.1093/mnras/staa1843}, \href
  {https://doi.org/10.1093/mnras/staa1843} {496, 4366}

\bibitem[\protect\citeauthoryear{Berezinsky}{Berezinsky}{1991}]{BEREZINSKY1991375}
Berezinsky V.,  1991, \mn@doi [Nucl. Phys. B Proc. Suppl.]
  {10.1016/0920-5632(91)90215-Z}, \href
  {https://www.sciencedirect.com/science/article/pii/092056329190215Z} {19,
  375}

\bibitem[\protect\citeauthoryear{Blandford \& K{\"o}nigl}{Blandford \&
  K{\"o}nigl}{1979}]{blandford1979relativistic}
Blandford R.,  K{\"o}nigl A.,  1979, \mn@doi [ApJ] {10.1086/157262}, \href
  {https://ui.adsabs.harvard.edu/abs/1979ApJ...232...34B} {232, 34}

\bibitem[\protect\citeauthoryear{Blandford \& Payne}{Blandford \&
  Payne}{1982}]{Blandford1982hydromagnetic}
Blandford R.~D.,  Payne D.~G.,  1982, \mn@doi [MNRAS]
  {10.1093/mnras/199.4.883}, \href {https://doi.org/10.1093/mnras/199.4.883}
  {199, 883}

\bibitem[\protect\citeauthoryear{Blandford \& Znajek}{Blandford \&
  Znajek}{1977}]{blandford1977extraction}
Blandford R.~D.,  Znajek R.~L.,  1977, \mn@doi [MNRAS]
  {10.1093/mnras/179.3.433}, \href {https://doi.org/10.1093/mnras/179.3.433}
  {179, 433}

\bibitem[\protect\citeauthoryear{Blasi}{Blasi}{2013a}]{Blasi2013}
Blasi P.,  2013a, \mn@doi [The A\&A Review] {10.1007/s00159-013-0070-7}, \href
  {https://doi.org/10.1007/s00159-013-0070-7} {21, 70}

\bibitem[\protect\citeauthoryear{{Blasi}}{{Blasi}}{2013b}]{Blasi2013origin}
{Blasi} P.,  2013b, \mn@doi [Nuclear Physics B Proceedings Supplements]
  {10.1016/j.nuclphysbps.2013.05.023}, \href
  {https://ui.adsabs.harvard.edu/abs/2013NuPhS.239..140B} {239, 140}

\bibitem[\protect\citeauthoryear{Blattnig, Swaminathan, Kruger, Ngom  \&
  Norbury}{Blattnig et~al.}{2000}]{Blattnig2000Parametrizations}
Blattnig S.~R.,  Swaminathan S.~R.,  Kruger A.~T.,  Ngom M.,   Norbury J.~W.,
  2000, \mn@doi [Phys. Rev. D] {10.1103/PhysRevD.62.094030}, \href
  {https://link.aps.org/doi/10.1103/PhysRevD.62.094030} {62, 094030}

\bibitem[\protect\citeauthoryear{B\"{o}ttcher, Reimer, Sweeney  \&
  Prakash}{B\"{o}ttcher et~al.}{2013}]{boettcher2013leptohadronic}
B\"{o}ttcher M.,  Reimer A.,  Sweeney K.,   Prakash A.,  2013, \mn@doi [ApJ]
  {10.1088/0004-637x/768/1/54}, \href
  {https://doi.org/10.1088%2F0004-637x%2F768%2F1%2F54} {768, 54}

\bibitem[\protect\citeauthoryear{Buitink et~al.,}{Buitink
  et~al.}{2016}]{buitink2016large}
Buitink S.,  et~al., 2016, \mn@doi [Nature] {10.1038/nature16976}, \href
  {https://doi.org/10.1038/nature16976} {531, 70}

\bibitem[\protect\citeauthoryear{Cangemi et~al.,}{Cangemi
  et~al.}{2021}]{Cangemi2020longterm}
Cangemi F.,  et~al., 2021, \mn@doi [A\&A] {10.1051/0004-6361/202038604}, \href
  {https://doi.org/10.1051/0004-6361/202038604} {650, A93}

\bibitem[\protect\citeauthoryear{Cao, Lucchini, Markoff, Connors  \&
  Grinberg}{Cao et~al.}{2021a}]{cao2021evidence}
Cao Z.,  Lucchini M.,  Markoff S.,  Connors R. M.~T.,   Grinberg V.,  2021a,
  \mn@doi [MNRAS] {10.1093/mnras/stab3080}, \href
  {https://doi.org/10.1093/mnras/stab3080} {509, 2517}

\bibitem[\protect\citeauthoryear{Cao, Aharonian, An, {Axikegu}, Bai, Bai, Bao
  \& Bastieri}{Cao et~al.}{2021b}]{Cao2021PeV}
Cao Z.,  Aharonian F.~A.,  An Q.,  {Axikegu} Bai L.~X.,  Bai Y.~X.,  Bao Y.~W.,
    Bastieri D.,  2021b, \mn@doi [Nature] {10.1038/s41586-021-03498-z}, \href
  {https://doi.org/10.1038/s41586-021-03498-z} {594, 33}

\bibitem[\protect\citeauthoryear{Caprioli}{Caprioli}{2012}]{Caprioli2012nldsa}
Caprioli D.,  2012, \mn@doi [J. Cosmology Astropart. Phys.]
  {10.1088/1475-7516/2012/07/038}, 2012, 038

\bibitem[\protect\citeauthoryear{Caprioli \& Spitkovsky}{Caprioli \&
  Spitkovsky}{2014}]{Caprioli2014i}
Caprioli D.,  Spitkovsky A.,  2014, \mn@doi [ApJ] {10.1088/0004-637x/783/2/91},
  \href {https://doi.org/10.1088%2F0004-637x%2F783%2F2%2F91} {783, 91}

\bibitem[\protect\citeauthoryear{Caprioli, Blasi  \& Amato}{Caprioli
  et~al.}{2011}]{CAPRIOLI2011nonlinear}
Caprioli D.,  Blasi P.,   Amato E.,  2011, \mn@doi [Astroparticle Physics]
  {https://doi.org/10.1016/j.astropartphys.2010.10.011}, \href
  {https://www.sciencedirect.com/science/article/pii/S0927650510002082} {34,
  447}

\bibitem[\protect\citeauthoryear{Carulli, Reynoso  \& Romero}{Carulli
  et~al.}{2021}]{carulli2021neutrino}
Carulli A.~M.,  Reynoso M.~M.,   Romero G.~E.,  2021, \mn@doi [Astroparticle
  Physics] {10.1016/j.astropartphys.2021.102557}, \href
  {https://www.sciencedirect.com/science/article/pii/S0927650521000013} {128,
  102557}

\bibitem[\protect\citeauthoryear{Chang et~al.,}{Chang
  et~al.}{2008}]{chang2008excess}
Chang J.,  et~al., 2008, \mn@doi [Nature] {10.1038/nature07477}, \href
  {https://doi.org/10.1038/nature07477} {456, 362}

\bibitem[\protect\citeauthoryear{Chatzis, Petropoulou  \& Vasilopoulos}{Chatzis
  et~al.}{2021}]{Chatzis2022radio}
Chatzis M.,  Petropoulou M.,   Vasilopoulos G.,  2021, \mn@doi [MNRAS]
  {10.1093/mnras/stab3098}, \href {https://doi.org/10.1093/mnras/stab3098}
  {509, 2532}

\bibitem[\protect\citeauthoryear{Chauhan et~al.,}{Chauhan
  et~al.}{2019}]{Chauhan2019J1535}
Chauhan J.,  et~al., 2019, \mn@doi [MNRAS: Letters] {10.1093/mnrasl/slz113},
  \href {https://doi.org/10.1093/mnrasl/slz113} {488, L129}

\bibitem[\protect\citeauthoryear{Cooper, Gaggero, Markoff  \& Zhang}{Cooper
  et~al.}{2020}]{cooper2020xrbcrs}
Cooper A.~J.,  Gaggero D.,  Markoff S.,   Zhang S.,  2020, \mn@doi [MNRAS]
  {10.1093/mnras/staa373}, \href {https://doi.org/10.1093/mnras/staa373} {493,
  3212}

\bibitem[\protect\citeauthoryear{Corbel \& Fender}{Corbel \&
  Fender}{2002}]{Corbel2002NIR}
Corbel S.,  Fender R.~P.,  2002, \mn@doi [ApJ] {10.1086/341870}, \href
  {https://doi.org/10.1086%2F341870} {573, L35}

\bibitem[\protect\citeauthoryear{Corbel et~al.,}{Corbel
  et~al.}{2013}]{corbel2013formation}
Corbel S.,  et~al., 2013, \mn@doi [MNRAS: Letters] {10.1093/mnrasl/slt018},
  \href {https://doi.org/10.1093/mnrasl/slt018} {431, L107}

\bibitem[\protect\citeauthoryear{Coriat, Fender, Tasse, Smirnov, Tzioumis  \&
  Broderick}{Coriat et~al.}{2019}]{Coriat2019twisted}
Coriat M.,  Fender R.~P.,  Tasse C.,  Smirnov O.,  Tzioumis A.~K.,   Broderick
  J.~W.,  2019, \mn@doi [\mnras] {10.1093/mnras/stz099}, \href
  {https://ui.adsabs.harvard.edu/abs/2019MNRAS.484.1672C} {484, 1672}

\bibitem[\protect\citeauthoryear{{Crumley}, {Ceccobello}, {Connors}  \&
  {Cavecchi}}{{Crumley} et~al.}{2017}]{crumley2017symbiosis}
{Crumley} P.,  {Ceccobello} C.,  {Connors} R.~M.~T.,   {Cavecchi} Y.,  2017,
  \mn@doi [\aap] {10.1051/0004-6361/201630229}, \href
  {http://adsabs.harvard.edu/abs/2017A%26A...601A..87C} {601, A87}

\bibitem[\protect\citeauthoryear{Crumley, Caprioli, Markoff  \&
  Spitkovsky}{Crumley et~al.}{2019}]{crumley2019kinetic}
Crumley P.,  Caprioli D.,  Markoff S.,   Spitkovsky A.,  2019, \mn@doi [MNRAS]
  {10.1093/mnras/stz232}, \href {https://doi.org/10.1093/mnras/stz232} {485,
  5105}

\bibitem[\protect\citeauthoryear{{DAMPE collaboration} et~al.,}{{DAMPE
  collaboration} et~al.}{2019}]{dampe2019measurement}
{DAMPE collaboration} et~al., 2019, \mn@doi [Science advances]
  {10.1126/sciadv.aax3793}, \href
  {https://www.science.org/doi/abs/10.1126/sciadv.aax3793} {5, eaax3793}

\bibitem[\protect\citeauthoryear{De~La Torre~Luque, Gaggero, Grasso  \&
  Marinelli}{De~La Torre~Luque et~al.}{2022}]{Luque2022prospects}
De~La Torre~Luque P.,  Gaggero D.,  Grasso D.,   Marinelli A.,  2022, \mn@doi
  [Frontiers in Astronomy and Space Sciences] {10.3389/fspas.2022.1041838},
  \href {https://www.frontiersin.org/articles/10.3389/fspas.2022.1041838} {9}

\bibitem[\protect\citeauthoryear{Deegan, Combet  \& Wynn}{Deegan
  et~al.}{2009}]{Deegan2009duty}
Deegan P.,  Combet C.,   Wynn G.~A.,  2009, \mn@doi [MNRAS]
  {10.1111/j.1365-2966.2009.15573.x}, \href
  {https://doi.org/10.1111/j.1365-2966.2009.15573.x} {400, 1337}

\bibitem[\protect\citeauthoryear{Dermer}{Dermer}{1986}]{Dermer1986secondary}
Dermer C.~D.,  1986, \mn@doi [\aap]
  {https://ui.adsabs.harvard.edu/abs/1986A&A...157..223D}, \href
  {https://ui.adsabs.harvard.edu/abs/1986A&A...157..223D} {157, 223}

\bibitem[\protect\citeauthoryear{Distefano, Guetta, Waxman  \&
  Levinson}{Distefano et~al.}{2002}]{Distefano_2002}
Distefano C.,  Guetta D.,  Waxman E.,   Levinson A.,  2002, \mn@doi [ApJ]
  {10.1086/341144}, \href {https://doi.org/10.1086%2F341144} {575, 378}

\bibitem[\protect\citeauthoryear{Drenkhahn \& Spruit}{Drenkhahn \&
  Spruit}{2002}]{Drenkhahn2002}
Drenkhahn G.,  Spruit H.~C.,  2002, \mn@doi [A\&A]
  {10.1051/0004-6361:20020839}, \href
  {https://doi.org/10.1051/0004-6361:20020839} {391, 1141}

\bibitem[\protect\citeauthoryear{Drury}{Drury}{1983}]{drury1983introduction}
Drury L.~O.,  1983, \mn@doi [Rep. Prog. Phys.] {10.1088/0034-4885/46/8/002},
  \href {https://doi.org/10.1088%2F0034-4885%2F46%2F8%2F002} {46, 973}

\bibitem[\protect\citeauthoryear{Dundovic, Evoli, Gaggero  \& Grasso}{Dundovic
  et~al.}{2021}]{Dundovic2021hermes}
Dundovic A.,  Evoli C.,  Gaggero D.,   Grasso D.,  2021, \mn@doi [A\&A]
  {10.1051/0004-6361/202140801}, \href
  {https://doi.org/10.1051/0004-6361/202140801} {653, A18}

\bibitem[\protect\citeauthoryear{Evoli, Gaggero, Vittino, Bernardo, Mauro,
  Ligorini, Ullio  \& Grasso}{Evoli et~al.}{2017}]{Evoli_2017}
Evoli C.,  Gaggero D.,  Vittino A.,  Bernardo G.~D.,  Mauro M.~D.,  Ligorini
  A.,  Ullio P.,   Grasso D.,  2017, \mn@doi [J. Cosmology Astropart. Phys.]
  {10.1088/1475-7516/2017/02/015}, \href
  {https://doi.org/10.1088/1475-7516/2017/02/015} {2017, 015}

\bibitem[\protect\citeauthoryear{Evoli, Gaggero, Vittino, Mauro, Grasso  \&
  Mazziotta}{Evoli et~al.}{2018}]{Evoli_2018}
Evoli C.,  Gaggero D.,  Vittino A.,  Mauro M.~D.,  Grasso D.,   Mazziotta
  M.~N.,  2018, \mn@doi [J. Cosmology Astropart. Phys.]
  {10.1088/1475-7516/2018/07/006}, \href
  {https://doi.org/10.1088/1475-7516/2018/07/006} {2018, 006}

\bibitem[\protect\citeauthoryear{{Falcke} \& {Biermann}}{{Falcke} \&
  {Biermann}}{1995}]{falcke1995jet}
{Falcke} H.,  {Biermann} P.~L.,  1995, \mn@doi [\aap]
  {10.48550/ARXIV.ASTRO-PH/9411096}, \href
  {http://adsabs.harvard.edu/abs/1995A%26A...293..665F} {293, 665}

\bibitem[\protect\citeauthoryear{Fender, Pooley, Durouchoux, Tilanus  \&
  Brocksopp}{Fender et~al.}{2000}]{Fender2000flat}
Fender R.~P.,  Pooley G.~G.,  Durouchoux P.,  Tilanus R. P.~J.,   Brocksopp C.,
   2000, \mn@doi [MNRAS] {10.1046/j.1365-8711.2000.03219.x}, \href
  {https://doi.org/10.1046/j.1365-8711.2000.03219.x} {312, 853}

\bibitem[\protect\citeauthoryear{Fender, Homan  \& Belloni}{Fender
  et~al.}{2009}]{Fender2009jets}
Fender R.~P.,  Homan J.,   Belloni T.~M.,  2009, \mn@doi [MNRAS]
  {10.1111/j.1365-2966.2009.14841.x}, \href
  {https://doi.org/10.1111/j.1365-2966.2009.14841.x} {396, 1370}

\bibitem[\protect\citeauthoryear{Feyereisen, Tamborra  \& Ando}{Feyereisen
  et~al.}{2017}]{feyereisen2017one}
Feyereisen M.~R.,  Tamborra I.,   Ando S.,  2017, \mn@doi [J. Cosmology
  Astropart. Phys.] {10.1088/1475-7516/2017/03/057}, \href
  {https://doi.org/10.1088%2F1475-7516%2F2017%2F03%2F057} {2017, 057}

\bibitem[\protect\citeauthoryear{Fornieri, Gaggero  \& Grasso}{Fornieri
  et~al.}{2020}]{Fornieri2020}
Fornieri O.,  Gaggero D.,   Grasso D.,  2020, \mn@doi [JCAP]
  {10.1088/1475-7516/2020/02/009}, \href
  {https://doi.org/10.1088/1475-7516/2020/02/009} {2020, 009}

\bibitem[\protect\citeauthoryear{{Gaggero}, {Grasso}, {Marinelli}, {Urbano}  \&
  {Valli}}{{Gaggero} et~al.}{2015}]{gaggero2015KRAgamma}
{Gaggero} D.,  {Grasso} D.,  {Marinelli} A.,  {Urbano} A.,   {Valli} M.,  2015,
  \mn@doi [\apjl] {10.1088/2041-8205/815/2/L25}, \href
  {https://ui.adsabs.harvard.edu/abs/2015ApJ...815L..25G} {815, L25}

\bibitem[\protect\citeauthoryear{{Gaia Collaboration} et~al.,}{{Gaia
  Collaboration} et~al.}{2021}]{GaiaDR3201early}
{Gaia Collaboration} et~al., 2021, \mn@doi [A\&A]
  {10.1051/0004-6361/202039657}, \href
  {https://doi.org/10.1051/0004-6361/202039657} {649, A1}

\bibitem[\protect\citeauthoryear{Gaisser, Engel  \& Resconi}{Gaisser
  et~al.}{2016}]{gaisser2016cosmic}
Gaisser T.~K.,  Engel R.,   Resconi E.,  2016, Cosmic rays and particle
  physics.
Cambridge University Press, \mn@doi{10.1017/CBO9781139192194}

\bibitem[\protect\citeauthoryear{Gandhi et~al.,}{Gandhi
  et~al.}{2011}]{Gandhi2011synchbreak}
Gandhi P.,  et~al., 2011, \mn@doi [ApJ] {10.1088/2041-8205/740/1/l13}, \href
  {https://doi.org/10.1088\%2F2041-8205\%2F740\%2F1\%2Fl13} {740, L13}

\bibitem[\protect\citeauthoryear{{Ginzburg} \& {Syrovatskii}}{{Ginzburg} \&
  {Syrovatskii}}{1964}]{Ginzburg1964}
{Ginzburg} V.~L.,  {Syrovatskii} S.~I.,  1964, {The Origin of Cosmic Rays}.
Elsevier

\bibitem[\protect\citeauthoryear{Gomel et~al.,}{Gomel
  et~al.}{2022}]{Gomel2022dr3}
Gomel R.,  et~al., 2022, arXiv:2206.06032, \href
  {https://arxiv.org/abs/2206.06032} {}

\bibitem[\protect\citeauthoryear{Guo, Li, Daughton  \& Liu}{Guo
  et~al.}{2014}]{Guo2014relativisticreconnection}
Guo F.,  Li H.,  Daughton W.,   Liu Y.-H.,  2014, \mn@doi [Phys. Rev. Lett.]
  {10.1103/PhysRevLett.113.155005}, \href
  {https://link.aps.org/doi/10.1103/PhysRevLett.113.155005} {113, 155005}

\bibitem[\protect\citeauthoryear{Guo et~al.,}{Guo et~al.}{2016}]{Guo2016}
Guo F.,  et~al., 2016, \mn@doi [ApJ] {10.3847/2041-8205/818/1/l9}, \href
  {https://doi.org/10.3847%2F2041-8205%2F818%2F1%2Fl9} {818, L9}

\bibitem[\protect\citeauthoryear{{H.E.S.S. Collaboration} et~al.,}{{H.E.S.S.
  Collaboration} et~al.}{2018a}]{HESS2018GalacticCenter}
{H.E.S.S. Collaboration} et~al., 2018a, \mn@doi [A\&A]
  {10.1051/0004-6361/201730824}, \href
  {https://doi.org/10.1051/0004-6361/201730824} {612, A9}

\bibitem[\protect\citeauthoryear{{H.E.S.S. Collaboration} et~al.,}{{H.E.S.S.
  Collaboration} et~al.}{2018b}]{HESSCollaboration2018Vela}
{H.E.S.S. Collaboration} et~al., 2018b, \mn@doi [A\&A]
  {10.1051/0004-6361/201732153}, \href
  {https://doi.org/10.1051/0004-6361/201732153} {620, A66}

\bibitem[\protect\citeauthoryear{{Hailey}, {Mori}, {Bauer}, {Berkowitz}, {Hong}
   \& {Hord}}{{Hailey} et~al.}{2018}]{Hailey2018cusp}
{Hailey} C.~J.,  {Mori} K.,  {Bauer} F.~E.,  {Berkowitz} M.~E.,  {Hong} J.,
  {Hord} B.~J.,  2018, \mn@doi [\nat] {10.1038/nature25029}, \href
  {https://ui.adsabs.harvard.edu/abs/2018Natur.556...70H} {556, 70}

\bibitem[\protect\citeauthoryear{Hakobyan, Philippov  \& Spitkovsky}{Hakobyan
  et~al.}{2019}]{Hakobyan_2019}
Hakobyan H.,  Philippov A.,   Spitkovsky A.,  2019, \mn@doi [ApJ]
  {10.3847/1538-4357/ab191b}, \href {https://doi.org/10.3847/1538-4357/ab191b}
  {877, 53}

\bibitem[\protect\citeauthoryear{Hillas}{Hillas}{1984}]{hillas1984origin}
Hillas A.~M.,  1984, \mn@doi [Annual review of A\&A]
  {10.1146/annurev.aa.22.090184.002233}, \href
  {https://ui.adsabs.harvard.edu/abs/1984ARA&A..22..425H} {22, 425}

\bibitem[\protect\citeauthoryear{Hinton et~al.,}{Hinton
  et~al.}{2008}]{Hinton_2008}
Hinton J.~A.,  et~al., 2008, \mn@doi [ApJ] {10.1088/0004-637x/690/2/l101},
  \href {https://doi.org/10.1088/0004-637x/690/2/l101} {690, L101}

\bibitem[\protect\citeauthoryear{Hjellming \& Johnston}{Hjellming \&
  Johnston}{1988}]{hjellming1988radio}
Hjellming R.,  Johnston K.,  1988, \mn@doi [ApJ] {10.1086/166318}, \href
  {https://ui.adsabs.harvard.edu/abs/1988ApJ...328..600H} {328, 600}

\bibitem[\protect\citeauthoryear{{IceCube-Gen2 Collaboration}
  et~al.,}{{IceCube-Gen2 Collaboration} et~al.}{2014}]{IceCube-Gen2}
{IceCube-Gen2 Collaboration} et~al., 2014, \mn@doi [arXiv:1412.5106]
  {arXiv:1412.5106}

\bibitem[\protect\citeauthoryear{Jokipii}{Jokipii}{1987}]{jokipii1987rate}
Jokipii J.,  1987, \mn@doi [ApJ]
  {https://adsabs.harvard.edu/pdf/1987ApJ...313..842J}, \href
  {https://adsabs.harvard.edu/pdf/1987ApJ...313..842J} {313, 842}

\bibitem[\protect\citeauthoryear{Jourdain, Roques, Chauvin  \& Clark}{Jourdain
  et~al.}{2012}]{jourdain2012separation}
Jourdain E.,  Roques J.,  Chauvin M.,   Clark D.,  2012, \mn@doi [ApJ]
  {10.1088/0004-637x/761/1/27}, \href
  {https://doi.org/10.1088%2F0004-637x%2F761%2F1%2F27} {761, 27}

\bibitem[\protect\citeauthoryear{Kafexhiu, Aharonian, Taylor  \& Vila}{Kafexhiu
  et~al.}{2014}]{kafexhiu2014parametrization}
Kafexhiu E.,  Aharonian F.,  Taylor A.,   Vila G.,  2014, \mn@doi [Physical
  Review D] {10.1103/PhysRevD.90.123014}, \href
  {https://link.aps.org/doi/10.1103/PhysRevD.90.123014} {90, 123014}

\bibitem[\protect\citeauthoryear{Kagan, Sironi, Cerutti  \& Giannios}{Kagan
  et~al.}{2015}]{kagan2015relativistic}
Kagan D.,  Sironi L.,  Cerutti B.,   Giannios D.,  2015, \mn@doi
  [Space~Sci.~Rev.] {10.1007/s11214-014-0132-9}, \href
  {https://doi.org/10.1007/s11214-014-0132-9} {191, 545}

\bibitem[\protect\citeauthoryear{Kalamkar, Casella, Uttley, O'Brien, Russell,
  Maccarone, van~der Klis  \& Vincentelli}{Kalamkar
  et~al.}{2016}]{Kalamkar2016detection}
Kalamkar M.,  Casella P.,  Uttley P.,  O'Brien K.,  Russell D.,  Maccarone T.,
  van~der Klis M.,   Vincentelli F.,  2016, \mn@doi [\mnras]
  {10.1093/mnras/stw1211}, \href {https://doi.org/10.1093/mnras/stw1211} {460,
  3284}

\bibitem[\protect\citeauthoryear{Kantzas et~al.,}{Kantzas
  et~al.}{2021}]{kantzas2020cyg}
Kantzas D.,  et~al., 2021, \mn@doi [MNRAS] {10.1093/mnras/staa3349}, \href
  {https://doi.org/10.1093/mnras/staa3349} {500, 2112}

\bibitem[\protect\citeauthoryear{Kantzas, Markoff, Lucchini, Ceccobello,
  Grinberg, Connors  \& Uttley}{Kantzas et~al.}{2022}]{kantzas2022gx}
Kantzas D.,  Markoff S.,  Lucchini M.,  Ceccobello C.,  Grinberg V.,  Connors
  R. M.~T.,   Uttley P.,  2022, \mn@doi [MNRAS] {10.1093/mnras/stac004}, \href
  {https://doi.org/10.1093/mnras/stac004} {510, 5187}

\bibitem[\protect\citeauthoryear{{Kantzas}, {Markoff}, {Lucchini}, {Ceccobello}
   \& {Chatterjee}}{{Kantzas} et~al.}{2023}]{Kantzas2023MassLoading}
{Kantzas} D.,  {Markoff} S.,  {Lucchini} M.,  {Ceccobello} C.,   {Chatterjee}
  K.,  2023, \mn@doi [\mnras] {10.1093/mnras/stad521}, \href
  {https://ui.adsabs.harvard.edu/abs/2023MNRAS.520.6017K} {520, 6017}

\bibitem[\protect\citeauthoryear{Keivani et~al.,}{Keivani
  et~al.}{2018}]{Keivani_2018}
Keivani A.,  et~al., 2018, \mn@doi [ApJ] {10.3847/1538-4357/aad59a}, \href
  {https://doi.org/10.3847/1538-4357/aad59a} {864, 84}

\bibitem[\protect\citeauthoryear{Kelner \& Aharonian}{Kelner \&
  Aharonian}{2008}]{kelner2008energy}
Kelner S.,  Aharonian F.,  2008, \mn@doi [Physical Review D]
  {10.1103/PhysRevD.78.034013}, \href
  {https://link.aps.org/doi/10.1103/PhysRevD.78.034013} {78, 034013}

\bibitem[\protect\citeauthoryear{Kelner \& Aharonian}{Kelner \&
  Aharonian}{2010}]{kelner2008erratum}
Kelner S.~R.,  Aharonian F.~A.,  2010, \mn@doi [Phys. Rev. D]
  {10.1103/PhysRevD.82.099901}, \href
  {https://link.aps.org/doi/10.1103/PhysRevD.82.099901} {82, 099901}

\bibitem[\protect\citeauthoryear{Kelner, Aharonian  \& Bugayov}{Kelner
  et~al.}{2006}]{kelner2006energy}
Kelner S.,  Aharonian F.~A.,   Bugayov V.,  2006, \mn@doi [Physical Review D]
  {10.1103/PhysRevD.74.034018}, \href
  {https://link.aps.org/doi/10.1103/PhysRevD.74.034018} {74, 034018}

\bibitem[\protect\citeauthoryear{K\"onigl}{K\"onigl}{1980}]{Koenigl1980}
K\"onigl A.,  1980, \mn@doi [The Physics of Fluids] {10.1063/1.863110}, \href
  {https://aip.scitation.org/doi/abs/10.1063/1.863110} {23, 1083}

\bibitem[\protect\citeauthoryear{Kovalev, Plavin  \& Troitsky}{Kovalev
  et~al.}{2022}]{Kovalev_2022}
Kovalev Y.~Y.,  Plavin A.~V.,   Troitsky S.~V.,  2022, \mn@doi [APJ Letters]
  {10.3847/2041-8213/aca1ae}, \href
  {https://dx.doi.org/10.3847/2041-8213/aca1ae} {940, L41}

\bibitem[\protect\citeauthoryear{{Laurent}, {Rodriguez}, {Wilms}, {Cadolle
  Bel}, {Pottschmidt}  \& {Grinberg}}{{Laurent}
  et~al.}{2011}]{2011Sci...332..438L}
{Laurent} P.,  {Rodriguez} J.,  {Wilms} J.,  {Cadolle Bel} M.,  {Pottschmidt}
  K.,   {Grinberg} V.,  2011, \mn@doi [Science] {10.1126/science.1200848},
  \href {http://adsabs.harvard.edu/abs/2011Sci...332..438L} {332, 438}

\bibitem[\protect\citeauthoryear{Levinson \& Waxman}{Levinson \&
  Waxman}{2001}]{levinson2001probing}
Levinson A.,  Waxman E.,  2001, \mn@doi [Phys. Rev. Lett.]
  {10.1103/PhysRevLett.87.171101}, \href
  {https://link.aps.org/doi/10.1103/PhysRevLett.87.171101} {87, 171101}

\bibitem[\protect\citeauthoryear{Lewis, Russell, Fender, Roche  \& Clark}{Lewis
  et~al.}{2008}]{lewis2008continued}
Lewis F.,  Russell D.~M.,  Fender R.~P.,  Roche P.,   Clark J.~S.,  2008,
  \mn@doi [arXiv:0811.2336] {0811.2336}

\bibitem[\protect\citeauthoryear{L{\'o}pez-Coto, de O{\~n}a~Wilhelmi,
  Aharonian, Amato  \& Hinton}{L{\'o}pez-Coto et~al.}{2022}]{lopez2022gamma}
L{\'o}pez-Coto R.,  de O{\~n}a~Wilhelmi E.,  Aharonian F.,  Amato E.,   Hinton
  J.,  2022, \mn@doi [Nature Astronomy] {10.1038/s41550-021-01580-0}, \href
  {https://doi.org/10.1038/s41550-021-01580-0} {pp~1--8}

\bibitem[\protect\citeauthoryear{Lorimer et~al.,}{Lorimer
  et~al.}{2006}]{Lorimer2006}
Lorimer D.~R.,  et~al., 2006, \mn@doi [MNRAS]
  {10.1111/j.1365-2966.2006.10887.x}, \href
  {https://doi.org/10.1111/j.1365-2966.2006.10887.x} {372, 777}

\bibitem[\protect\citeauthoryear{Lucchini, Markoff, Crumley, Krau{\ss}  \&
  Connors}{Lucchini et~al.}{2018}]{lucchini2019breaking}
Lucchini M.,  Markoff S.,  Crumley P.,  Krau{\ss} F.,   Connors R. M.~T.,
  2018, \mn@doi [MNRAS] {10.1093/mnras/sty2929}, \href
  {https://doi.org/10.1093/mnras/sty2929} {482, 4798}

\bibitem[\protect\citeauthoryear{{Lucchini}, {Russell}, {Markoff},
  {Vincentelli}, {Gardenier}, {Ceccobello}  \& {Uttley}}{{Lucchini}
  et~al.}{2021}]{Lucchini2021correlation}
{Lucchini} M.,  {Russell} T.~D.,  {Markoff} S.~B.,  {Vincentelli} F.,
  {Gardenier} D.,  {Ceccobello} C.,   {Uttley} P.,  2021, \mn@doi [\mnras]
  {10.1093/mnras/staa3957}, \href
  {https://ui.adsabs.harvard.edu/abs/2021MNRAS.501.5910L} {501, 5910}

\bibitem[\protect\citeauthoryear{Lucchini et~al.,}{Lucchini
  et~al.}{2022}]{lucchini2022BHJet}
Lucchini M.,  et~al., 2022, \mn@doi [MNRAS] {10.1093/mnras/stac2904}, \href
  {https://doi.org/10.1093/mnras/stac2904} {517, 5853}

\bibitem[\protect\citeauthoryear{Luque, Mazziotta, Loparco, Gargano  \&
  Serini}{Luque et~al.}{2021}]{Luque_2021Markov}
Luque P. D. L.~T.,  Mazziotta M.,  Loparco F.,  Gargano F.,   Serini D.,  2021,
  \mn@doi [JCAP] {10.1088/1475-7516/2021/07/010}, \href
  {https://dx.doi.org/10.1088/1475-7516/2021/07/010} {2021, 010}

\bibitem[\protect\citeauthoryear{Luque, Gaggero, Grasso, Fornieri, Egberts,
  Steppa  \& Evoli}{Luque et~al.}{2022}]{luque2022galactic}
Luque P. D. l.~T.,  Gaggero D.,  Grasso D.,  Fornieri O.,  Egberts K.,  Steppa
  C.,   Evoli C.,  2022, \mn@doi [arXiv:2203.15759]
  {https://arxiv.org/abs/2203.15759}

\bibitem[\protect\citeauthoryear{{MAGIC Collaboration} et~al.,}{{MAGIC
  Collaboration} et~al.}{2020}]{MAGICCollaboration2020Geminga}
{MAGIC Collaboration} et~al., 2020, \mn@doi [A\&A]
  {10.1051/0004-6361/202039131}, \href
  {https://doi.org/10.1051/0004-6361/202039131} {643, L14}

\bibitem[\protect\citeauthoryear{Malkov \& Drury}{Malkov \&
  Drury}{2001}]{malkov2001nonlinear}
Malkov M.,  Drury L.~O.,  2001, \mn@doi [Rep. Prog. Phys.]
  {10.1088/0034-4885/64/4/201}, \href
  {https://doi.org/10.1088%2F0034-4885%2F64%2F4%2F201} {64, 429}

\bibitem[\protect\citeauthoryear{Malyshev, Zdziarski  \& Chernyakova}{Malyshev
  et~al.}{2013}]{malyshev2013high}
Malyshev D.,  Zdziarski A.~A.,   Chernyakova M.,  2013, \mn@doi [MNRAS]
  {10.1093/mnras/stt1184}, \href {https://doi.org/10.1093/mnras/stt1184} {434,
  2380}

\bibitem[\protect\citeauthoryear{Mannheim}{Mannheim}{1993}]{mannheim1993proton}
Mannheim K.,  1993, \mn@doi [A\&A]
  {https://ui.adsabs.harvard.edu/abs/1993A&A...269...67M}, \href
  {https://ui.adsabs.harvard.edu/abs/1993A&A...269...67M} {269, 67}

\bibitem[\protect\citeauthoryear{Mannheim \& Schlickeiser}{Mannheim \&
  Schlickeiser}{1994}]{mannheim1994interactions}
Mannheim K.,  Schlickeiser R.,  1994, \mn@doi [A\&A]
  {https://cds.cern.ch/record/259140}, \href
  {https://cds.cern.ch/record/259140} {286, 983}

\bibitem[\protect\citeauthoryear{Markoff, Falcke  \& Fender}{Markoff
  et~al.}{2001}]{markoff2001jet}
Markoff S.,  Falcke H.,   Fender R.,  2001, \mn@doi [A\&A]
  {10.1051/0004-6361:20010420}, \href
  {https://ui.adsabs.harvard.edu/abs/2001A&A...372L..25M} {372, L25}

\bibitem[\protect\citeauthoryear{Markoff, Nowak, Corbel, Fender  \&
  Falcke}{Markoff et~al.}{2003}]{markoff2003exploring}
Markoff S.,  Nowak M.,  Corbel S.,  Fender R.,   Falcke H.,  2003, \mn@doi
  [A\&A] {10.1051/0004-6361:20021497}, \href
  {https://doi.org/10.1051/0004-6361:20021497} {397, 645}

\bibitem[\protect\citeauthoryear{Markoff, Nowak  \& Wilms}{Markoff
  et~al.}{2005}]{Markoff2005}
Markoff S.,  Nowak M.~A.,   Wilms J.,  2005, \mn@doi [ApJ] {10.1086/497628},
  \href {https://doi.org/10.1086%2F497628} {635, 1203}

\bibitem[\protect\citeauthoryear{Migliari, Fender  \& Méndez}{Migliari
  et~al.}{2002}]{Migliari2002iron}
Migliari S.,  Fender R.,   Méndez M.,  2002, \mn@doi [Science]
  {10.1126/science.1073660}, \href
  {https://www.science.org/doi/abs/10.1126/science.1073660} {297, 1673}

\bibitem[\protect\citeauthoryear{Miller-Jones et~al.,}{Miller-Jones
  et~al.}{2021}]{JMJ2021cygx1}
Miller-Jones J. C.~A.,  et~al., 2021, \mn@doi [Science]
  {10.1126/science.abb3363}, \href
  {https://www.science.org/doi/abs/10.1126/science.abb3363} {371, 1046}

\bibitem[\protect\citeauthoryear{Mori et~al.,}{Mori et~al.}{2021}]{Mori_2021}
Mori K.,  et~al., 2021, \mn@doi [ApJ] {10.3847/1538-4357/ac1da5}, \href
  {https://doi.org/10.3847/1538-4357/ac1da5} {921, 148}

\bibitem[\protect\citeauthoryear{M\"ucke \& Protheroe}{M\"ucke \&
  Protheroe}{2001}]{MUCKE2001121}
M\"ucke A.,  Protheroe R.,  2001, \mn@doi [Astroparticle Physics]
  {https://doi.org/10.1016/S0927-6505(00)00141-9}, \href
  {https://www.sciencedirect.com/science/article/pii/S0927650500001419} {15,
  121}

\bibitem[\protect\citeauthoryear{M{\"u}cke, Protheroe, Engel, Rachen  \&
  Stanev}{M{\"u}cke et~al.}{2003}]{MUCKE2003protonBLLac}
M{\"u}cke A.,  Protheroe R.,  Engel R.,  Rachen J.,   Stanev T.,  2003, \mn@doi
  [Astroparticle Physics] {https://doi.org/10.1016/S0927-6505(02)00185-8},
  \href {http://www.sciencedirect.com/science/article/pii/S0927650502001858}
  {18, 593 }

\bibitem[\protect\citeauthoryear{Murase, Inoue  \& Dermer}{Murase
  et~al.}{2014}]{Murase2014diffuse}
Murase K.,  Inoue Y.,   Dermer C.~D.,  2014, \mn@doi [Phys. Rev. D]
  {10.1103/PhysRevD.90.023007}, \href
  {https://link.aps.org/doi/10.1103/PhysRevD.90.023007} {90, 023007}

\bibitem[\protect\citeauthoryear{Oda et~al.,}{Oda et~al.}{2019}]{Oda2019J1828}
Oda S.,  et~al., 2019, \mn@doi [PASJ] {10.1093/pasj/psz091}, \href
  {https://doi.org/10.1093/pasj/psz091} {71, 108}

\bibitem[\protect\citeauthoryear{Olejak, Belczynski, Bulik  \&
  Sobolewska}{Olejak et~al.}{2020}]{Olejak2019synthesis}
Olejak A.,  Belczynski K.,  Bulik T.,   Sobolewska M.,  2020, \mn@doi [A\&A]
  {10.1051/0004-6361/201936557}, \href
  {https://doi.org/10.1051/0004-6361/201936557} {638, A94}

\bibitem[\protect\citeauthoryear{Orosz et~al.,}{Orosz
  et~al.}{2001}]{Orosz_2001}
Orosz J.~A.,  et~al., 2001, \mn@doi [ApJ] {10.1086/321442}, \href
  {https://doi.org/10.1086/321442} {555, 489}

\bibitem[\protect\citeauthoryear{Park, Caprioli  \& Spitkovsky}{Park
  et~al.}{2015}]{park2015simultaneous}
Park J.,  Caprioli D.,   Spitkovsky A.,  2015, \mn@doi [Phys. Rev. Lett.]
  {10.1103/PhysRevLett.114.085003}, \href
  {https://link.aps.org/doi/10.1103/PhysRevLett.114.085003} {114, 085003}

\bibitem[\protect\citeauthoryear{Pepe, Vila  \& Romero}{Pepe
  et~al.}{2015}]{pepe2015lepto}
Pepe C.,  Vila G.~S.,   Romero G.~E.,  2015, \mn@doi [A\&A]
  {10.1051/0004-6361/201527156}, \href
  {https://doi.org/10.1051/0004-6361/201527156} {584, A95}

\bibitem[\protect\citeauthoryear{Petropoulou \& Sironi}{Petropoulou \&
  Sironi}{2018}]{petropoulou2018steady}
Petropoulou M.,  Sironi L.,  2018, \mn@doi [MNRAS] {10.1093/mnras/sty2702},
  \href {https://doi.org/10.1093/mnras/sty2702} {481, 5687}

\bibitem[\protect\citeauthoryear{Petropoulou, Dimitrakoudis, Padovani,
  Mastichiadis  \& Resconi}{Petropoulou
  et~al.}{2015}]{Petropoulou2015photohadronic}
Petropoulou M.,  Dimitrakoudis S.,  Padovani P.,  Mastichiadis A.,   Resconi
  E.,  2015, \mn@doi [MNRAS] {10.1093/mnras/stv179}, \href
  {https://doi.org/10.1093/mnras/stv179} {448, 2412}

\bibitem[\protect\citeauthoryear{Pierre-Auger}{Pierre-Auger}{2023}]{Deligny_2023Auger}
Pierre-Auger C.,  2023, \mn@doi [Journal of Physics: Conference Series]
  {10.1088/1742-6596/2429/1/012009}, \href
  {https://doi.org/10.1088%2F1742-6596%2F2429%2F1%2F012009} {2429, 012009}

\bibitem[\protect\citeauthoryear{{Prodanovi{\'c}}, {Fields}  \&
  {Beacom}}{{Prodanovi{\'c}} et~al.}{2007}]{prodanovic2007diffuse}
{Prodanovi{\'c}} T.,  {Fields} B.~D.,   {Beacom} J.~F.,  2007, \mn@doi
  [Astroparticle Physics] {10.1016/j.astropartphys.2006.08.007}, \href
  {https://ui.adsabs.harvard.edu/abs/2007APh....27...10P} {27, 10}

\bibitem[\protect\citeauthoryear{Rachen \& Biermann}{Rachen \&
  Biermann}{1993}]{rachen1993extragalactic}
Rachen J.~P.,  Biermann P.~L.,  1993, \mn@doi [A\&A]
  {https://doi.org10.48550/arXiv.astro-ph/9301010}, \href
  {https://doi.org/10.48550/arXiv.astro-ph/9301010} {272, 161}

\bibitem[\protect\citeauthoryear{Rachen \& M{\'e}sz{\'a}ros}{Rachen \&
  M{\'e}sz{\'a}ros}{1998}]{rachen1998photohadronic}
Rachen J.~P.,  M{\'e}sz{\'a}ros P.,  1998, \mn@doi [\prd]
  {10.1103/PhysRevD.58.123005}, \href
  {https://ui.adsabs.harvard.edu/abs/1998PhRvD..58l3005R} {58, 123005}

\bibitem[\protect\citeauthoryear{Reynoso, Romero  \& Christiansen}{Reynoso
  et~al.}{2008}]{reynoso2008ss433}
Reynoso M.~M.,  Romero G.~E.,   Christiansen H.~R.,  2008, \mn@doi [MNRAS]
  {10.1111/j.1365-2966.2008.13364.x}, \href
  {https://doi.org/10.1111/j.1365-2966.2008.13364.x} {387, 1745}

\bibitem[\protect\citeauthoryear{Ripperda, Liska, Chatterjee, Musoke,
  Philippov, Markoff, Tchekhovskoy  \& Younsi}{Ripperda
  et~al.}{2022}]{Ripperda_2022}
Ripperda B.,  Liska M.,  Chatterjee K.,  Musoke G.,  Philippov A.~A.,  Markoff
  S.~B.,  Tchekhovskoy A.,   Younsi Z.,  2022, \mn@doi [ApJ Letters]
  {10.3847/2041-8213/ac46a1}, \href {https://doi.org/10.3847/2041-8213/ac46a1}
  {924, L32}

\bibitem[\protect\citeauthoryear{Rodriguez et~al.,}{Rodriguez
  et~al.}{2015}]{0004-637X-807-1-17}
Rodriguez J.,  et~al., 2015, \mn@doi [ApJ] {10.1088/0004-637x/807/1/17}, \href
  {https://doi.org/10.1088%2F0004-637x%2F807%2F1%2F17} {807, 17}

\bibitem[\protect\citeauthoryear{Romero \& Orellana}{Romero \&
  Orellana}{2005}]{romero2005misaligned}
Romero G.~E.,  Orellana M.,  2005, \mn@doi [A\&A] {10.1051/0004-6361:20052664},
  \href {https://doi.org/10.1051/0004-6361:20052664} {439, 237}

\bibitem[\protect\citeauthoryear{Romero \& Vila}{Romero \&
  Vila}{2008}]{Romero2008proton}
Romero G.~E.,  Vila G.~S.,  2008, \mn@doi [A\&A] {10.1051/0004-6361:200809563},
  \href {https://doi.org/10.1051/0004-6361:200809563} {485, 623}

\bibitem[\protect\citeauthoryear{Romero, Torres, Bernad{\'o}  \&
  Mirabel}{Romero et~al.}{2003}]{romero2003hadronic}
Romero G.~E.,  Torres D.~F.,  Bernad{\'o} M.~K.,   Mirabel I.,  2003, \mn@doi
  [A\&A] {10.1051/0004-6361:20031314-1}, \href
  {https://doi.org/10.1051/0004-6361:20031314-1} {410, L1}

\bibitem[\protect\citeauthoryear{Rushton et~al.,}{Rushton
  et~al.}{2012}]{rushton2012weak}
Rushton A.,  et~al., 2012, \mn@doi [MNRAS] {10.1111/j.1365-2966.2011.19959.x},
  \href {https://doi.org/10.1111/j.1365-2966.2011.19959.x} {419, 3194}

\bibitem[\protect\citeauthoryear{Russell, Soria, Motch, Pakull, Torres, Curran,
  Jonker  \& Miller-Jones}{Russell et~al.}{2014}]{Russell2014J1836}
Russell T.~D.,  Soria R.,  Motch C.,  Pakull M.~W.,  Torres M. A.~P.,  Curran
  P.~A.,  Jonker P.~G.,   Miller-Jones J. C.~A.,  2014, \mn@doi [MNRAS]
  {10.1093/mnras/stt2480}, \href {https://doi.org/10.1093/mnras/stt2480} {439,
  1381}

\bibitem[\protect\citeauthoryear{Russell et~al.,}{Russell
  et~al.}{2019}]{russell2019maxij1535571}
Russell T.~D.,  et~al., 2019, \mn@doi [\apj] {10.3847/1538-4357/ab3d36}, \href
  {https://ui.adsabs.harvard.edu/abs/2019ApJ...883..198R} {883, 198}

\bibitem[\protect\citeauthoryear{Sabatini et~al.,}{Sabatini
  et~al.}{2013}]{Sabatini_2013}
Sabatini S.,  et~al., 2013, \mn@doi [ApJ] {10.1088/0004-637x/766/2/83}, \href
  {https://doi.org/10.1088/0004-637x/766/2/83} {766, 83}

\bibitem[\protect\citeauthoryear{Safi-Harb et~al.,}{Safi-Harb
  et~al.}{2022}]{Safi-Harb_2022}
Safi-Harb S.,  et~al., 2022, \mn@doi [The Astrophysical Journal]
  {10.3847/1538-4357/ac7c05}, \href
  {https://dx.doi.org/10.3847/1538-4357/ac7c05} {935, 163}

\bibitem[\protect\citeauthoryear{Schneider}{Schneider}{2019}]{schneider2019characterization}
Schneider A.,  2019, \mn@doi [arXiv:1907.11266] {10.48550/arXiv.1907.11266}

\bibitem[\protect\citeauthoryear{Sironi \& Spitkovsky}{Sironi \&
  Spitkovsky}{2009}]{sironi2009particle}
Sironi L.,  Spitkovsky A.,  2009, \mn@doi [ApJ] {10.1088/0004-637x/698/2/1523},
  \href {https://doi.org/10.1088%2F0004-637x%2F698%2F2%2F1523} {698, 1523}

\bibitem[\protect\citeauthoryear{Sironi \& Spitkovsky}{Sironi \&
  Spitkovsky}{2014}]{Sironi_2014}
Sironi L.,  Spitkovsky A.,  2014, \mn@doi [ApJ] {10.1088/2041-8205/783/1/l21},
  \href {https://doi.org/10.1088%2F2041-8205%2F783%2F1%2Fl21} {783, L21}

\bibitem[\protect\citeauthoryear{Sironi, Spitkovsky  \& Arons}{Sironi
  et~al.}{2013}]{Sironi2013}
Sironi L.,  Spitkovsky A.,   Arons J.,  2013, \mn@doi [ApJ]
  {10.1088/0004-637x/771/1/54}, \href
  {https://doi.org/10.1088%2F0004-637x%2F771%2F1%2F54} {771, 54}

\bibitem[\protect\citeauthoryear{Sironi, Petropoulou  \& Giannios}{Sironi
  et~al.}{2015}]{sironi2015relativistic}
Sironi L.,  Petropoulou M.,   Giannios D.,  2015, \mn@doi [MNRAS]
  {10.1146/annurev.astro.37.1.409}, \href
  {https://doi.org/10.1146/annurev.astro.37.1.409} {450, 183}

\bibitem[\protect\citeauthoryear{Stephens \& Badhwar}{Stephens \&
  Badhwar}{1981}]{stephens1981production}
Stephens S.,  Badhwar G.,  1981, \mn@doi [Astrophysics and Space Science]
  {10.1007/BF00651256}, \href {https://doi.org/10.1007/BF00651256} {76, 213}

\bibitem[\protect\citeauthoryear{Stettner}{Stettner}{2019}]{stettner2019measurement}
Stettner J.,  2019, \mn@doi [arXiv:1908.09551] {10.48550/arXiv.1908.09551}

\bibitem[\protect\citeauthoryear{TA}{TA}{2022}]{ta2022Highlights}
TA C.,  2022, Highlights from the Telescope Array Experiments,
  \mn@doi{10.48550/ARXIV.2209.03591}

\bibitem[\protect\citeauthoryear{Tavani et~al.,}{Tavani
  et~al.}{2009}]{Tavani_2009CygnusX3}
Tavani M.,  et~al., 2009, \mn@doi [Nature] {10.1038/nature08578}, \href
  {http://dx.doi.org/10.1038/nature08578} {462, 620}

\bibitem[\protect\citeauthoryear{Tetarenko, Sivakoff, Heinke  \&
  Gladstone}{Tetarenko et~al.}{2016}]{Tetarenko2016}
Tetarenko B.~E.,  Sivakoff G.~R.,  Heinke C.~O.,   Gladstone J.~C.,  2016,
  \mn@doi [ApJ Supplement Series] {10.3847/0067-0049/222/2/15}, \href
  {https://doi.org/10.3847%2F0067-0049%2F222%2F2%2F15} {222, 15}

\bibitem[\protect\citeauthoryear{Tetarenko et~al.,}{Tetarenko
  et~al.}{2021}]{tetarenko2021measuring}
Tetarenko A.~J.,  et~al., 2021, \mn@doi [MNRAS] {10.1093/mnras/stab820}, \href
  {https://doi.org/10.1093/mnras/stab820} {504, 3862}

\bibitem[\protect\citeauthoryear{{Thoudam, S.}, {Rachen, J. P.}, {van Vliet,
  A.}, {Achterberg, A.}, {Buitink, S.}, {Falcke, H.}  \& {H\"orandel, J.
  R.}}{{Thoudam, S.} et~al.}{2016}]{Thoudam2016spectrum}
{Thoudam, S.} {Rachen, J. P.} {van Vliet, A.} {Achterberg, A.} {Buitink, S.}
  {Falcke, H.}  {H\"orandel, J. R.} 2016, \mn@doi [A\&A]
  {10.1051/0004-6361/201628894}, \href
  {https://doi.org/10.1051/0004-6361/201628894} {595, A33}

\bibitem[\protect\citeauthoryear{Torres, Romero  \& Mirabel}{Torres
  et~al.}{2005}]{Torres_2005}
Torres D.~F.,  Romero G.~E.,   Mirabel F.,  2005, \mn@doi [Chinese Astron.
  Astrophys.] {10.1088/1009-9271/5/s1/183}, \href
  {https://doi.org/10.1088%2F1009-9271%2F5%2Fs1%2F183} {5, 183}

\bibitem[\protect\citeauthoryear{{Uzdensky}}{{Uzdensky}}{2011}]{Uzdensky2011magnetic}
{Uzdensky} D.~A.,  2011, \mn@doi [\ssr] {10.1007/s11214-011-9744-5}, \href
  {https://ui.adsabs.harvard.edu/abs/2011SSRv..160...45U} {160, 45}

\bibitem[\protect\citeauthoryear{Vieyro \& Romero}{Vieyro \&
  Romero}{2012}]{vieyro2012magnetized}
Vieyro F.~L.,  Romero G.~E.,  2012, \mn@doi [A\&A]
  {10.1051/0004-6361/201218886}, \href
  {https://doi.org/10.1051/0004-6361/201218886} {542, A7}

\bibitem[\protect\citeauthoryear{Yoon et~al.,}{Yoon et~al.}{2017}]{Yoon_2017}
Yoon Y.~S.,  et~al., 2017, \mn@doi [ApJ] {10.3847/1538-4357/aa68e4}, \href
  {https://doi.org/10.3847/1538-4357/aa68e4} {839, 5}

\bibitem[\protect\citeauthoryear{Zanin, Fern{\'a}ndez-Barral, de
  O{\~n}a~Wilhelmi, Aharonian, Blanch, Bosch-Ramon  \& Galindo}{Zanin
  et~al.}{2016}]{zanin2016detection}
Zanin R.,  Fern{\'a}ndez-Barral A.,  de O{\~n}a~Wilhelmi E.,  Aharonian F.,
  Blanch O.,  Bosch-Ramon V.,   Galindo D.,  2016, \mn@doi [A\&A]
  {10.1051/0004-6361/201628917}, \href
  {https://doi.org/10.1051/0004-6361/201628917} {596, A55}

\bibitem[\protect\citeauthoryear{Zdziarski, Pjanka, Sikora  \&
  Stawarz}{Zdziarski et~al.}{2014}]{zdziarski2014jet}
Zdziarski A.~A.,  Pjanka P.,  Sikora M.,   Stawarz {\L}.,  2014, \mn@doi
  [MNRAS] {10.1093/mnras/stu1009}, \href
  {https://doi.org/10.1093/mnras/stu1009} {442, 3243}

\bibitem[\protect\citeauthoryear{Zdziarski, Malyshev, Chernyakova  \&
  Pooley}{Zdziarski et~al.}{2017}]{zdziarski2017high}
Zdziarski A.~A.,  Malyshev D.,  Chernyakova M.,   Pooley G.~G.,  2017, \mn@doi
  [MNRAS] {10.1093/mnras/stx1846}, \href
  {https://doi.org/10.1093/mnras/stx1846} {471, 3657}

\bibitem[\protect\citeauthoryear{Zhang, Feng, Lei, Tang  \& Tian}{Zhang
  et~al.}{2010}]{Zhang2010neutrinoLMXRBs}
Zhang J.~F.,  Feng Y.~G.,  Lei M.~C.,  Tang Y.~Y.,   Tian Y.~P.,  2010, \mn@doi
  [MNRAS] {10.1111/j.1365-2966.2010.17072.x}, \href
  {https://doi.org/10.1111/j.1365-2966.2010.17072.x} {407, 2468}

\bibitem[\protect\citeauthoryear{and RALF~WISCHNEWSKI et~al.,}{and
  RALF~WISCHNEWSKI et~al.}{2005}]{WISCHNEWSKI2005Baikal}
and RALF~WISCHNEWSKI et~al., 2005, \mn@doi [International Journal of Modern
  Physics A] {10.1142/s0217751x0503051x}, \href
  {https://doi.org/10.1142%2Fs0217751x0503051x} {20, 6932}

\bibitem[\protect\citeauthoryear{van~den Eijnden, Degenaar, Russell, Wijnands,
  Miller-Jones, Sivakoff  \& Hern{\'a}ndez~Santisteban}{van~den Eijnden
  et~al.}{2018}]{Eijnden2018evolving}
van~den Eijnden J.,  Degenaar N.,  Russell T.,  Wijnands R.,  Miller-Jones J.,
  Sivakoff G.,   Hern{\'a}ndez~Santisteban J.,  2018, \mn@doi [Nature]
  {10.1038/s41586-018-0524-1}, \href
  {https://doi.org/10.1038/s41586-018-0524-1} {562, 233}

\makeatother
\end{thebibliography}




\appendix


\section{Detected \bhs}\label{app: known sources}
In Table~\ref{table: 35 known sources}, we show the 35 Galactic \bhs we examine in this work. We use the WATCHDOG catalogue of \citet[][]{Tetarenko2016} for the 31 sources that were detected before 2016. We include 4 more sources detected by the Faulkes Telescopes project \citep[][]{lewis2008continued} after 2016. In the same table, we indicate whether the \bhs is low- (L) or high-mass (H).   

\begin{table*}
\setlength{\tabcolsep}{6pt} 
\renewcommand{\arraystretch}{1.3} 
\begin{tabular}{lcccccc}
\textbf{Name}                & \textbf{\begin{tabular}[c]{@{}c@{}}Dec \\ (J2000)\end{tabular}} & \textbf{\begin{tabular}[c]{@{}c@{}}RA \\ (J2000)\end{tabular}} & \textbf{\begin{tabular}[c]{@{}c@{}}Distance \\ (kpc)\end{tabular}} & \textbf{\begin{tabular}[c]{@{}c@{}}M$_{\rm BH}$\\ (M$_{\odot}$)\end{tabular}} & \textbf{\begin{tabular}[c]{@{}c@{}}Inclination \\ (deg)\end{tabular}} & \textbf{Companion} \\ \hline
1A~0620-00                   & -00 20 44.72                                                    & 06 22 44.503                                                   & 1.06                                                               & 6.6               & 51                                                                    & L                  \\
XTE~J1118+480                & +48 02 12.6                                                     & 11 18 10.80                                                    & 1.7                                                               & 7.3               & 75.0                                                                  & L                  \\
SWIFT~J1357.2-0933           & -09 19 12.00                                                    & 13 57 16.818                                                   & 3.9                                                                & \textit{8.0}               & \textit{60.0}                                                                  & L                  \\
MAXI~J1543-564               & -56 24 48.35                                                    & 15 43 17.336                                                   & \textit{5}                                                                  & \textit{8.0}               & \textit{60.0}                                                                  & L                  \\
XTE~J1550-564                & -56 28 35.0                                                     & 15 50 58.78                                                    & 4.4                                                                & 10.4              & 67.4                                                                  & L                  \\
4U~1630-472                  & -47 23 34.8                                                     & 16 34 01.61                                                    & 5                                                                  & \textit{8.0}               & \textit{60.0}                                                                  & L                  \\
XTE~J1650-500                & -49 57 43.6                                                     & 16 50 00.98                                                    & 2.6                                                                & 4.7               & 75.2                                                                  & L                  \\
GRO~J1655-40                 & -39 50 44.90                                                    & 16 54 00.137                                                   & 3.2                                                                & 5.4               & 69                                                                    & L                  \\
GX~339--4                     & -48 47 22.8                                                     & 17 02 49.36                                                    & \textit{8}                                                                  & \textit{8.0}               & \textit{60.0}                                                                  & L                  \\
IGR~J17091-3624              & -36 24 24.                                                      & 17 09 08.                                                      & \textit{5}                                                                  & \textit{8.0}               & \textit{60.0}                                                                  & L                  \\
GRS~1716-249                 & -25 01 03.4                                                     & 17 19 36.93                                                    & 2.4                                                                & 8.0               & 60.0                                                                  & L                  \\
XTE~J1720-318                & -31 45 01.25                                                    & 17 19 58.994                                                   & \textit{5}                                                                  & \textit{8.0}               & \textit{60.0}                                                                  & L                  \\
1E~1740.7-2942               & -29 44 42.6                                                     & 17 43 54.83                                                    & \textit{5}                                                                  & \textit{8.0}               & \textit{60.0}                                                                  & L                  \\
Swift~J174510.8-262411       & -26 24 12.60                                                    & 17 45 10.849                                                   & \textit{5}                                                                  & \textit{8.0}               & \textit{60.0}                                                                  & L                  \\
CXOGC~J174540.0-290031       & -29 00 31.0                                                     & 17 45 40.03                                                    & 8                                                                  & \textit{8.0}               & \textit{60.0}                                                                  & L                  \\
H~1743-322                   & -32 14 00.60                                                    & 17 46 15.608                                                   & 10.4                                                               & \textit{8.0}               & \textit{60.0}                                                                  & L                  \\
XTE~J1752-223                & -22 20 32.782                                                   & 17 52 15.095                                                   & 3.5                                                                & 9.6               & \textit{60.0}                                                                  & L                  \\
Swift~J1753.5-0127           & -01 27 06.2                                                     & 17 53 28.29                                                    & \textit{5}                                                                  & \textit{8.0}               & \textit{60.0}                                                                  & L                  \\
GRS~1758-258                 & -25 44 36.1                                                     & 18 01 12.40                                                    & \textit{5}                                                                  & \textit{8.0}               & \textit{60.0}                                                                  & L                  \\
SAX~J1819.3-2525 (V4641 Sgr) & -25 24 25.8                                                     & 18 19 21.63                                                    & 6.2                                                                & 6.4               & 72.3                                                                  & H                  \\
MAXI~J1836-194               & -19 19 12.1                                                     & 18 35 43.43                                                    & \textit{5}                                                                  & \textit{8.0}               & 9.5                                                                   & L                  \\
XTE~J1859+226                & +22 39 29.4                                                     & 18 58 41.58                                                    & 8                                                                  & 10.8              & 60                                                                    & L                  \\
XTE~J1908+094                & +09 23 04.90                                                    & 19 08 53.077                                                   & 6.5                                                                & \textit{8.0}               & \textit{60.0}                                                                  & L                  \\
SS~433                       & +04 58 57.9                                                     & 19 11 49.57                                                    & 5.5                                                                & \textit{8.0}               & \textit{60.0}                                                                  & L                  \\
GRS~1915+105                 & +10 56 44.8                                                     & 19 15 11.55                                                    & 8.6                                                                & 12.4              & 70                                                                    & L                  \\
4U~1956+350 (Cyg X--1)       & +35 12 05.778                                                   & 19 58 21.675                                                   & 2.22                                                               & 21.4              & 27.1                                                                 & H                  \\
GS~2000+251                  & +25 14 11.3                                                     & 20 02 49.58                                                    & 2.7                                                                & 8.4               & \textit{60.0}                                                                  & L                  \\
GS~2023+338 (V404~Cyg)  & +33 52 02.2                                                     & 20 24 03.83                                                    & 2.39                                                               & 7.2               & 80.1                                                                  & L                  \\
4U~2030+40 (Cyg~X--3)        & +40 57 27.9                                                     & 20 32 25.78                                                    & 8.3                                                                & 2.4               & 43                                                                    & H                  \\
MWC~656                      & +44 43 18.25                                                    & 22 42 57.30                                                    & 2.6                                                                & 5.4               & 66                                                                    & H                  \\
MAXI~J1820+070               & 7.18563                                                         & 275.091                                                        & 2.96                                                               & 9.2               & 63                                                                    & L                  \\
MAXI~J1348-630               & -63.274                                                         & 207.054                                                        & 5.3                                                                & 9.1               & 60                                                                    & L                  \\
MAXI~J1535-571               & -57.23                                                          & 233.832                                                        & 4.1                                                                & 10.4              & 45                                                                    & L                  \\
MAXI~J1828-249               & -25.029                                                         & 277.244                                                        & 7.0                                                                & 9.0               & 60                                                                    & L                 
\end{tabular}
\caption{The 35 known Galactic \bhs we use in this work. The values of the Distance, the Mass of the black hole and the inclination are from the WATCHDOG catalogue \citep[see][and references therein]{Tetarenko2016} for the first 31 sources, but we use the updated values of \citet[][]{JMJ2021cygx1} for \cyg. When the exact distance is unknown, \citet[][]{Tetarenko2016} assume 5\,kpc, when the mass is unknown, they assume 8\,$\rm M_{\odot}$, and when the inclination is unknown,  they assume $60^{\circ}$. We indicate these values with italics. MAXI~J1820+070 \citep[][]{Atri2020parallax}, MAXI~J1348-630 \citep[][]{Belloni2020J1348}, MAXI~J1535-571 \citep[][]{Chauhan2019J1535} and MAXI~J1828-249 \citep[][]{Oda2019J1828} were detected after 2016 and are not in the WATCHDOG catalogue. The last column indicates the nature of the companion, where L and H stand for low-mass and high-mass, respectively. 
}\label{table: 35 known sources}

\end{table*}

\section{Neutrino backgrounds}\label{app: neutrino backgrounds}
\subsection{Atmospheric neutrino background}

The spectrum of the lower neutrino energies is dominated by the so-called conventional atmospheric background, which is due to the decay of kaons and charged pions \citep[][]{gaisser2016cosmic}. The best fit of the atmospheric muon neutrino flux is \citep{feyereisen2017one}:
\begin{equation}\label{eq: flux of atmospheric neutrinos}
    \frac{d\Phi_{\nu}}{dE_{\nu}} = 2\times 10^{-14} \left( \frac{E_{\nu}}{10\,\rm{TeV}} \right)^{-3.7}\,\rm{GeV^{-1}\,cm^{-2}\,s^{-1}\, sr^{-1}}
\end{equation}
where the normalization is set by the $10\,$TeV $\nu_{\mu}$ flux. The power-law index changes to 3.9 for neutrino energies higher than $1\,$PeV.

\subsection{Astrophysical neutrino background}
The total all-flavour neutrino and anti-neutrino astrophysical spectrum was studied by the \ic collaboration \citep{aartsen2015combined} using both track-like (muon) and shower-like (electron) events in the TeV-PeV range, between 2009 (with the 59-string configuration) and 2012 (with the full 86-string configuration). It has recently been extended to 7.5\,yr of operation and the unbroken power-law best fit of the flux for the 7.5\,yr sample is \citep{schneider2019characterization}:
\begin{equation}
    \frac{d\Phi_{\nu}}{dE_{\nu}} = 6.45^{+1.46}_{-0.46}\times 10^{-18}\,\left( \frac{E_{\nu}}{100\,\rm{TeV}} \right)^{-\left(2.89^{+0.20}_{-0.19}\right)},
\end{equation}
where the uncertainties correspond to the $1\,\sigma$ confidence interval and the units are the same as equation~\ref{eq: flux of atmospheric neutrinos}.

The total astrophysical muon neutrino and anti-neutrino flux was presented by the \ic collaboration using track-like events in the neutrino energy range of $194\,$TeV and $7.8\,$PeV, between 2009 and 2015 \citep{aartsen2016observation}. It has recently been updated to the almost 10\,yr sample and the best fit flux is \citep{stettner2019measurement}:
\begin{equation}
    \frac{d\Phi_{\nu_{\mu}+\Bar{\nu_{\mu}}}}{dE_{\nu}} = 1.44^{+0.25}_{-0.24}\times 10^{-18}\, \left( \frac{E_{\nu}}{100\,\rm{TeV}} \right)^{-\left( 2.28^{+0.08}_{-0.09} \right)},
\end{equation}
with the same units as equation~\ref{eq: flux of atmospheric neutrinos}.

\subsection{Diffuse neutrino spectrum}
Cosmic rays in the Galactic plane interact with their ambient medium producing diffuse \gr emission \citep{fermilat2012galacticcentre,abramowski2016acceleration}. The \gr spectrum along the Galactic plane \citep{prodanovic2007diffuse,abdo2008measurement} can be explained by a
radial dependence of the diffusion of the CRs \citep{gaggero2015KRAgamma,luque2022galactic}. This model, referred to as KRA$_{\gamma}$, manages to reproduce the \gr spectrum and makes predictions about the diffuse neutrino emission from the Galactic plane \citep{gaggero2015KRAgamma,Albert_2018}. The primary CR distribution is assumed to have an exponential cutoff at a certain energy.

Based on this model, \ic and \ant released the upper limits at the 90 percent confidence level for the all-flavour neutrino flux \citep{ADRIANMARTINEZ2016143,aartsen2017constraints,Albert_2018}. According to \ic, the best fit is achieved with an exponential cutoff at $50\,$PeV (KRA$_{\gamma}^{50}$), whereas a cutoff at $5\,$PeV (KRA$_{\gamma}^5$) is found by \ant. 

\section{Neutrino detector effective areas}\label{app: neutrino effective areas}
\begin{figure}
    \includegraphics[width=1.05\columnwidth]{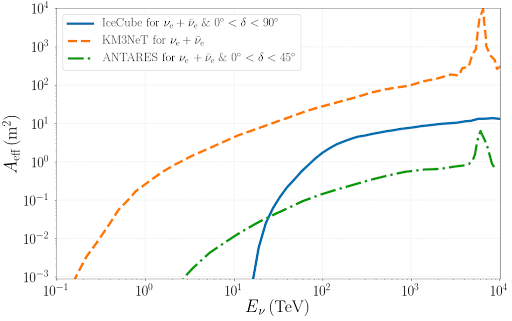}
    \caption{The effective areas of three  neutrino detectors for different declination angles as shown in the legend. 
    }
    \label{app:fig: effective areas electron neutrinos}

\end{figure}

To calculate the neutrino rates, we use the effective areas of \ic , \ant and \arca. These are plotted in Fig.~\ref{app:fig: effective areas electron neutrinos} for different declination angles as indicated in the legend. 

\section{Contribution from individual sources}\label{app: contribution from individual sources}

\begin{figure}
 \includegraphics[width=1.\columnwidth]{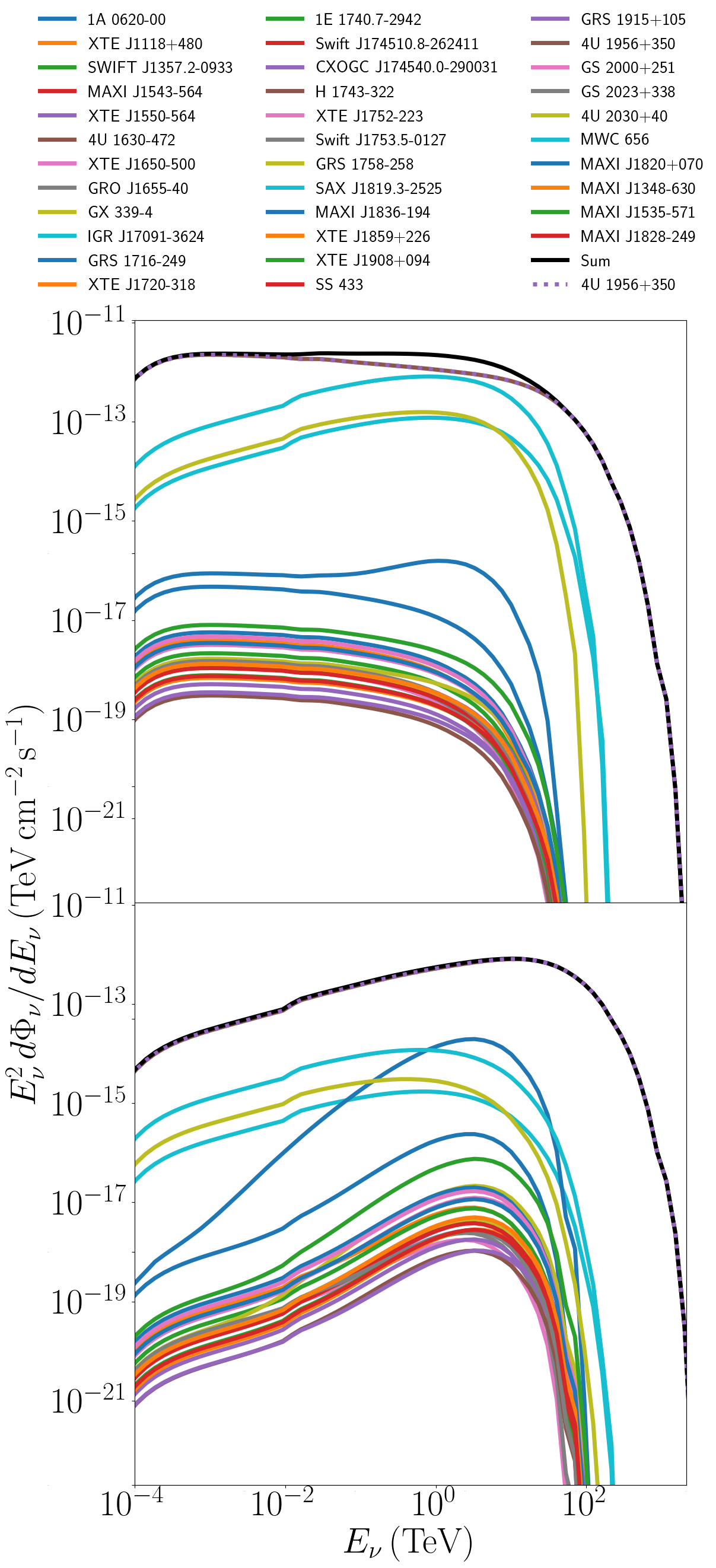}
 
    \caption{Contribution of all known sources to the total intrinsic $\nu_{\mu}+\bar{\nu}_{\mu}$ spectrum, assuming a soft power law of accelerated protons ($p=2.2$) for the top figure and a hard ($p=1.7$) for the bottom figure.
    }
    \label{fig: contribution from all known sources}
\end{figure}

For the soft proton power law (top panel of Fig.~\ref{fig: contribution from all known sources}), we see that the \hms \cyg, MWC~656 \citep[][]{Aleksic2015MagicObservations}, 
and SAX~J1819.3-2525 \citep[][]{Orosz_2001} are the sources that mainly contribute to the total neutrino spectrum. For the hard proton power law (bottom panel of Fig.~\ref{fig: contribution from all known sources}), \cyg dominates the neutrino emission.

\section{Jet model dependence}

\subsection{Neutrino flux per jet zone}\label{app: neutrino flux along the jet}

\begin{figure}
    \includegraphics[width=1.05\columnwidth]{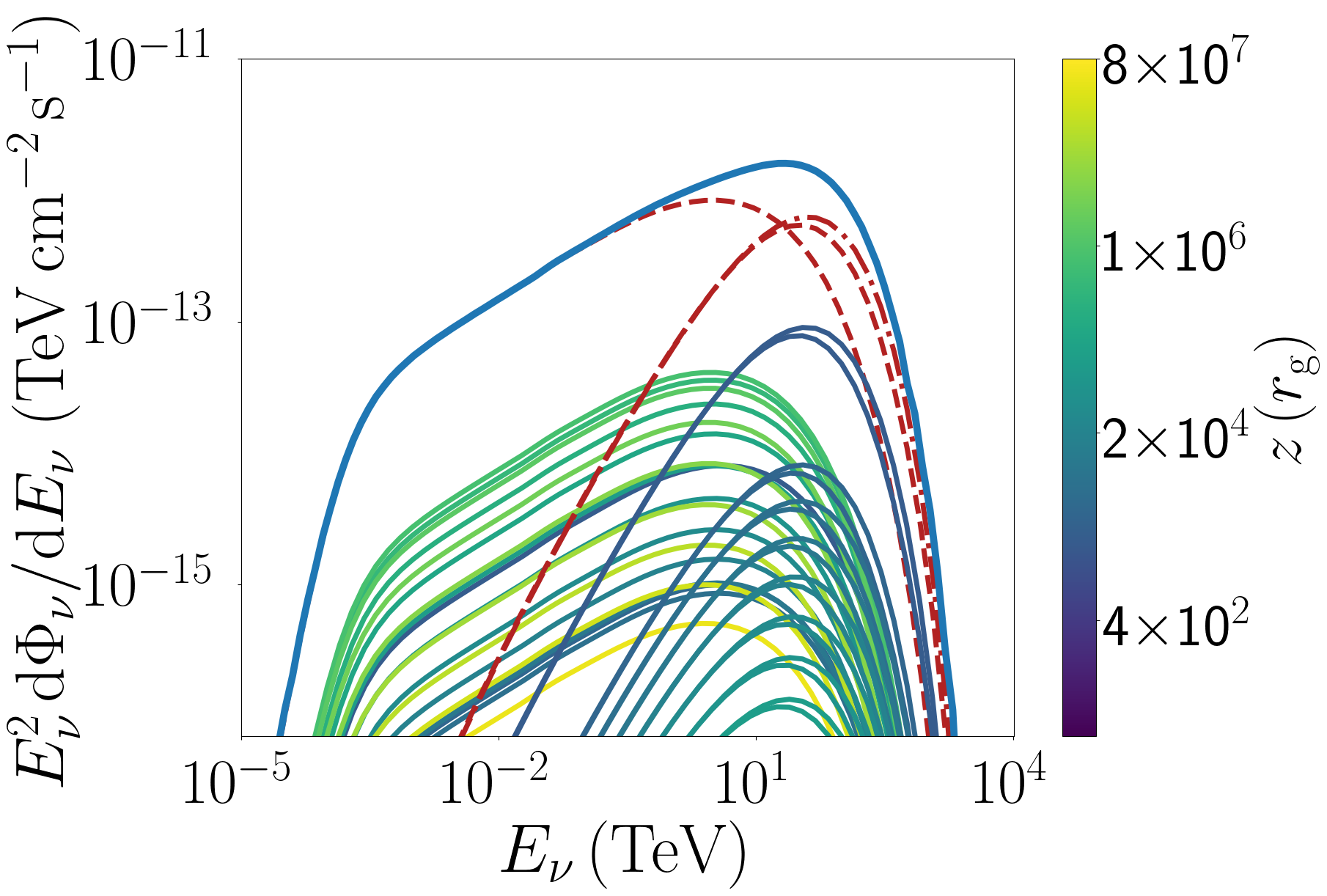}
    \caption{The intrinsic muon neutrino and antineutrino energy flux of the jets of \cyg for a proton power law with index $p=1.7$. The dashed lines are the same as the upper left panel of Fig.~\ref{fig: neutrinos Cyg X-1 with components} and the thin coloured lines correspond to individual jet zones as indicated in the colourmap.
    }
    \label{app:fig: muon neutrino flux of Cyg X-1 with each individual jet zone}

\end{figure}

\begin{figure}
    \includegraphics[width=1.05\columnwidth]{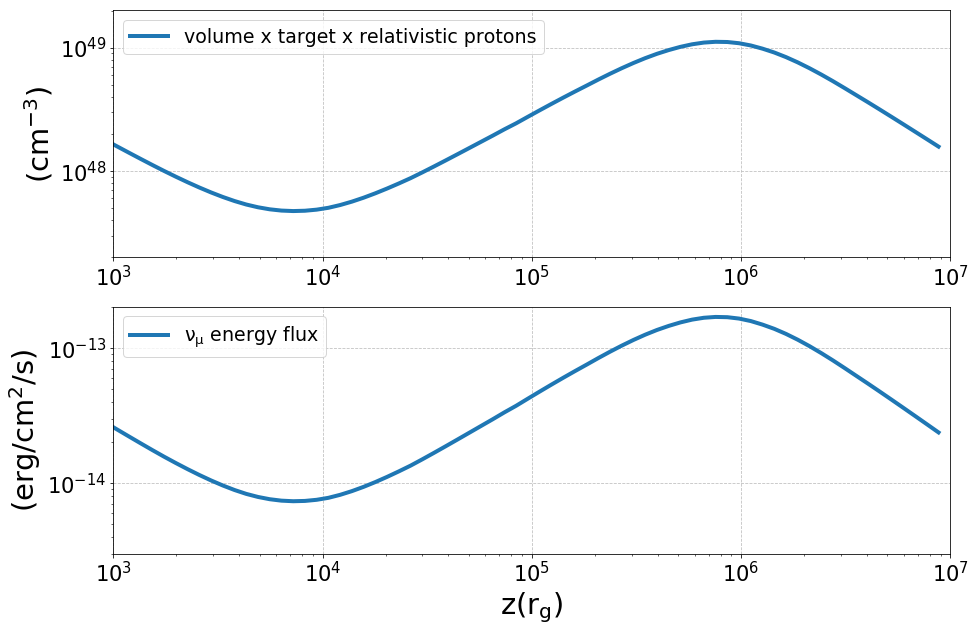}
    \caption{In the upper panel, we plot the number density of the accelerated protons multiplied by the density of the target protons and the volume of each jet segment along the jet. In the lower panel, we show the neutrino flux along the jets of \cyg.
    }
    \label{app:fig: pp neutrino flux along the jet axis}

\end{figure}

In Fig.~\ref{app:fig: muon neutrino flux of Cyg X-1 with each individual jet zone} we plot the total muon neutrino and antineutrino jet flux of \cyg similar to Fig.~\ref{fig: neutrinos Cyg X-1 with components} but we also include the flux of each individual jet segment as indicated by the colourmap. The neutrino flux due to \pg interactions drops along the jet height following the decrease of the density of the photon field that acts as target. The neutrino flux due to pp interactions, on the other hand, drops for the first few segments because the number density of the target protons is dominated by the cold protons of the jets. The neutrino flux increases at some distance of the order of $10^6\rm r_g$ which is approximately the separation distance between the companion star and the compact object because at that distance the target proton density is dominated by the cold protons of the wind of the companion star. In Fig.~\ref{app:fig: pp neutrino flux along the jet axis} we plot the number density of the accelerated protons multiplied by the density of the target protons and the volume. This quantity is proportional to the neutrino flux, which we show in the lower panel of the same figure.

\subsection{Physical processes timescales}\label{app: timescales for Cyg}
\begin{figure}
 \includegraphics[width=1.\columnwidth]{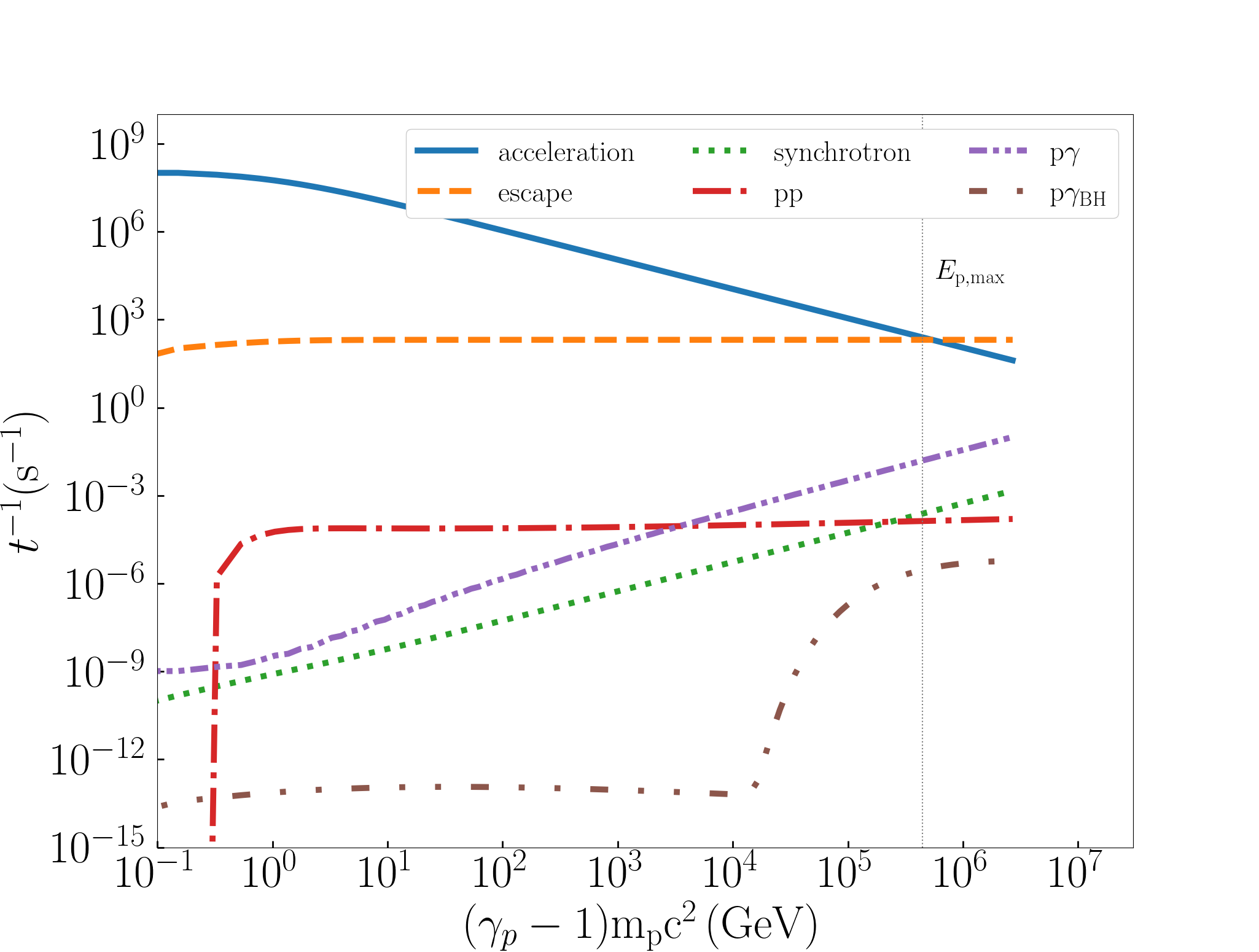}
 
    \caption{The inverse timescale  of the physical processes at the particle acceleration region of \cyg versus the proton kinetic energy. The vertical dashed line shows the maximum proton energy. 
    }
    \label{fig: timescales for Cyg}
\end{figure}

In Fig.~\ref{fig: timescales for Cyg} we plot the inverse timescale of the various physical processes in the jets of \cyg. The interface between the acceleration and the escape gives the maximum attainable energy. The rates of the hadronic process are at least three orders of magnitude below the escape rate, which means that CR protons escape the jets without loosing their energy due to hadronic interactions. 

\subsection{Maximum CR energy from a \hm}\label{app: CR Emax for high-mass bhs}
\begin{figure}
	\includegraphics[width=1.05\columnwidth]{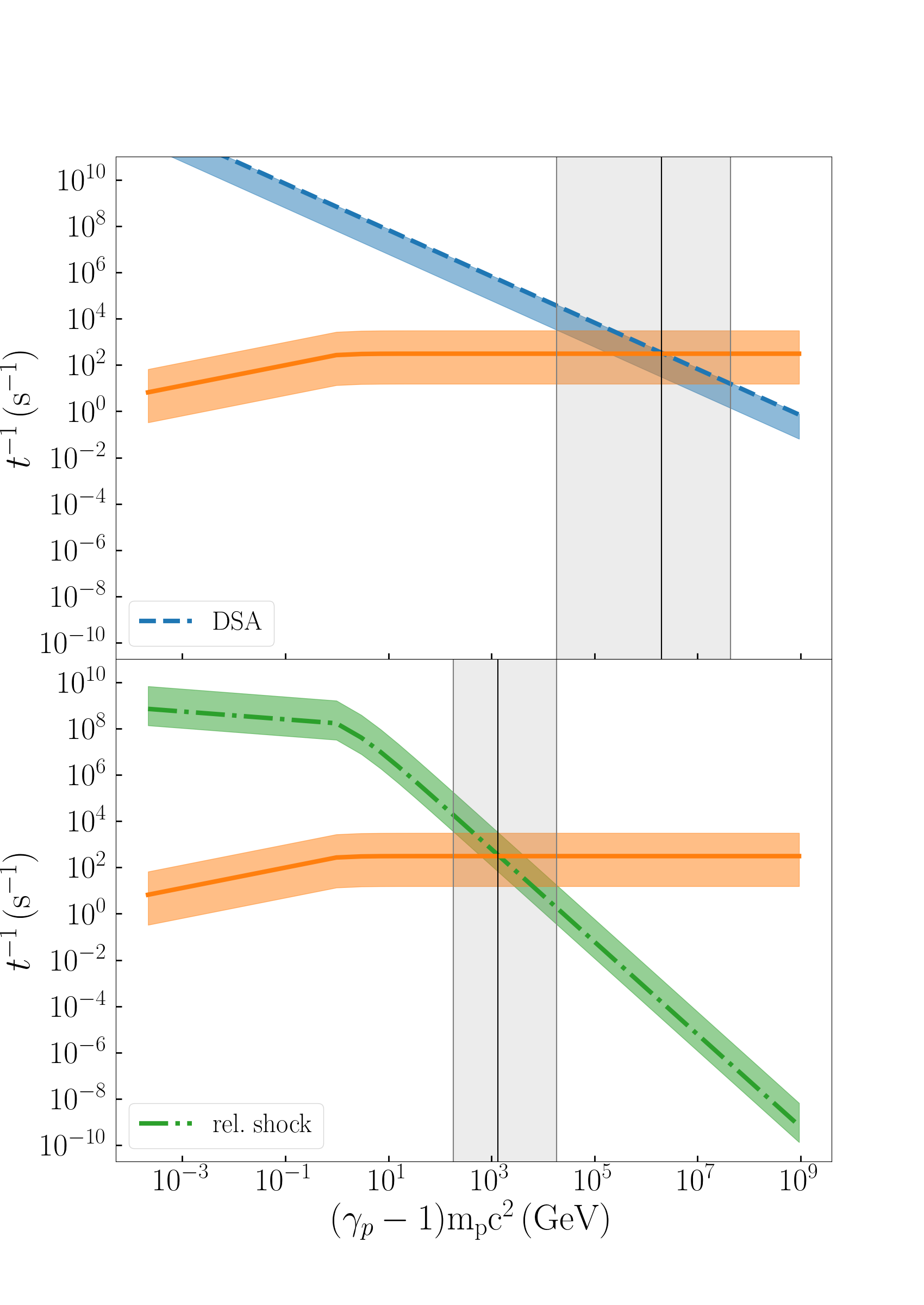}
    \caption{Similar to Fig.~\ref{fig: proton acceleration} but for \hms. Both mechanisms allow for maximum proton energy between 1\,TeV and 10\,PeV.
    }
    \label{app:fig: proton acceleration for high-mass bh}
\end{figure}

In Fig.~\ref{app:fig: proton acceleration for high-mass bh}, we show the inverse of the acceleration and escape timescale similar to Fig.~\ref{fig: proton acceleration} but for the case of a \hm based on \kcyg. For the acceleration due to magnetic reconnection, we use $n_e = 6\times 10^{11}\, \rm cm^{-3}$, which is the particle number density at $s_{\rm diss}$ for \cyg. Overall, the maximum CR energy ranges between 1\,TeV and 10\,PeV. See Section~\ref{sec: discussion} for a more coherent discussion.

\bsp	
\label{lastpage}
\end{document}